\RequirePackage{fix-cm}
\documentclass[pdflatex,sn-nature]{sn-jnl}
\usepackage{graphicx} 
\usepackage{multirow}%
\usepackage{amsmath,amssymb,amsfonts}%
\usepackage{amsthm}%
\usepackage{mathrsfs}%
\DeclareFontFamily{U}{rsfs}{\skewchar\font127 }
\DeclareFontShape{U}{rsfs}{m}{n}{<-6.5>rsfs5<6.5-8>rsfs7<8->rsfs10}{}
\usepackage[title]{appendix}%
\usepackage{xcolor}%
\usepackage{textcomp}%
\usepackage{manyfoot}%
\usepackage{booktabs}%
\usepackage{algorithm}%
\usepackage{algorithmicx}%
\usepackage{algpseudocode}%
\usepackage{listings}%

\usepackage{physics}%
\usepackage{makecell}%
\usepackage{placeins}%
\newtheorem{proposition}{Proposition}
\setcitestyle{super,sort&compress,open={},close={}}

\hypersetup{hidelinks,hypertexnames=false}
\begin{document}
	
	\title{Non-unitary and nonlinear quantum generative large model design and fast large-scale training without back-propagation}
	
	\author[1]{\fnm{Tianze} \sur{Zhu}}\email{tz\_19@mail.ustc.edu.cn}

    \author*[1, 3]{\fnm{Xiaofeng} \sur{Jiang}}\email{jxf@ustc.edu.cn}
	
	\author[1]{\fnm{Min} \sur{Liao}}\email{liaomin@mail.ustc.edu.cn}
	
	\author[2]{\fnm{Luying} \sur{Zhang}}\email{zly2024@mail.ustc.edu.cn}
	
	\author[1]{\fnm{Ruixi} \sur{Liu}}\email{lrx7262728@mail.ustc.edu.cn}

    \author[1]{\fnm{Sen} \sur{Kuang}}\email{skuang@ustc.edu.cn}

    \author[1]{\fnm{Yongzhe} \sur{Lin}}\email{lyz@mail.ustc.edu.cn}

    \author[1]{\fnm{Chenhao} \sur{Zhou}}\email{zhouchenhao@mail.ustc.edu.cn}

    \author[1, 3]{\fnm{Shuangwu} \sur{Chen}}\email{chensw@ustc.edu.cn}

    \author[1, 3]{\fnm{Huasen} \sur{He}}\email{hehuasen@ustc.edu.cn}

    \author[1]{\fnm{Xiaobin} \sur{Tan}}\email{xbtan@ustc.edu.cn}

    \author[1]{\fnm{Quan} \sur{Zheng}}\email{qzheng@ustc.edu.cn}
    
    \author[1, 3]{\fnm{Jian} \sur{Yang}}\email{jianyang@ustc.edu.cn}

	\affil[1]{\orgdiv{Department of Automation}, \orgname{University of Science and Technology of China}, \orgaddress{\city{Hefei}, \postcode{230027}, \state{Anhui}, \country{China}}}
    \affil[2]{\orgdiv{Institute of Advanced Technology}, \orgname{University of Science and Technology of China}, \orgaddress{\city{Hefei}, \postcode{230027}, \state{Anhui}, \country{China}}}

    \affil[3]{\orgdiv{Institute of Artificial Intelligence}, \orgname{Hefei Comprehensive National Science Center}, \orgaddress{\city{Hefei}, \postcode{230026}, \state{Anhui}, \country{China}}}

    \abstract{Transformer-based large models require huge computational and storage costs. Quantum computation is expected to address that due to its potential to speed up classical algorithms exponentially~\cite{bib57, bib58}. Key challenges of implementing quantum large models include the linear, unitary nature of qubit evolutions contrast with the non-unitary matrices and nonlinear functions in classical Transformers, as well as the difficulty of implementing gradient back-propagation on quantum hardware. Here we propose a fully quantum large model containing all the Transformer layers, implementing the non-unitary transformations through SVD-decomposition-based quantum circuits, and approximated nonlinear activation functions through pre-trained parameterized quantum circuits. We also propose a quantum circuit training algorithm based on multi-agent Reinforcement Learning, which can train large-scale parameterized quantum circuits without back-propagation. The dependency between the inference time complexity and the model’s hidden dimension has been reduced from polynomial to logarithmic, showing an exponential advantage. The inference executions on the quantum hardware verify our model’s 300-1000 times speed-up, and the measurements on the sequence-to-sequence tasks shows its learning ability. The training algorithm achieves only 25\% complexity of the best SOTA quantum circuit training algorithms, especially this ratio decreases fast as the parameter scale increases. Our work not only proposes a quantum large model that can significantly reduce the computational cost, but also promotes the quantum computation towards general computing.}
	
	\maketitle
	
	\section*{Main}
	The rapid development and widespread adoption of large models have significantly reshaped multiple industries~\cite{bick2026rapid,maslej2025artificial}. These models, based on the Transformer architecture, have demonstrated remarkable proficiency in tasks across a wide range. They have been integrated into various applications from content generations to advanced agents because of their versatility and scalability.
	
	Nevertheless, this success comes at a huge computational cost resulted directly by the structure of the Transformer architecture. Currently, the ability of the large models rely heavily on the scale of the parameter number~\cite{kaplan2020scaling,hoffmann2022training}, but the crucial components of a Transformer block, the self-attention layer and the feed-forward layer, incur time complexities that scale poorly with the model’s dominate factor of its parameter number, the hidden dimension. As a result, scaling large models leads to dramatically higher computation time, memory demand and energy consumption, bringing a fundamental obstacle to their development.
	
	Quantum computation is expected to address this problem due to its potential to exponentially speed up classical algorithms~\cite{bib57, bib58}. The probabilistic interpretation of quantum mechanics matches the principle of the large models naturally. The words in the sequences are represented by vectors that transform through the model, and the output vectors are interpreted as probability distributions across the dictionary. Besides, implementing quantum large models will significantly develop the universality of quantum computation. Currently existing quantum algorithms are all intended to solve certain problems, while quantum large models have powerful abilities to learn general tasks, promoting quantum computation to generalization.
	
	However, implementing quantum large models meets significant challenges. In the self-attention layer, quantum gates perform unitary matrix transformations, while the attention heads are non-unitary matrices. The weighted aggregation of vectors also cannot be turned into quantum circuits directly~\cite{bib9}, as it conflicts the non-cloning theorem. Meanwhile, nonlinear operations cannot be accurately performed by quantum circuits due to the linear nature of quantum gates. Another challenge is lacking efficient algorithm for quantum circuit training. The backpropagation, widely used for large model training, requires access to all intermediate states throughout the model. However, measuring the qubits halfway collapses their quantum states, disrupting the remaining operations of the model~\cite{bib52}. Two current alternatives is the parameter-shift~\cite{bib16} which only allows for sequential updates of quantum circuit parameters, and SPSA~\cite{spall1992spsa} which has difficulty to convergent near optimal, all highly inefficient for large quantum models.
	
	Recent researches have begun exploring quantum versions of Transformer models, but none have succeeded in constructing a fully quantum model that replicates all crucial components of the Transformer architecture~\cite{bib20}. Some works~\cite{bib21,bib22,bib23,bib24,bib25,bib26} partially convert the Transformer into quantum circuits, while remain other components classical. These works do not speed-up the overall time complexity of the large models dominantly, because the complexity of their models still depends on the classical parts. Other works~\cite{bib27,bib28,bib29,bib30,bib31,bib32,bib33,bib34,bib35} attempt to construct fully quantum models but omit certain parts of the Transformer architecture. The ability of these models is weakened compared to the classical ones, and thus having difficulty replacing the classical models in the coming quantum computation era. A few works~\cite{bib51} provide neither end-to-end implementation nor full execution experiments or simulations of their models.
	
\begin{figure}[htbp]
		\centering
		\includegraphics[width=\textwidth]{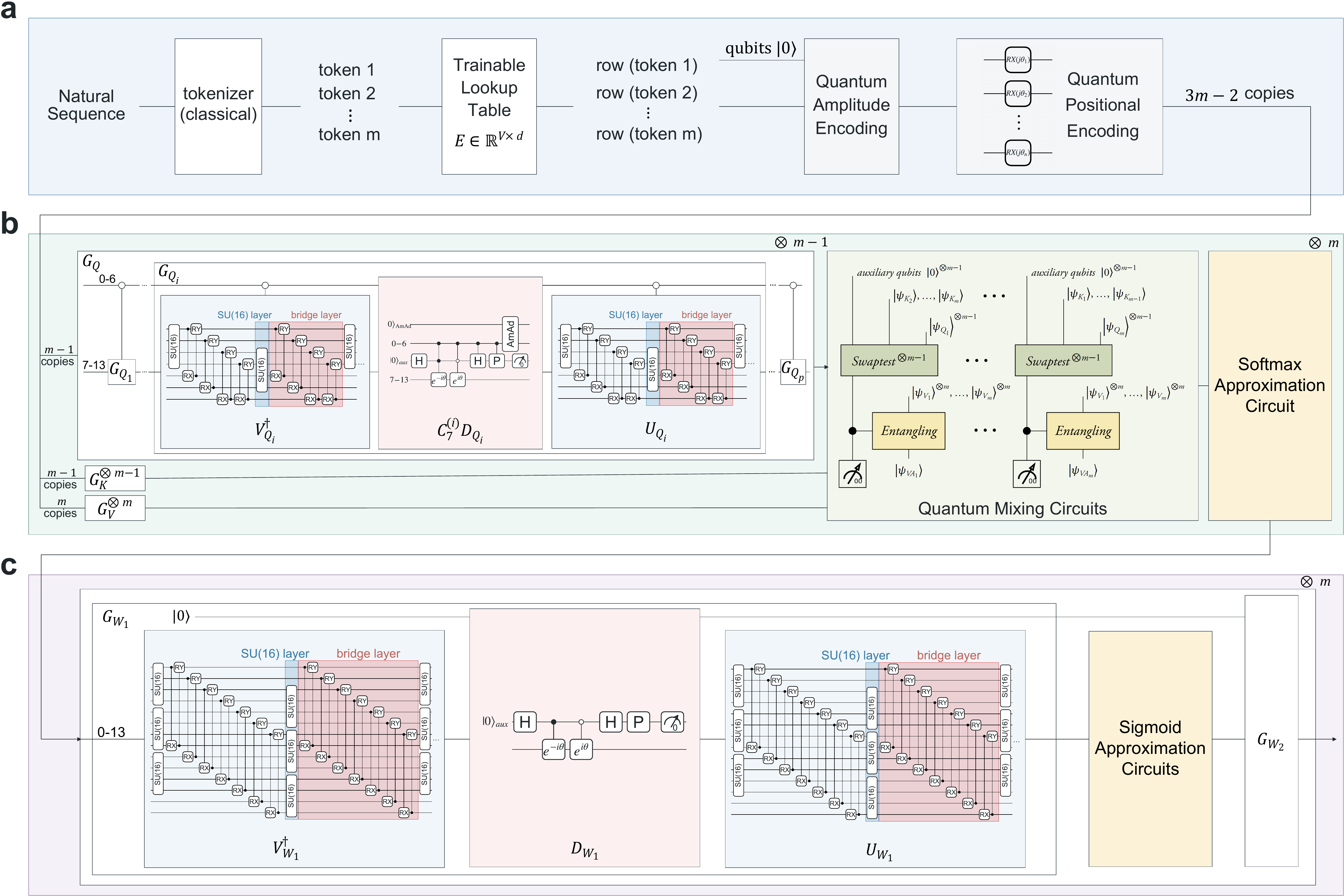}
		\caption{\textbf{Overview of the quantum Transformer architecture.} (a) The encoding layer. A classical tokenizer and trainable embedding table generate token vectors. Amplitude embedding encodes the vectors onto the qubits, and positional encoding supply the positional information. (b) The self-attention layer. $3m-2$ simultaneous attention projection circuits, grouped into $G_Q^{\otimes m-1}$, $G_K^{\otimes m-1}$ and $G_V^{\otimes m}$, processes the qubits, followed by quantum mixing circuits mixing the quantum states of the qubits depending on Q-K fidelity, as well as the Softmax approximation circuits. (c) The feed-forward layer. It uses a quantum circuit structure of SVD decomposition as its linear building blocks. The full layer contains two such circuits with a Sigmoid approximation circuits.}
		\label{fig:architecture}
	\end{figure}

	To construct a large model that can run on quantum hardware and achieve exponential acceleration while maintaining performance comparable to classical large models (especially the GPT series~\cite{bib8}), we present a quantum large model that fully implements the encoding layer, self-attention layer, and feed-forward layer of the Transformer architecture, and reduces the dependence of overall time complexity on hidden dimension from quadratic to logarithmic, as shown in Fig. \ref{fig:architecture}. In addition, we introduce an efficient training algorithm for quantum large models based on reinforcement learning, being able to update parameters simultaneously without difficulty to converge.
	

	\FloatBarrier

    \section{Encoding Layer with Quantum Word Embedding and Quantum Position Encoding}\label{sec:encoding}
	A practical large generative model should be able to process information commonly appears in daily life, such as natural language sentences, images, etc.. We refer to such information as natural sequences. However, neither classical nor quantum computers can directly process natural sequences. Therefore, the first step in large generative models is to convert the input natural sequences into sequences of integers called tokens that computers can handle. This transformation process is known as tokenization. After tokenization, the token sequence is embedded in a matrix $E_B$ with its row number equal to the sequence length $m$ and column number equals to the model dimension $d$. For example, in GPT-3\cite{bib8}, 2048 12288-dimensional selected embedding table rows are stacked in sequence order to form the matrix $E_B$. Another essential encoding technique is positional encoding, which encodes the order of the tokens and also produces a $2048 \times 12288$ matrix $E_p$ for GPT-3, the final output is $E_{O}=E_B+E_P$.

	For a quantum large model, we propose a quantum encoding layer for encoding natural sequences into qubits in Fig. \ref{fig:architecture}a. We use the classic tokenization, followed by the quantum embedding which embed the token sequence onto qubits according to a trainable looking-up table through quantum amplitude encoding, obtaining the qubit sequence $\ket{\psi_{E_{B_1}}}, \ket{\psi_{E_{B_2}}}, \ldots, \ket{\psi_{E_{B_m}}}$. Due to the high information density of the quantum amplitude embedding, we use only $mn=m\log_2d$ qubits to embed the whole sequence, compared with the $md$ elements of $E_B$. For example, quantum version of GPT-3 only uses $mn=2048\times14$ qubits. Next, we perform positional encoding. Instead of directly replicating the process of GPT-3, we implement it through positional encoding circuits. Specifically, for the $j$-th token position ($1\leq j\leq m$), we apply $R(j\theta_k)$ to the $k$-th qubit ($1\leq k\leq n$). This approach avoids mixing unknown quantum states which is challenging to implement. The output qubits of the positional encoding form the output of the encoding layer, denoted by $\ket{\psi_{E_1}},\ldots,\ket{\psi_{E_m}}$. The processes will repeat for $3m-2$ times simultaneously for each input sequence because of the non-cloning theorem.

	\section{Multi-head Quantum Self-attention Layer with Non-unitary Attention, Fidelity-guided Superposition, and Nonlinear Transformation}\label{sec:attention}
	One of the main reasons why Transformer large generative models are so powerful is the self-attention mechanism\cite{vaswani2017attention}. It allows the model to analyze the relationships between the words in a sequence, thus enhancing its ability to process long sequences effectively. Its core principle is to evaluate the correlation between the tokens within a sequence, enabling a more efficient understanding of the sequence's meaning. In GPT-3, $E_O$ from the encoding layer is multiplied by attention head matrices $W_{Q_i}$, $W_{K_i}$, and $W_{V_i}$, $1\leq i\leq96$, producing $Q_i$, $K_i$ and $V_i$. Attention head matrices serves as a guideline for evaluating the correlations between different token in the sequence. The rows of $V_i$ are then mixed with the softmaxed value of dot products of the row vectors from $Q_i$ and $K_i$, and the final output matrix $W_O$ is the concatenation row-mixed $V_i$.

	The self-attention layer heavily relies on large scale matrix computations which are computationally expensive, but quantum computing offers a significant advantage, as quantum gates perform unitary transformations on the quantum states of qubits. However, implementing a quantum circuit for the self-attention mechanism faces challenges, mainly due to the non-unitary nature of the self-attention heads, the process of mixing the unknown state vector with arbitrary proportions, and the nonlinear softmax function. We propose a quantum approach to the multi-head self-attention layer that addresses these challenges and provides an exponential speed-up compared to the classical method in Fig. \ref{fig:architecture}b. The qubits outputted by the encoding layer are grouped and pass the attention head circuits respectively to form the query states $\ket{\psi_{Q_j}}$, the key states $\ket{\psi_{K_j}}$ and the value states $\ket{\psi_{V_j}}$, $1\leq j\leq m$. The 3 groups of states then pass the mixing circuits where the value states are mixed according to the inner products of the query and key states. Finally, the mixed value states pass the non-linear softmax circuit to form the output states of the self-attention layer.

	\subsection{Multiple quantum attention heads with arbitrary non-unitary attention matrices}
    \label{sec:heads}
	There is in general no restriction for the classical attention head matrices, but ordinary quantum circuits are constrained by the fundamental principles of quantum mechanics, which allow only unitary evolutions. This restriction will limit the flexibility of the attention heads and weaken the self-attention mechanism. Here we present a scheme realizing non-unitary multiple attention heads by quantum circuits. 
    
    The $3m-2$ copies of the encoded qubits are divided into 3 groups, then pass through the attention projection circuits $G_Q$, $G_K$, and $G_V$, each performs a block-diagonal matrix transformation with each block representing an attention head. For example, $G_Q=\text{diag}(G_{Q_1},G_{Q_2},\ldots,G_{Q_p})$, so as $G_K$ and $G_V$. To address the unitary limitation, for each attention head circuit $G_{Q_i}$, we employ singular value decomposition (SVD) to decompose a non-unitary matrix into a product of 2 unitary matrices and a non-unitary diagonal matrix, then implementing them separately. 
    
	For the non-unitary diagonal part, we use the probabilistic circuit by Zylberman et al.~\cite{bib10} consuming 1 ancillary qubit, with the average success probability 0.5. To prevent resource wasting when failures occur, we design a repeat-until-success (RUS) strategy by the actual transformation in fail cases and the analysis of the relationship between it and the target transformation. By the RUS protocol, the target transformation can be recovered by the exactly same circuit with a different parameter set iterated from the previous one. On average, two executions are required to obtain a success.

	For the unitary parts, the main challenge is to accommodate as much as trainable parameters while control the circuit depth. Consequently, the quantum gates in the circuit should be as parallel as possible, and the parameter redundancy should be carefully avoided. To maximize the quantum gates parallelism, we propose a quantum circuit structure that use 4-qubit general circuits~\cite{bib36} ($SU(16)$ circuits) as components and arrange them as a layered pattern, inspired by the idea of modular designing of the classical integrated circuits. However, sequential $SU(16)$ circuits with shared operated qubits introduce parameter redundancy, caused by the sequential quantum gates in the same kind. To avoid that, we design bridge layers placed between the $SU(16)$ circuit layers. Each bridge layer consists of a sequence of controlled parameterized rotation gates across 6 qubits. By randomly sampled Jacobian decomposition~\cite{funcke2021expressivity}, we proved that the bridge layer effectively eliminate all parameter redundancy.

	To construct the block-diagonal $G_Q$, $G_K$, and $G_V$, we arrange the controlled SVD attention head circuits as a sequence. For example, in the quantum version of GPT-3, 128 7-qubit circuits $G_{Q_1},\ldots,G_{Q_{128}}$ are controlled by another 7 qubits with the controlled logic $C_7^{(1)}=\text{Bin}(0)$ to $C_7^{(128)}=\text{Bin}(127)$, forming the circuit  $G_Q=C_7^{(1)}G_{Q_1}C_7^{(2)}G_{Q_2}\ldots C_7^{(128)}G_{Q_{128}}$. So as $G_K$ and $G_V$. Because of the existence of post-selection in the non-unitary diagonal circuits, the implementation of controlled $G_{Q_i}$ is not trivial. We designed a controlled non-unitary diagonal circuit to address the implementation difficulty. $G_Q$, $G_K$, and $G_V$ output $m-1$ copies of query states $\ket{\psi_{Q_j}}$, key states $\ket{\psi_{K_j}}$, and $m$ copies of value states $\ket{\psi_{V_j}}$, $1\leq j\leq m$.

	\subsection{Superposition of the attention states by quantum mixing circuits}
    \label{sec:mixing}
	The mixing strategy used in classical Transformers cannot be directly implemented in quantum computation~\cite{bib9}, as it conflicts the non-cloning theorem. To address this, we designed a quantum mixing method based on the SwapTest circuit\cite{bib12,bib13} and the Strongly Entangling circuit\cite{bib11}. In classical Transformer, the query-key dot products act as the aggregation weights. Here, we use the query-key fidelity as the mixing guidance. For each token position $j$, every copy of $\ket{\psi_{Q_j}}$ passes through a SwapTest circuit with $\ket{\psi_{K_\ell}}$. The query-key fidelity is encoded in the SwapTest ancillary qubit. Imitating the weighted aggregation of classical Transformer, the corresponding value states then pass through Strongly Entangling circuits controlled by these auxiliary qubits, mixing the value states according to the query-key fidelity. After entanglement, the auxiliary qubits are post-selected by the outcome $0$. Performing all SwapTests and entangling operations simultaneously requires $(m-1)^2$ auxiliary qubits. Since measured auxiliary qubits can be recycled, this cost can be reduced by grouping the operations. We denote the resulting mixed value states by $\ket{\psi_{VA_1}},\ket{\psi_{VA_2}},\ldots,\ket{\psi_{VA_m}}$.

	\subsection{Approximated nonlinear transformation by pre-trained quantum circuits}
    \label{sec:activations}
	Classical Transformer uses Softmax function to slightly reshape the value state aggregations. Due to its nonlinear nature, implementing it directly on quantum states is challenging. To address this, we design a quantum circuit capable of approximating the softmax function using one auxiliary qubit. We put 2 unitary circuit that also consists $SU(16)$ circuits, first on the data qubits, second on the ancillary qubit as well as the first $n-1$ data qubits. 2 unitary circuits are isolated by a bridge layer. The ancillary qubit is measured finally, with post-selecting 0. The softmax circuit must be trained independently before being incorporated into the full quantum model. For training, we input random quantum states to the circuit, use the Softmax result of the states as target states, and use the fidelity between the target states and the output states as the metric need to be maximized. The final fidelity of the circuit after training can reach about 0.98.

	\FloatBarrier
	\section{The Feed-forward Layer with Non-unitary Neural Layers and Nonlinear Activation}\label{sec:ffn}
	In classical Transformer, the feed-forward layer is a fully connected neural network with 2 layers. In GPT-3, The number of neurons in both the input and output layers matches the model dimension $d=12288$, while the hidden layer contains 49152. Its function can be expressed as $FFN(x)=\phi(xW_1+b_1)W_2+b_2$ where $W_1$, $W_2$ and $b_1$, $b_2$ are trainable. While its structure is relatively simple, it actually contains most of the trainable parameters in the model. Since large-scale neural networks are computationally expensive, and quantumn circuits can effectively perform these matrix transformations as well as approximating the activation function, here we present the equivalent quantum version of the feed-forward layer of GPT-3 in Fig. \ref{fig:architecture}c. The layer consists 2 circuits $G_{W_1}$ and $G_{W_2}$ with an SVD structure, corresponding to the matrices $W_1$ and $W_2$, and a pre-trained Sigmoid approximation circuit, similar with the Softmax circuit, between them. 
	\FloatBarrier

	\section{Training Algorithm based on Multi-agent Reinforcement Learning}\label{sec:training}
	GPT-3 uses a modified version of the Common Crawl dataset, combined with some other texts from daily cases for training\cite{bib8}. Due to the auto-regressive characteristic of GPT-3, the sequence cross-entropy loss $L_t=-\frac{1}{m}\sum\limits_{t=1}^m\sum\limits_{i=1}^{V}y_{ti}\log_2\hat{y}_{ti}$ is chosen as the cost function, where $t$, $V$, $\boldsymbol{y}_t$ and $\hat{\boldsymbol{y}}_t$ are the token position, vocabulary size, the target probability distribution and the output probability distribution from the model. The parameter update of the GPT-3 model uses the widely-used back-propagation algorithm with an Adam optimizer.

	We use the sequence fidelity $F_Y=\frac{1}{m} \sum\limits_{k=1}^{m}|\langle y_k\vert\hat{y}_k\rangle|^2=\frac{1}{m} \sum\limits_{k=1}^{m}\operatorname{Tr}(\rho_k\hat{\rho}_k)$ as the cost function, which has similar mathematical properties with the cross-entropy. However, the backpropagation algorithm cannot be directly implemented on quantum hardware in a reasonable time and resource consuming for a large model~\cite{bib52} because of its requirement to access all the intermediate quantum states, which is impossible without destroying the quantum state and interrupting the execution of the circuit. Therefore, gradients of parameters in quantum circuits must be estimated without relying on intermediate states.

	\subsection{Reinforcement Learning Training mechanism without gradient back-propagation}
	To address the problem, we formulate the parameter adjustment as a Reinforcement Learning problem~\cite{bib37}. The state is defined as $S_t=\left[\sin\boldsymbol{\theta}_t,\cos\boldsymbol{\theta}_t,\boldsymbol{d}^{\mathrm{fresh}}_t,\boldsymbol{d}^{\mathrm{mom}}_t,\boldsymbol{z}_t\right]$, where $\boldsymbol{\theta}_t$ is the current quantum circuit parameters, 
	 $\boldsymbol{d}^{\mathrm{fresh}}_t$ is the gradient direction estimated by simultaneous perturbation, $\boldsymbol{d}^{\mathrm{mom}}_t$ is the processed $\boldsymbol{d}^{\mathrm{fresh}}_t$ by Adam optimizer. $\boldsymbol{z}_t$ contains the current fidelity, step scale, previous action and reward. The action set is $\mathcal{A}_t=\left\{(\boldsymbol{\kappa},s)| \boldsymbol{\kappa}\in\{\boldsymbol{0},\boldsymbol{d}^{\mathrm{fresh}}_t,\boldsymbol{d}^{\mathrm{mom}}_t\}, s\in\{0.25,0.5,1,2\}\right\}$. For an action $a_t=(\boldsymbol{\kappa}(a_t),s(a_t))\in\mathcal{A}_t$, $\boldsymbol{\kappa}({a_t})$ denotes the selected direction and $s({a_t})$ denotes its step multiplier. The parameters are updated by $\boldsymbol{\theta}'_t(a_t)=\operatorname{wrap}_{[-\pi,\pi)}\!\left(\boldsymbol{\theta}_t+\alpha_t s(a_t)\boldsymbol{\kappa}(a_t)\right)$ if $\boldsymbol{\theta}'_t(a_t)$ increases the fidelity, otherwise the current parameters are kept. The single step reward is $r(a_t)=\tanh\!\left[c_R\dfrac{F_{Y}(\boldsymbol{\theta}'_t(a_t))-F_{Y}(\boldsymbol{\theta}_t)}{\alpha_t}\right]$ with idle penalty in low fidelity. The learning objective is maximizing $J(\pi)=\mathbb{E}_{\pi}\!\left[\sum_{t=0}^{+\infty}\gamma^tr(a_t)\right]$ where $\gamma$ is the discount factor. A dueling DQN\cite{wang2016dueling} is used to find the best parameter tuning policy.

	\subsection{Multi-agent expansion for large scale circuits}
	To scale this formulation to circuits with large parameter sets, we use SPSA to generate rough tuned parameters with moderate fidelity, then partition the parameter indices into $M$ disjoint blocks $\mathcal{B}_1,\ldots,\mathcal{B}_M$, and assign one independent dueling DQN to each block, in each iteration, only the parameters of the rotated block are updated. The complete process forms the Multi-Agent Reinforcement Learning for Quantum Circuit Training (MARL-QCT) algorithm. See Chapter \ref{methods:MARL-QCT1}, \ref{methods:MARL-QCT2} and Algorithm \ref{methods:attention-head} in Methods for detail.

	\section{Time Complexity Analysis}\label{sec:complexity}
    We examined our quantum large model's time complexity and compare it with the classical GPT model. The time complexity of our model is subject to the quantum circuit depth. When computing the circuit depth, we first decompose our circuit into the universal quantum gate set plus the controlled rotation gates, as they are explicitly marked as supported gates by the quantum chips of Tianyan platform~\cite{bib50}.

    The quantum circuit depth of the encoding layer of our model is subject to the specific quantum hardware. In worse cases, it can result into a time complexity of $O(m + (\log_2{d})^2)$. The depth of the self-attention layer and the feed-forward layer is $\left[m+2(n_{SUa}+3)p+2n_{SUs}-1\right]n+3m+2(1-n_{SUa})p\log_2 p+\left[\left(253-(-1)^{n-\log_2 p}\right)n_{SUa}+(-1)^{n-\log_2 p}+23\right]p+\left(251+(-1)^n\right)n_{SUs}+\frac{3}{2}\left(1-(-1)^n\right), n>4,4<n-\log_2p<8$ and $(4n_{SUf}+4n_{SUs}+2)n+2\left(251+(-1)^n\right)(n_{SUf}+n_{SUs})+31-5(-1)^n,n>4$, resulting into the time complexity of  $O((m+p)\log_2d-p\log_2p)$ and $O(\log_2d)$ respectively. In comparison, the time complexity of the 3 layers of a classical transformer is $O(md)$, $O(m^2 d)$ and $O(md^2)$. Our quantum large model thus showing an exponential time complexity reduction in the dependency of the model's hidden dimension $d$, which is the dominate factor of both the classical and the quantum models' parameter number.

    \section{Experiments}\label{sec:experiments}

    \subsection{Inference execution on quantum hardware}
    We executed our quantum large model on the quantum computer Tianyan-287 of Tianyan Quantum Computing Platform~\cite{bib50} and tested the inference time consuming. It is the most advanced quantum chip in China, containing the largest amount of data qubit number (105 qubits) among the universal quantum chips. We executed a quantum large model with the largest scale capable with Tianyan-287: 4 qubits per token with sequence length $m=3$, consuming 92 qubits. We collected 2 types of data: the quantum circuit depth and the quantum circuit execution time on Tianyan-287, as shown in Tab. \ref{tab_tianyan_time}. We also timed the self-attention layer and the FFN layer of a classical Transformer with exactly the same scale on a classical server for comparison.

    \begin{table}[htbp]
		\centering
		\begin{tabular}{lllll}
			\toprule
			Layer & \multicolumn{2}{c}{Theoretical calculation number} & \multicolumn{2}{c}{Measured execution time}\\ 
             & classical & quantum & classical & quantum\\
            \midrule
			Self-Attention & 206208 & 315 & 0.321 & 0.452\\
			FFN & 385536 & 1044 & 0.183 & 0.303\\
			\bottomrule
		\end{tabular}
		\caption{The time complexity and time consuming of the implemented quantum large model on the Tianyan-287 quantum chip, with the scale of 4 qubits per token (corresponding to hidden dimension $d=16$) and sequence length $m=3$, compared with the classical Transformer with the same scale.}
		\label{tab_tianyan_time}
	\end{table}

    The quantum circuit depth of the self-attention layer and the FFN layer of our model after scale reduction is 315 and 1044 respectively. In comparison, the corresponding calculation number in the classical layers are 206208 and 385536, excluding simultaneous calculations. The theoretical inference speed of our model is thus 300 to 1000 times faster than the classical under in this scale. The measured inference time of the 2 models are similar, mainly due to that the time length of a single quantum gate on Tianyan-287 varies from 28ns to 60ns, which is about $10^3$ times of those logic gates in modern classical integrated circuits.

    \subsection{Training Quantum Circuits with MARL-QCT}
    To evaluate the proposed training approach and its scalability, we apply the MARL-QCT algorithm to some important component circuits in the self-attention layer and the FFN layer, and compared the performance with 2 SOTA algorithms, SPSA~\cite{spall1992spsa} and parameter-shift~\cite{bib16}. Specifically, we tuned the parameters of an $\mathrm{SU}(16)$ circuit and a $G_Q$ circuit with qubit and attention head number $n=6,p=4$, let the circuits approximate given transformations. We compared 2 matrices: the final fidelity the algorithms reach by a fixed tuning iteration number, and the circuit calling numbers the algorithms need to reach a given target fidelity. These experiments verified the scaling ability and the distinct advantage of MARL-QCT algorithm we proposed compared with the SOTA algorithms. 

    \begin{figure}[htbp]
		\centering
		\includegraphics[width=\linewidth]{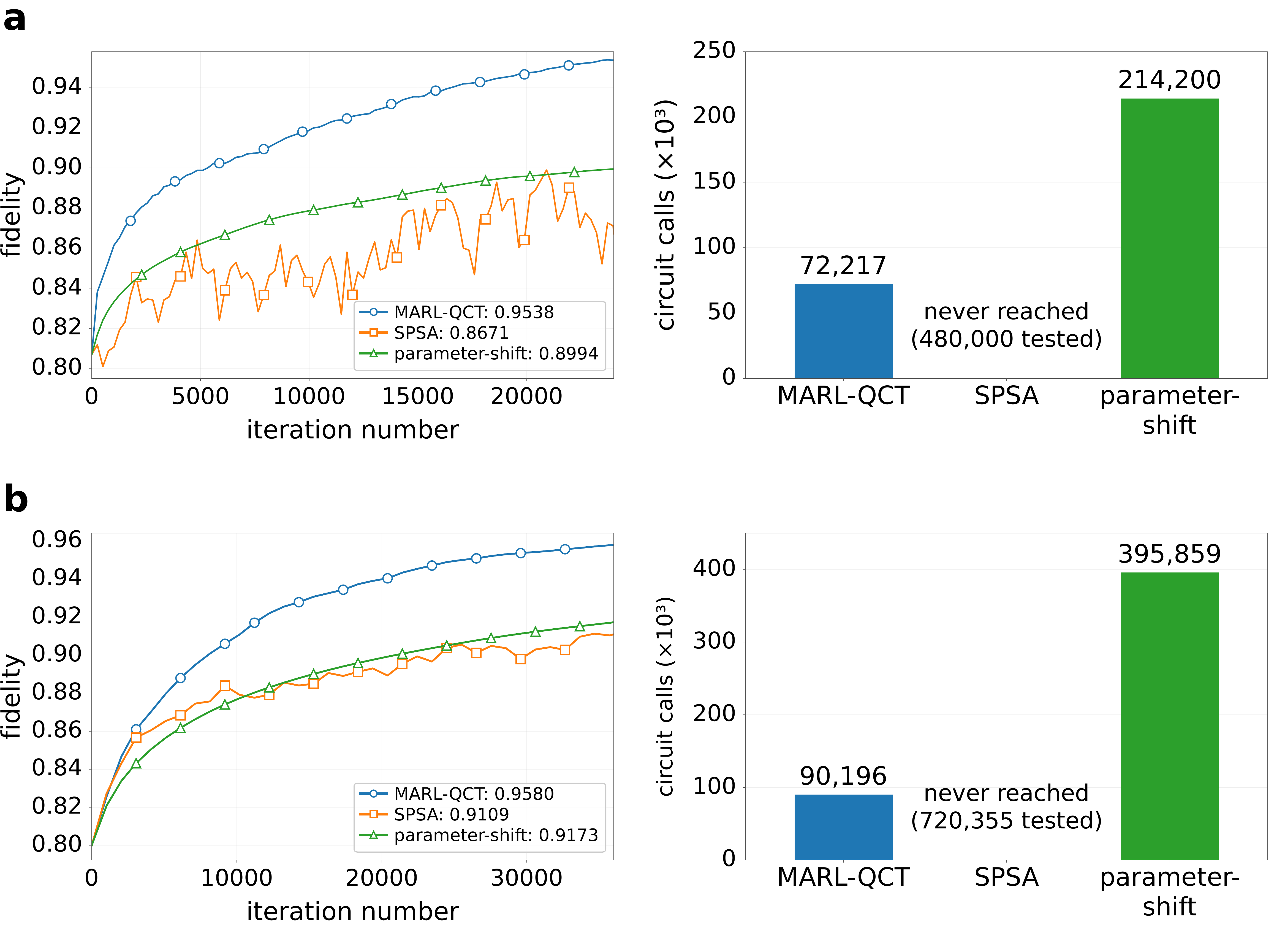}
		\caption{\textbf{Comparisons of the performance of MARL-QCT, SPSA and parameter-shift on quantum circuit parameter tuning}. The experiment results show that MARL-QCT is able to reach a higher fidelity with sufficiently lower time consuming, establishing a distinct advantage compared to the SOTA algorithm. (a) Results for the $\mathrm{SU}(16)$ circuit task, tuning 255 parameters. Starting from the rough tuned parameters with fidelity 0.806949, MARL-QCT reaches the fidelity 0.953760 in 24000 iterations, while SPSA and parameter-shift reach 0.867119 and 0.899941. MARL-QCT thus gives a 143.99\% and 57.87\% more fidelity improvement. Meanwhile, MARL-QCT takes only 33.71\% circuit calls of the parameter-shift to reach the target fidelity 0.95, while SPSA cannot reach the target fidelity because of the gradient estimation noise. (b) Results for the $G_Q$ circuit task, tuning 1020 parameters. Starting from the rough tuned parameters with fidelity 0.800053, MARL-QCT can reach the fidelity 0.957973 in 36000 iterations, while SPSA and parameter-shift reach 0.910945 and 0.917277. MARL-QCT thus gives a 42.41\% and 34.72\% more fidelity improvement. Meanwhile, MARL-QCT takes only 22.78\% circuit callings of the parameter-shift to reach the target fidelity 0.95, while SPSA still cannot reach the target fidelity.}
		\label{fig_training_comparison}
	\end{figure}

    As shown in Fig. \ref{fig_training_comparison}, after training, MARL-QCT managed to reach the fidelity of 0.953 in 24000 iterations starting from rough tuned parameters in the $SU(16)$ task, making 143.99\% and 57.87\% more fidelity improvement compared with SPSA and parameter-shift in the same iteration number. Meanwhile, to reach the same fidelity 0.95, MARL-QCT only takes 33.71\% circuit callings of the parameter-shift, while SPSA never reached this fidelity throughout the experiments. These results confirmed the advantage of MARL-QCT compared to the SOTA algorithms, achieving higher fidelity than SPSA while significantly reducing the high circuit calling time of parameter-shift caused by sequential parameter gradient estimation. In the $G_Q$ task, MARL-QCT reached the fidelity of 0.956 in 36000 iterations, making 42.41\% and 34.72\% more fidelity improvement, and reached the fidelity of 0.95 by only 22.78\% circuit callings of the parameter-shift. These results show the excellent scaling ability of our algorithm: while the growth of the parameter number scale slows down all the algorithms, MARL-QCT propose stronger advantage on quickly achieving a target fidelity than parameter-shift.

    \subsection{Sequence-to-Sequence Task Tests}

\begin{figure}[htbp]
		\centering
		\includegraphics[width=\linewidth]{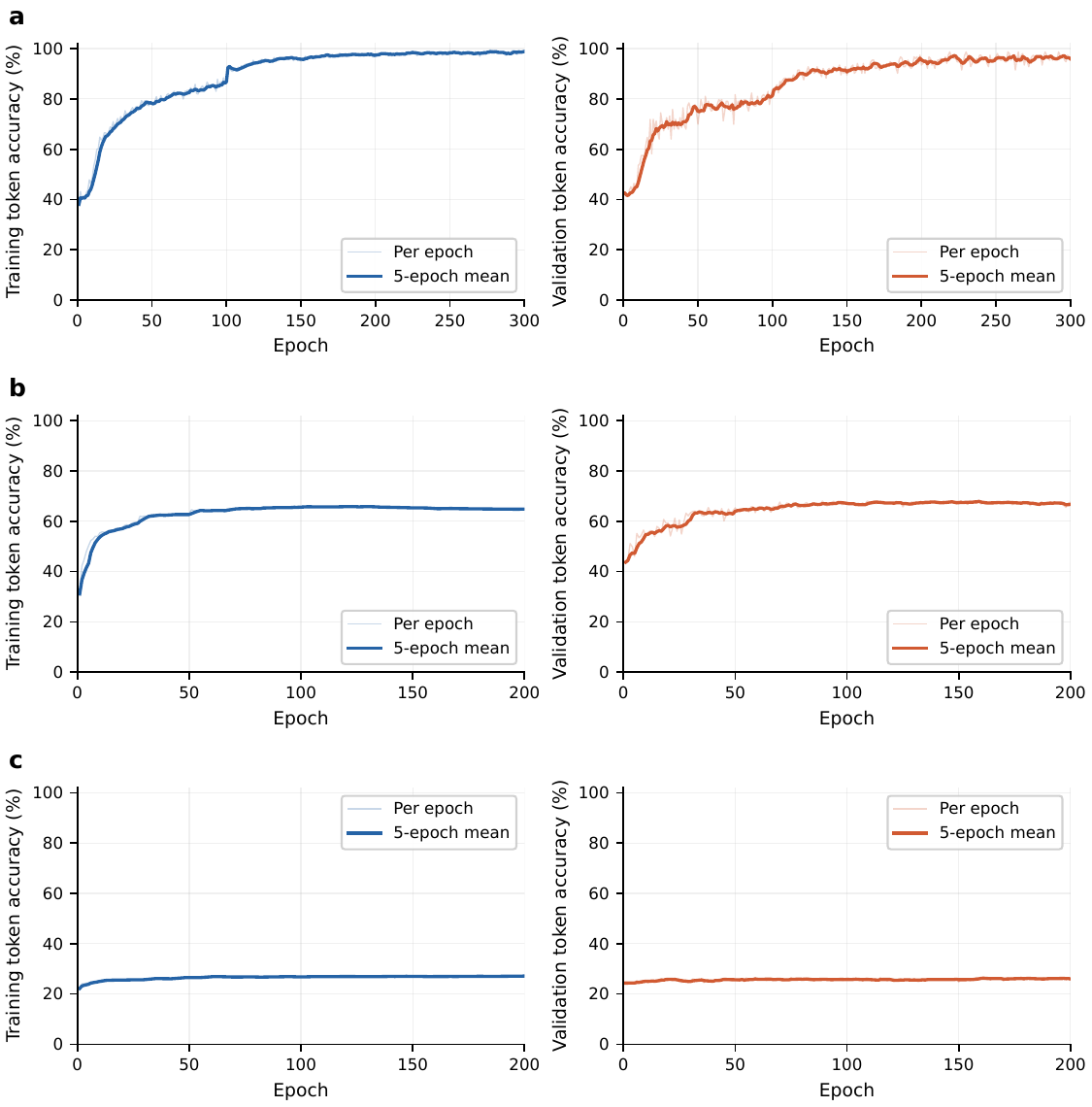}
		\caption{\textbf{Token accuracy for Dyck, SCAN-length and PCFG task}. Figure a, b and c show Dyck-4, SCAN-length and PCFG, respectively, with training on the left and validation on the right. The validation token accuracy for the 3 tasks reaches 96.69\%, 67.56\% and 25.76\% respectively. The results shows that our quantum large model under the tested scale can learn the Dyck task effectively, sometimes produce error when encountering complex input sequence analysis in SCAN-length task, while the accuracy is limited by the small model scale in PCFG.}
		\label{fig_sequence_task_results}
	\end{figure}

    To verify the expressivity and generalization ability, we test the proposed quantum large model with 3 qubits per token and sequence length $m=6$ on the Dyck-4 Balanced Brackets task, the SCAN-length task and the Probabilistic Context-Free Grammar (PCFG) task. The model uses one self-attention layer with 2 heads, one FFN layer and three entangling layers, with 877 trainable parameters. As shown in Fig.~\ref{fig_sequence_task_results}(a), with training on the left and validation on the right, our quantum large model obtains 96.69\% validation token accuracy in Dyck-4, showing that the model can learn to track nested dependencies, and remembering the incomplete sequence as a stack. As shown in Fig.~\ref{fig_sequence_task_results}(b), our model reaches 67.56\% validation token accuracy in SCAN-length, which is lower than Dyck-4 since SCAN-length requires quantity analysis and re-ordering. As shown in Fig.~\ref{fig_sequence_task_results}(c), our model can reach 25.76\% validation token accuracy in PCFG task, which is probably resulted by the relatively high difficulty of the task and the small scale of the trainable parameter number.

	\FloatBarrier

    \section{Conclusion}
    We proposed a quantum large model based on the Transformer architecture and a training algorithm for training large-scale quantum circuits. The problem of the unitary constraint and the difficulty of performing nonlinear transformations of quantum circuits have been effectively addressed, and the exclusiveness of high fidelity and low time consuming in SOTA quantum circuit algorithms has been canceled by MARL-QCT, our training algorithm.
    
    Furthermore, our work makes contributions to promote quantum computation toward generalization. Existing quantum algorithms are all designed for solving specific problems, but large models can learn knowledges from training, and generalize it through task executions. Therefore, the quantum large model we proposed can serve as a general quantum algorithm, showing the evidence that quantum computation can achieve exponential speed advantage compared with classical computation in general tasks.
    
    Some challenges remain. Deploying large-scale quantum algorithms on recent NISQ quantum hardware faces severe difficulty. As a result, executing our full-scale model as well as MARL-QCT directly on quantum hardware remain as future works. With the development of large-scale quantum devices with strong error-correction ability, these challenges will eventually disappear, and we will finally see the significant advantages of the quantum large model as well as the generalization of quantum computation.

\clearpage

\renewcommand{\refname}{References}


\clearpage
\section{Methods}\label{methods:sec3}

	\subsection{Encoding Layer with Quantum Word Embedding and Quantum Position Encoding}\label{methods:subsec2}
	A practical large generative model should be able to process information commonly appears in daily life, such as natural language sentences, images, and other forms of data. We refer to such information as natural sequences. However, neither classical nor quantum computers can directly process natural sequences. Therefore, the first step in large generative models is to convert the input natural sequences into sequences of integers that computers can handle. This transformation process is known as tokenization. 

	A widely used tokenization method is byte-level Byte Pair Encoding (BPE)\cite{bib7,radford2019gpt2}, which processes groups of characters as units rather than entire words. These character groups represent frequently occurring patterns in natural sequences and do not necessarily correspond to complete words, although many of them do. Tokenization relies on a predefined vocabulary that maps each character group to a unique positive integer, referred to as a token. For example, the title of this paper, ``Non-unitary and nonlinear quantum generative large model design and fast large-scale training without back-propagation'', is tokenized by GPT-3 into the following character groups: ``Non'', ``-'', ``unit'', ``ary'', `` and'', `` non'', ``linear'', `` quantum'', `` gener'', ``ative'', `` large'', `` model'', `` design'', `` and'', `` fast'', `` large'', ``-'', ``scale'', `` training'', `` without'', `` back'', ``-'', ``prop'', ``ag'', ``ation''. Then it is converted into the token sequence [15419, 12, 403, 9331, 290, 1729, 29127, 14821, 1152, 876, 1588, 2746, 1486, 290, 3049, 1588, 12, 9888, 3047, 1231, 736, 12, 22930, 363, 341].

	After tokenization, the input token sequence is embedded in a matrix $E_B$ with $m$ rows and $d$ columns. Each row of $E_B$ represents a token. The $m$ is the sequence length that the model takes in, and $d$ is the model dimension. This conversion process is implemented by embedding lookup. It is, at its core, a row gather: each integer token in the sequence is used directly as an index into an embedding table $E$, and the selected rows are stacked in sequence order to form the matrix $E_B$. The embedding table $E$ is trainable, and its row number is about 200000 in GPT-3. Because the forward pass reads $m$ specific rows, back-propagation deposits gradient into exactly those rows of $E$. The embedding gradient is actually sparse across the rows of $E$, because a token that never appears in a batch receives zero gradient that step.

	Another essential encoding technique in Transformer-based generative models is positional encoding, which encodes the order of character groups within the sequence. Positional encoding is typically performed simultaneously with word embedding and produces a $ 2048 \times 12288 $ matrix. For example, in GPT-3, the positional encoding matrix consists of 12,288 trainable parameters $ \omega_1, \omega_2, ..., \omega_{12288} $, arranged as follows:
	\begin{equation}
		E_P=
		\begin{bmatrix}
			\sin(\omega_1) & \sin(\omega_2) & \dots & \sin(\omega_{12288}) \\
			\sin(2\omega_1) & \sin(2\omega_2) & \dots & \sin(2\omega_{12288}) \\
			\vdots & \vdots & \ddots & \vdots \\
			\sin(2048\omega_1) & \sin(2048\omega_2) & \dots & \sin(2048\omega_{12288})
		\end{bmatrix}
	\end{equation}
	Most large generative models sum the embedding result $ E_B $ and the positional encoding result $ E_P $, forming the final output of the encoding layer:
	\begin{equation}
		E_{O}=E_B+E_P
	\end{equation}

\begin{figure}[ht]
		\centering
		\includegraphics[width=0.75\textwidth]{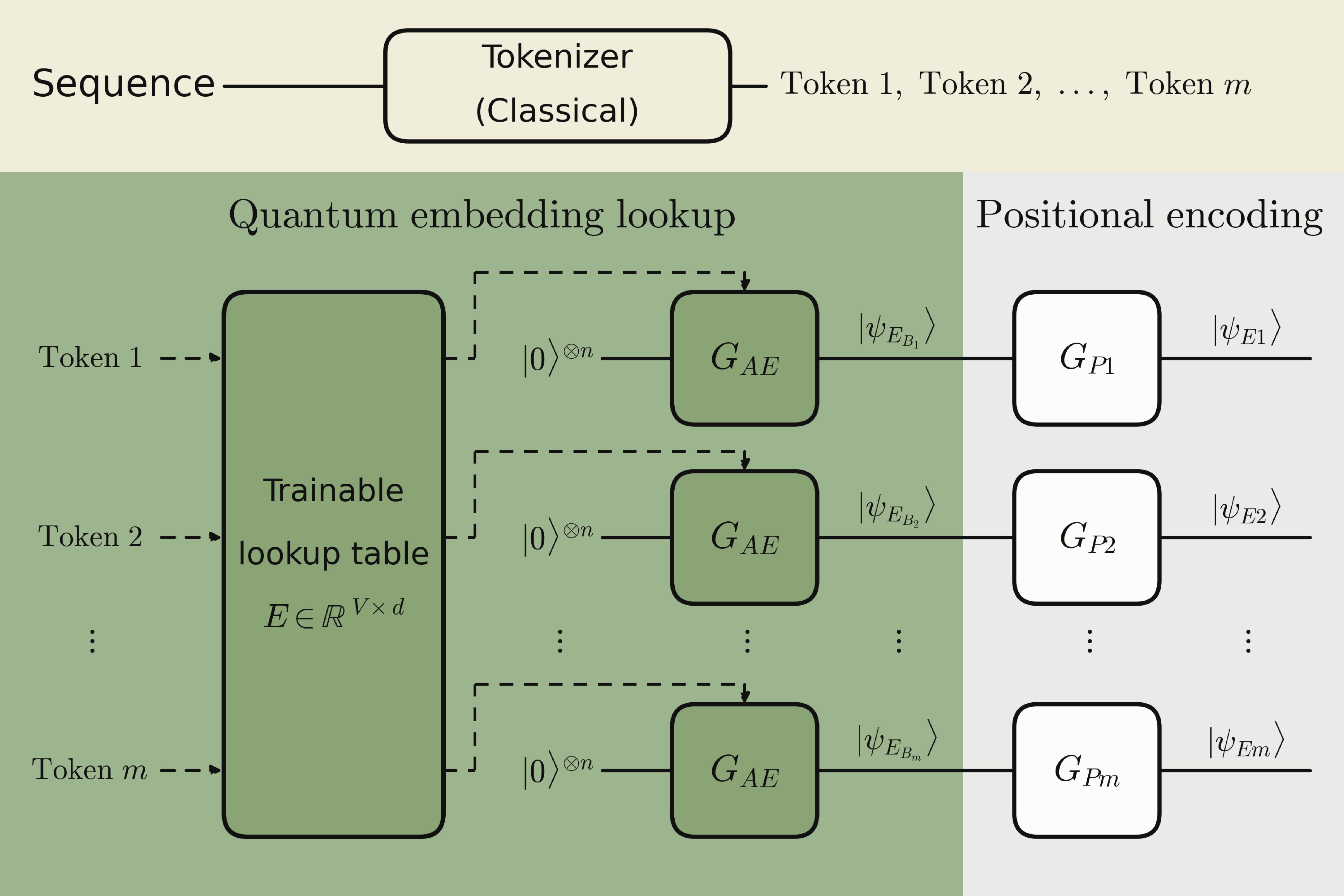}
		\caption{The structure of the quantum encoding layer. A classical tokenizer convert the input natural sequence into token sequence, followed by the quantum embedding lookup consist of a trainable lookup table and the quantum amplitude encoding circuits $G_{AE}$. $G_{AE}$ encodes the corresponding rows in the lookup table into the quantum state of the qubits. Finally, positional encoding is applied.}
		\label{methods:fig_encoding}
	\end{figure}

	For a quantum large model, the encoding processes of natural sequences are also needed. Here, we propose a quantum encoding layer for encoding natural sequences into qubits, as shown in Fig. \ref{methods:fig_encoding}. We use the classic tokenizer to tokenize the natural sequences, followed by the quantum embedding lookup. We store a trainable lookup table $E$ as the classical large models do, look up the corresponding rows of the token sequence, and embed the rows onto the qubits through quantum amplitude embedding, obtaining the qubit sequence $\ket{\psi_{E_{B_1}}}, \ket{\psi_{E_{B_2}}}, \ldots, \ket{\psi_{E_{B_m}}}$. This qubit sequence is equivalent to the matrix $E_B$. Let $n=\lceil\log_2d\rceil$ denote the number of qubits used to encode each $d$-dimensional token vector. Due to the high information density of quantum amplitude embedding, we use only $mn$ qubits to embed the whole sequence, compared with the $md$ elements of $E_B$. For example, for a large model with a scale similar to GPT-3, $E_B$ contains $m\times d=2048\times12288$ elements, while its quantum version uses $mn=2048\times14$ qubits. The quantum amplitude embedding circuits depend on the quantum device and its supported gate set. Generally, their depth can be controlled between $O(1)$ and $O\left(n^2\right)=O\left(\left(\log_2d\right)^2\right)$, as discussed in Chapter 4.

	\begin{figure}[ht]
		\centering
		\includegraphics[width=0.95\textwidth]{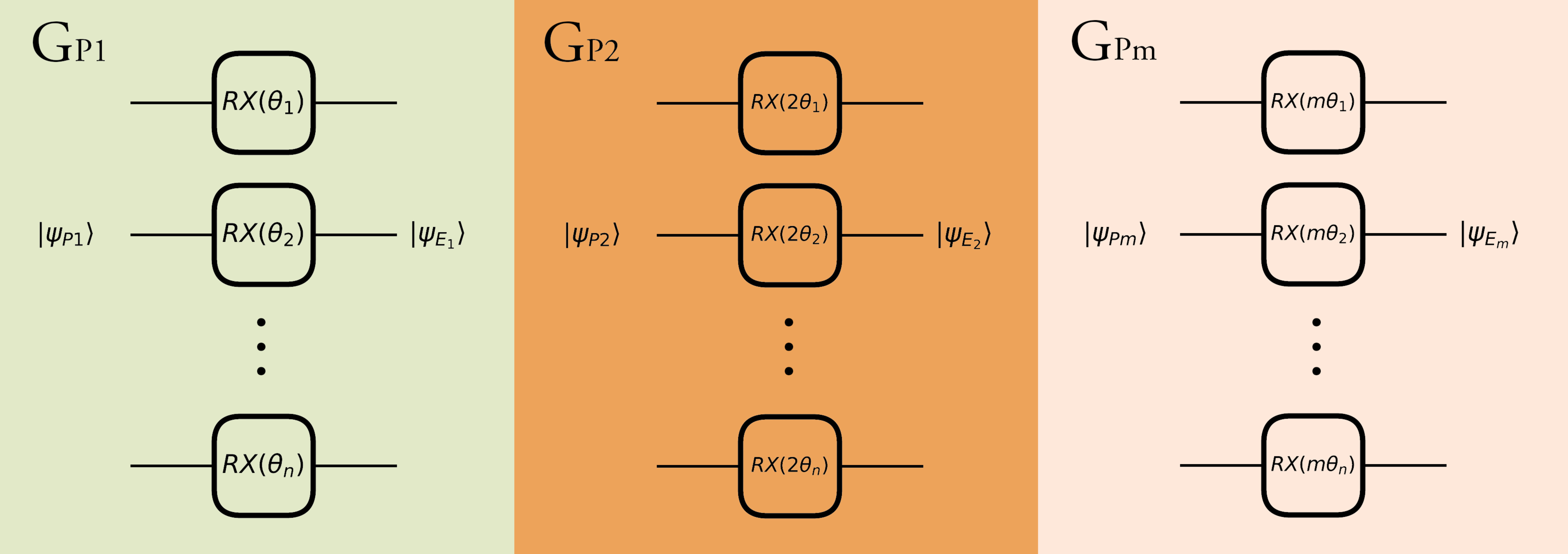}
		\caption{The structure of the positional encoding circuit $G_{P1}$ to $G_{Pm}$. They consist of X-axis rotation gates applied to each qubit. For the qubits corresponding to the $k$-th group of characters, the rotation angles are $k$ times the angles used in the positional encoding circuit for the first group of characters.}
		\label{methods:fig_positional_encoding}
	\end{figure}
	Next, we perform positional encoding. Instead of directly replicating the process of GPT-3, we implement positional encoding through quantum rotations. The positional encoding circuits are denoted by $G_{P_1},\ldots,G_{P_m}$. Specifically, for the $j$-th token position, we apply $R(j\theta_k)$ to the $k$-th qubit, where $1\leq k\leq n$ and $1\leq j\leq m$. Thus, $G_{P_1}$ applies $R(\theta_k)$ to qubit $k$, and the rotation angles are scaled by the token position in the subsequent circuits. This approach avoids mixing quantum states with arbitrary weights, which is challenging to implement in quantum circuits. The output states of the positional encoding form the output of the encoding layer and are denoted by $\ket{\psi_{E_1}},\ldots,\ket{\psi_{E_m}}$, where $m$ is the sequence length. The positional encoding method is shown in Fig. \ref{methods:fig_positional_encoding}.

	The processes in the encoding layer will repeat for $3m-2$ times for each input sequence because of the non-cloning theorem in quantum mechanics that forbids copying arbitrary quantum states\cite{wootters1982cloning}, and the self-attention layer of our model needs to mix the states of the qubits that multiple copies of the encoded qubits are necessary. Fortunately, the repeating of the processes will not affect the time complexity because the operations in the encoding layer can be fully parallel.

	\subsection{Multi-head Quantum Self-attention Layer}
	\label{methods:section_self_attention}
	One of the main reasons why large generative models based on the Transformer architecture are so powerful is the self-attention mechanism\cite{vaswani2017attention}. This mechanism allows the model to analyze the relationships between groups of characters in a sequence, thus enhancing its ability to process long sequences effectively. The core principle of the self-attention mechanism is to evaluate the correlation between character groups within a sequence, enabling a more efficient understanding of the sequence's meaning.

	Continue using GPT-3\cite{bib8} as an example. The output matrix from the encoding layer, denoted as $E_O$, has dimensions $2048 \times 12288$. $E_O$ is then multiplied by matrix groups $W_{Q_i}$, $W_{K_i}$, and $W_{V_i}$, $1\leq i\leq96$ each with dimensions $12288 \times 128$:
	\begin{equation}
		\begin{matrix}
			Q_i = E_{O}W_{Q_i} \\
			K_i = E_{O}W_{K_i} \\
			V_i = E_{O}W_{V_i}. \\
		\end{matrix}
	\end{equation}
	The sets of three matrices $W_{Q_i}$, $W_{K_i}$, and $W_{V_i}$, $1\leq i\leq96$ is referred to as attention heads. The letters Q, K, and V stand for query, key, and value, respectively. The attention head serves as a guideline for evaluating the correlations between different groups of characters in the sequence represented by $E_O$. Each attention head focuses on different aspects of these relationships. For instance, one attention head might detect stronger correlations between groups of characters that are close to each other, while another might focus on the correlations between the nouns and the verb.  The dot products of the row vectors from $Q_i$ and $K_i$ are then computed, producing $2048^2 = 4.19 \times 10^6$ results. These results are organized into matrices $A_i$, where $1 \leq i \leq 96$:
	\begin{equation}
		A_i = Q_i K_i^T.
	\end{equation}
	The matrix element $[A_i]_{jk}$ represents the correlation between the $j$-th and $k$-th group of characters. Row-wise softmax is then applied to the matrix $A_i$ to normalize the correlations between the group of characters corresponding to the row and the other groups, which is referred to the attention between the groups of characters. Following this, the attention matrices $\text{row\_softmax}(A_i), 1\leq i\leq 96$ are multiplied by $V_i$ respectively to generate the probability distribution of the groups of characters:
	\begin{equation}
		Att_i = \text{row\_softmax}(A_i) V_i
	\end{equation}
	where the size of $Att_i$ is $2048\times128$. Finally, the matrix pieces $Att_1, Att_2, ..., Att_{96}$ are concatenated to restore the original matrix size of $2048 \times 12288$. The final output of the self-attention layer is:
	\begin{equation}
		Att = \text{concat}_{i=1}^{96}(Att_i)W_O
	\end{equation}
	where $W_O$ is a $12288 \times 12288$ matrix. The mechanism involving multiple attention heads is commonly referred to as multi-head self-attention. For GPT-3, there are $3 \times 12288\times128 = 4.7186 \times 10^6$ trainable parameters in each attention head, so the total number of parameters in a single multi-head attention layer is $96 \times 4.7186 \times 10^6+12288^2 = 6.0398 \times 10^8$.

\begin{figure}[ht]
		\centering
		\includegraphics[width=0.95\textwidth]{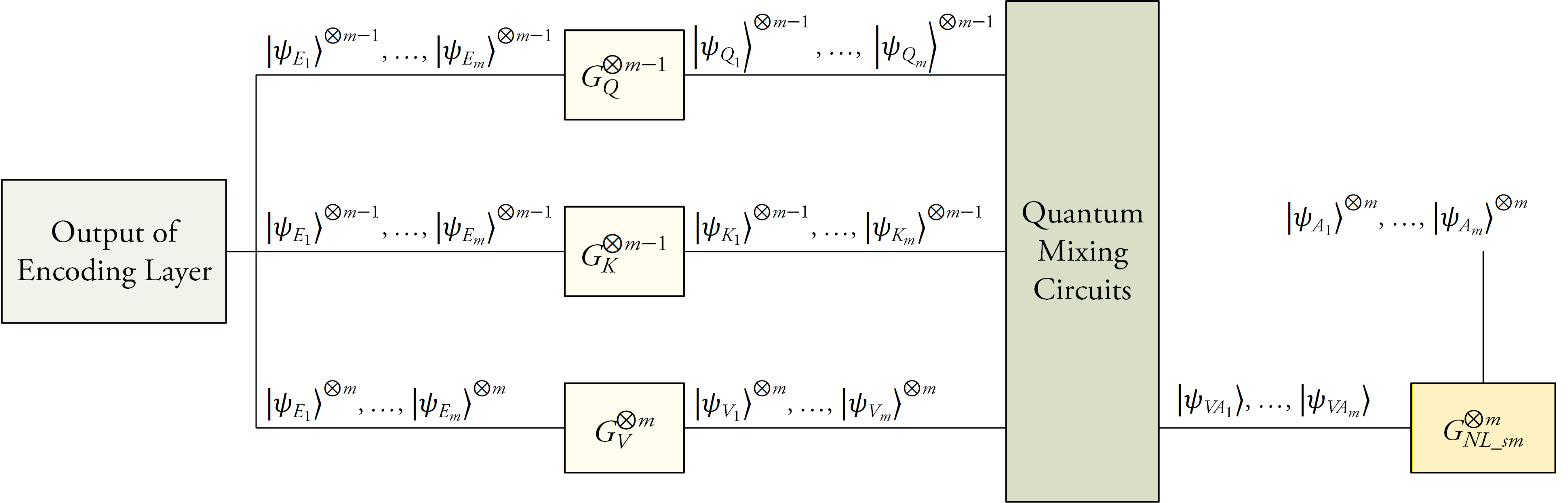}
		\caption{Diagram of the structure of the self-attention layer. The $3m-2$ copies of the encoded qubits $\ket{\psi_{E_1}}$ to $\ket{\psi_{E_m}}$ pass through the attention-head-circuits $G_Q$, $G_K$ and $G_V$ to acquire the query states, the key states and the value states relatively, with query These states pass through the mixing circuits consist of SwapTest circuits\cite{bib12,bib13} and Strong Entangling circuits\cite{bib11} to mix the value states according to the inner product of the corresponding query states and key states. Finally, measurements are performed on the SwapTest auxiliary qubits to eliminate the entanglement between the auxiliary qubits and value states and reset the auxiliary qubits to $\ket{0}$ to let the following operations reuse them. The mixed value states finally pass the Softmax circuits $G_{NL\_sm}$ to form the output states of the attention layer $\ket{\psi_{A_1}}$ to $\ket{\psi_{A_m}}$.}
		\label{methods:fig_attention}
	\end{figure}

	The self-attention layer heavily relies on matrix computations, particularly matrix multiplications. Each row vector in the output matrix of the encoding layer, $E_O$, is multiplied by 96 different attention heads $W_{Q_i}, W_{K_i}, W_{V_i}$, where $1 \leq i \leq 96$, and then by the matrix $W_O$ before being passed through the self-attention layer. Matrix computations are computationally expensive, but quantum computing offers a significant advantage in terms of time complexity for these operations, due to the fact that the minimum computation unit in quantum computers are quantum gates performing unitary transformations on the quantum states of qubits, compared with the classical computers that use logical gates as computation unit. However, implementing a quantum circuit for the self-attention mechanism faces challenges, mainly due to the non-unitary nature of the self-attention heads, the process of mixing the unknown state vector with arbitrary proportions, and the softmax activation function. In this work, we propose a quantum approach to the multi-head self-attention layer that addresses these challenges and provides an exponential speed-up compared to the classical method. The whole structure of the self-attention layer is shown in Fig \ref{methods:fig_attention}. The $3m-2$ copies of the encoded qubits from the encoding layer are divided into 3 groups, and pass the attention head circuits respectively to form the query states $\ket{\psi_{Q_j}}$, the key states $\ket{\psi_{K_j}}$ and the value states $\ket{\psi_{V_j}}$, $1\leq j\leq m$. The 3 groups of states then pass the mixing circuits where the value states are mixed according to the inner products of the query and key states. Finally, the mixed value states pass the nonlinear softmax circuit to form the output states of the self-attention layer.

	\subsubsection{Multiple quantum attention heads with arbitrary non-unitary attention matrices}
	The encoding layer outputs $3m-2$ copies of the encoded qubit $\ket{\psi_{E_1}}$ to $\ket{\psi_{E_m}}$ simultaneously, as mentioned earlier. These qubits will pass through three parameterized quantum circuits $G_Q$, $G_K$, and $G_V$, each acting as one part in the multiple attention heads $W_{Q_i}$, $W_{K_i}$, and $W_{V_i}$. In our quantum large model, the number of qubits representing each token is 14, resulting in a model dimension of $2^{14}=16384$. We use a multi-head self-attention mechanism with 128 attention heads in our quantum model, similar to GPT-3’s multi-head self-attention with 96 attention heads. The overall matrices for the quantum circuits $G_Q$, $G_K$, and $G_V$ are:
	\begin{equation}
		\begin{matrix}
			G_Q = \begin{bmatrix}
				G_{Q_1} \\
				& G_{Q_2} \\
				& & \dots \\
				& & & G_{Q_{128}}
			\end{bmatrix}, & 
			G_K = \begin{bmatrix}
				G_{K_1} \\
				& G_{K_2} \\
				& & \dots \\
				& & & G_{K_{128}}
			\end{bmatrix} \\
			G_V = \begin{bmatrix}
				G_{V_1} \\
				& G_{V_2} \\
				& & \dots \\
				& & & G_{V_{128}}
			\end{bmatrix}
		\end{matrix}
		\label{methods:attention-head}
	\end{equation}
	Take $G_Q$ as an example to explain how we implement multiple quantum attention heads. The first step is to implement the blocks $G_{Q_i}, 1\leq i\leq128$. In classical large models, the blocks are arbitrary matrices. However, the ordinary parameterized quantum circuits are constrained by the fundamental principles of quantum mechanics, which allow only unitary evolutions. This restriction limits the flexibility of the self-attention heads, since unitary transformations represent only a subset of all possible matrix transformations. If we are restricted to unitary transformations on qubit states, the performance of the quantum self-attention mechanism layer will be severely limited and may not achieve the same capability as classical self-attention mechanism.
	\begin{figure}[ht]
		\centering
		\includegraphics[width=0.4\linewidth]{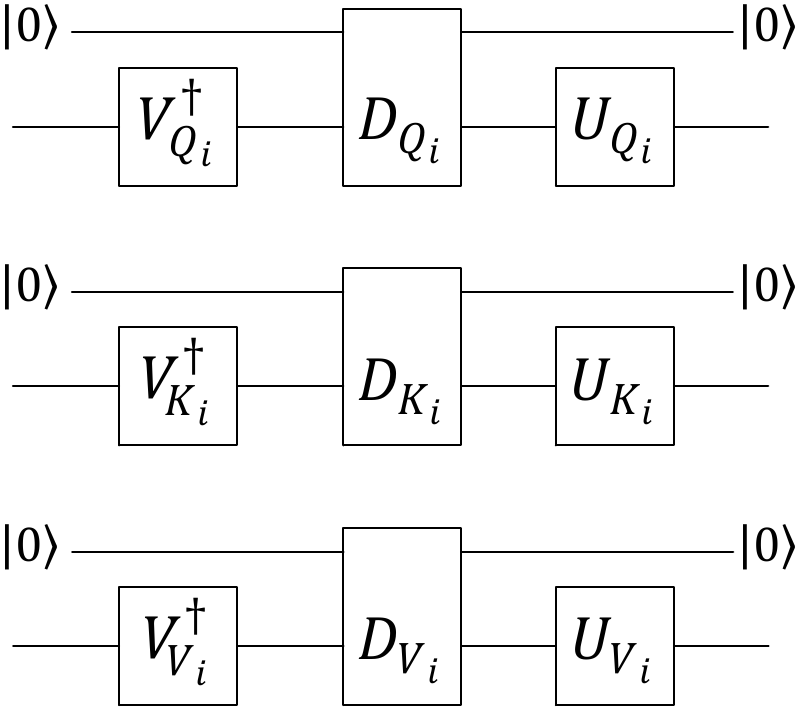}
		\caption{Detail of preparing the states $\ket{\psi_{Q_j}}$, $\ket{\psi_{K_j}}$ and $\ket{\psi_{V_j}}$, $1\leq j\leq m$, showing the general structure of the quantum circuits $G_{Q_i}$, $G_{K_i}$ and $G_{V_i}$, $1\leq i\leq 128$. The structure of $U_Q$, $U_K$, $U_V$, $V^\dagger_Q$, $V^\dagger_K$ and $V^\dagger_V$ is shown in Fig. \ref{methods:fig_unitary_layer_7} and \ref{methods:fig_unitary_layer_14} while the structure of $D_Q$, $D_K$ and $D_V$ is shown in Fig. \ref{methods:fig_diag_circuit}.}
		\label{methods:fig_attention_svd}
	\end{figure}

	To address this limitation, we employ singular value decomposition (SVD) to decompose a non-unitary matrix into a product of unitary matrices and a non-unitary diagonal matrix, then implementing them separately. The general of the quantum circuit structure of this method is shown in Fig. \ref{methods:fig_attention_svd}. The SVD allows us to express $G_{Q_i}$ as a product of two unitary matrices, $U$ and $V$, and a diagonal matrix $D$ with non-negative real numbers on the diagonal:
	\begin{equation}
		G_{Q_i} = U^{(i)} D^{(i)} V^{\dagger(i)}.
	\end{equation}
	After the decomposition, we implement the quantum circuit performing $U^{(i)}$, $V^{\dagger(i)}$ and $D^{(i)}$, then apply them on the qubits sequentially in the order of $V^{\dagger(i)}$, $D^{(i)}$ and $U^{(i)}$ to construct the block $G_{Q_i}$. 

	\begin{figure}[ht]
		\centering
		\includegraphics[width=0.75\textwidth]{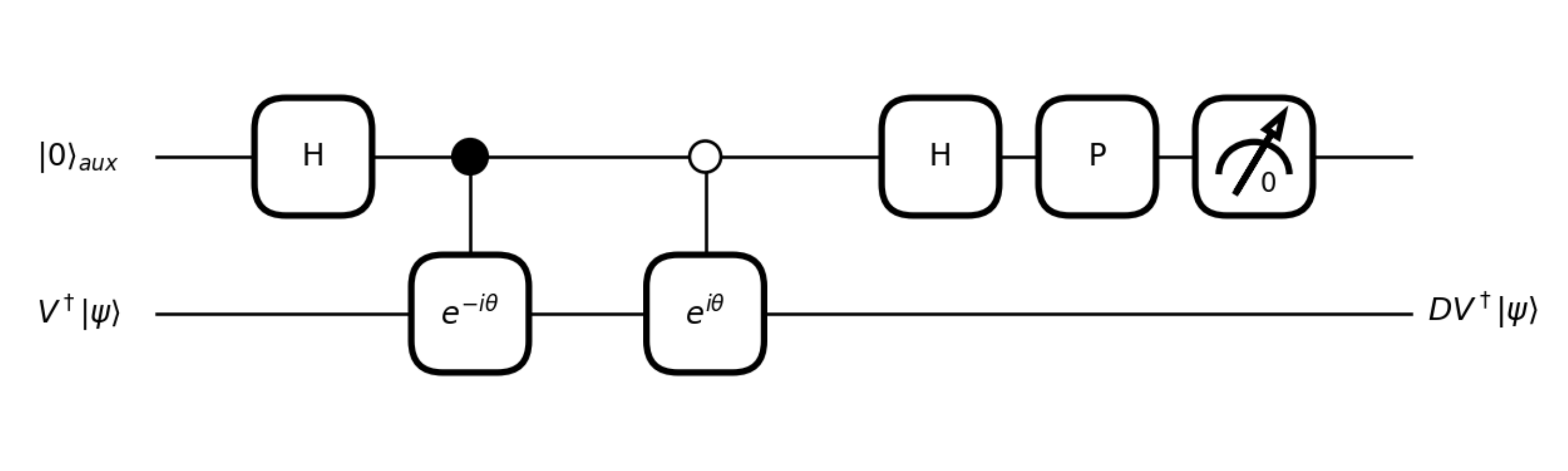}
		\caption{The quantum circuit implementing the non-unitary diagonal operation with matrix form $D=\operatorname{diag}(d_1,d_2,\ldots,d_{2^q})$ on $q$ data qubits. The Hadamard gate on the ancillary qubit turns the input state $\ket{\psi}\otimes\ket{0}$ into $\ket{\psi}\otimes(\ket{0}+\ket{1})/\sqrt{2}$. The controlled phase operations and the second Hadamard gate produce $\cos(\theta)\ket{\psi}\otimes\ket{0}+i\sin(\theta)\ket{\psi}\otimes\ket{1}$. The phase gate $P_{\mathrm{anc}}$ removes the factor $i$, giving $\cos(\theta)\ket{\psi}\otimes\ket{0}+\sin(\theta)\ket{\psi}\otimes\ket{1}$. The operation succeeds when measurement of the ancillary qubit gives 0.}
		\label{methods:fig_diag_circuit}
	\end {figure}
	First, we implement the non-unitary diagonal circuit $D^{(i)}$. Let $q$ denote the number of data qubits on which this diagonal block acts; for an attention-head block, $q=n-\log_2p$. We use a quantum circuit capable of probabilistically implementing the non-unitary diagonal operations by Zylberman et al.~\cite{bib10}. This circuit is illustrated in Fig. \ref{methods:fig_diag_circuit}. Given the parameters $\{\theta_1,\theta_2,\ldots,\theta_{2^q}\}$ and the input $\ket{\psi}\otimes\ket{0}$, the circuit outputs $\cos(\theta)\ket{\psi}\otimes\ket{0}$ after measurement and post-selection of the ancillary state $\ket{0}$. Here $\theta=\operatorname{diag}(\theta_1,\theta_2,\ldots,\theta_{2^q})$, and its matrix cosine is
	\begin{equation}
		\cos(\theta)=\sum\limits_{s=0}^{+\infty}\frac{(-1)^s}{(2s)!}\theta^{2s}=\operatorname{diag}\left(\cos(\theta_1),\cos(\theta_2),\ldots,\cos(\theta_{2^q})\right).
	\end{equation}
	The non-unitary diagonal operation is $D^{(i)}=\operatorname{diag}(d_1,d_2,\ldots,d_{2^q})$, acting on $V^{\dagger(i)}\ket{\psi_{E_j}}$ for token position $j$. We define $\theta=\arccos(D^{(i)})=\operatorname{diag}(\arccos(d_1),\arccos(d_2),\ldots,\arccos(d_{2^q}))$, where each $\theta_k\in[0,\pi]$. The circuit acts as described in Eq. \ref{methods:diagcir}:
	\begin{equation*}
		\begin{aligned}
			\ket{\psi}\otimes \ket{0}
			& \stackrel{H_{anc}}\rightarrow \ket{\psi}\otimes \frac{\ket{0}+\ket{1}}{\sqrt{2}}\\
			& \stackrel{C(e^{-i\theta})}\rightarrow
			\frac{\ket{\psi}\otimes\ket{0}+e^{-i\theta}\ket{\psi}\otimes\ket{1}}{\sqrt{2}} \\
			& \stackrel{\overline{C}(e^{i\theta})}\rightarrow
			\frac{e^{i\theta}\ket{\psi}\otimes\ket{0}+e^{-i\theta}\ket{\psi}\otimes\ket{1}}{\sqrt{2}}\\
			& \stackrel{H_{anc}}\rightarrow
			\frac{e^{i\theta}\ket{\psi}\otimes(\ket{0}+\ket{1})+e^{-i\theta}\ket{\psi}\otimes(\ket{0}-\ket{1})}{2} \\
			& =\frac{(e^{i\theta}+e^{-i\theta})\ket{\psi}\otimes\ket{0}+(e^{i\theta}-e^{-i\theta})\ket{\psi}\otimes\ket{1}}{2}\\
		\end{aligned}
	\end{equation*}
	The ancillary phase gate then gives
	\begin{equation}
		\begin{aligned}
			&\frac{(e^{i\theta}+e^{-i\theta})\ket{\psi}\otimes\ket{0}+(e^{i\theta}-e^{-i\theta})\ket{\psi}\otimes\ket{1}}{2}\\
			& \stackrel{P_{anc}}\rightarrow\frac{1}{2}\left[(e^{i\theta}+e^{-i\theta})\ket{\psi}\otimes\ket{0}\right.\\
			&\qquad\left.+\frac{1}{i}(e^{i\theta}-e^{-i\theta})\ket{\psi}\otimes\ket{1}\right]\\
			& =\cos(\theta)\ket{\psi}\otimes\ket{0}+\sin(\theta)\ket{\psi}\otimes\ket{1}
		\end{aligned}
		\label{methods:diagcir}
	\end{equation}
	where
	\begin{equation}
		P_{anc}=
		\begin{bmatrix}
			1 & 0 \\
			0 & -i \\
		\end{bmatrix}, 
		H_{anc}=\frac{1}{\sqrt{2}}
		\begin{bmatrix}
			1 & 1 \\
			1 & -1 \\
		\end{bmatrix}. 
	\end{equation}
	Before outputting the target qubits, a measurement is made on the ancillary qubit. We will successfully turn the input qubits $V^{\dagger(i)}\ket{\psi}\otimes\ket{0}$ into $\cos (\theta)V^{\dagger(i)}\ket{\psi}\otimes\ket{0}=D^{(i)} V^{\dagger(i)}\ket{\psi}\otimes\ket{0}$ if the result of the measurement is 0. However, if the result is 1, the state will be turned into $\sin (\theta)V^{\dagger(i)}\ket{\psi}\otimes\ket{1}$ which is not expected, and a failure occurs.

	To prevent wasting time and qubits when failures occur, we adopt a repeat-until-success (RUS) strategy. We define $\theta^{(0)}=\theta$. After the first failed attempt, the accumulated failure operator is
	\begin{equation}
		D_{\mathrm{fail}}^{(1)}=\sin(\theta^{(0)})=\operatorname{diag}\left(\sqrt{1-d_1^2},\sqrt{1-d_2^2},\ldots,\sqrt{1-d_{2^q}^2}\right).
	\end{equation}
	The normalized recovery operation after this failure is proportional to
	\begin{equation}
		D_{\mathrm{reco}}^{(1)}=\frac{1}{N^{(1)}}D^{(i)}\left(D_{\mathrm{fail}}^{(1)}\right)^{-1}=\frac{1}{N^{(1)}}\operatorname{diag}\left(\frac{d_1}{\sqrt{1-d_1^2}},\ldots,\frac{d_{2^q}}{\sqrt{1-d_{2^q}^2}}\right),
	\end{equation}
	where
	\begin{equation}
		N^{(1)}=\max_{1\leq k\leq2^q}\left|\frac{d_k}{\sin\theta_k^{(0)}}\right|,\qquad
		\theta_k^{(1)}=\arccos\left(\frac{d_k}{N^{(1)}\sin\theta_k^{(0)}}\right).
	\end{equation}
	The factor $N^{(1)}$ ensures that every argument of $\arccos$ lies in $[-1,1]$; the associated global scaling is physically irrelevant after state normalization. If the first recovery attempt also fails, the accumulated failure operator becomes
	\begin{equation}
		D_{\mathrm{fail}}^{(2)}=\sin(\theta^{(0)})\sin(\theta^{(1)}),
	\end{equation}
	and the next recovery operation is
	\begin{equation}
		D_{\mathrm{reco}}^{(2)}=\frac{1}{N^{(2)}}D^{(i)}\left(D_{\mathrm{fail}}^{(2)}\right)^{-1}.
	\end{equation}
	More generally, after $r\geq1$ consecutive failures, we define
	\begin{equation}
		\begin{aligned}
			D_{\mathrm{fail}}^{(r)}&=\prod_{s=0}^{r-1}\sin\left(\theta^{(s)}\right),\\
			N^{(r)}&=\max_{1\leq k\leq2^q}\left|\frac{d_k}{\prod_{s=0}^{r-1}\sin\left(\theta_k^{(s)}\right)}\right|,\\
			\theta_k^{(r)}&=\arccos\left(\frac{d_k}{N^{(r)}\prod_{s=0}^{r-1}\sin\left(\theta_k^{(s)}\right)}\right),\qquad 1\leq k\leq2^q.
		\end{aligned}
	\end{equation}
	The recovery circuit at attempt $r$ implements $\cos(\theta^{(r)})$, which is proportional to $D^{(i)}\left(D_{\mathrm{fail}}^{(r)}\right)^{-1}$. Zylberman J. et al. pointed out that, for $D^{(i)}=\operatorname{diag}(d_1,d_2,\ldots,d_{2^q})$ and $\ket{\psi}=(\psi_1,\psi_2,\ldots,\psi_{2^q})^{\mathsf T}$, the success probability is
	\begin{equation}
		P_{\mathrm{succ}}=\sum\limits_{k=1}^{2^q}\frac{d_k^2}{d_{\max}^2}|\psi_k|^2,\qquad d_{\max}=\max_{1\leq k\leq2^q}|d_k|.
	\end{equation}
	This probability is 0.5 on average if we have no prior knowledge of the elements $|\psi_k|^2$ and $d_k$. We can increase the probability of success at the cost of repetitions; on average, two executions are required to obtain a success.

\begin{figure}[ht]
		\centering
		\includegraphics[width=0.85\textwidth]{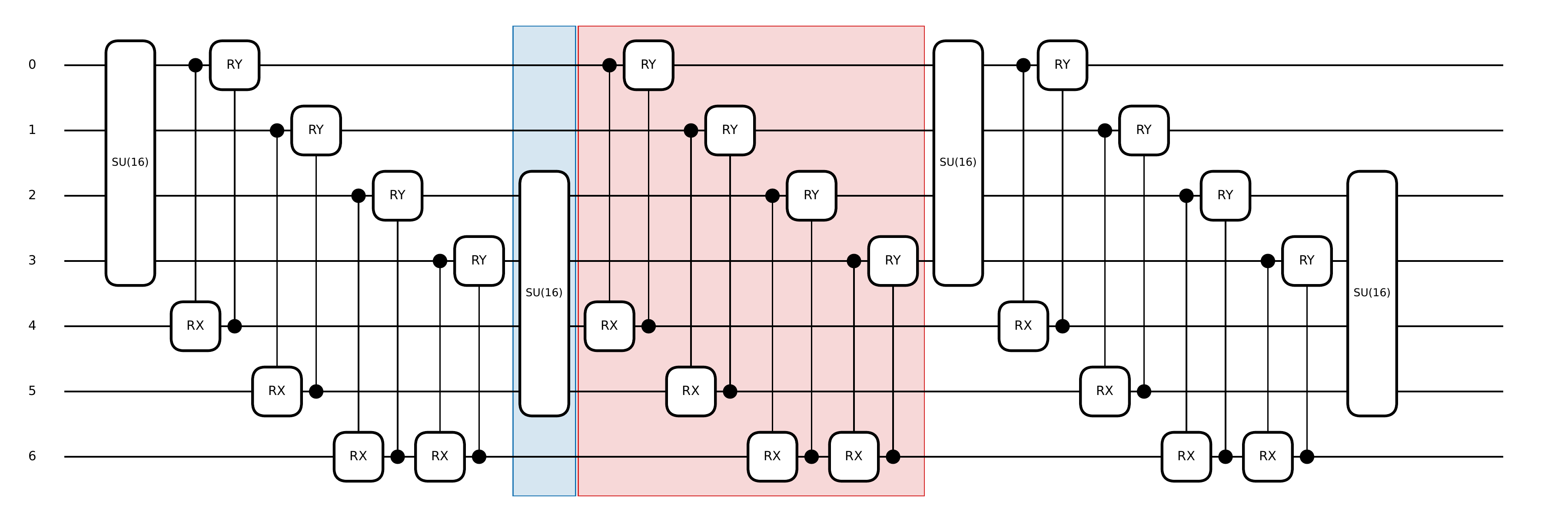}
		\caption{The quantum circuit diagram of the unitary part in the circuit $G_{Q_i}$, $G_{K_i}$, $G_{V_i}$, $1\leq i\leq 128$. The circuit follows the pattern of $SU(16)$ circuit layers and bridge layers, marked by the blue box and the red box on the diagram respectively. The $SU(16)$ circuits in a single layer are arranged to operate in parallel to accommodate more trainable parameters in lower circuit depth, and bridge layers locate at the middle of the $SU(16)$ circuit layers to isolate the $SU(16)$ circuits, thus avoiding any parameter redundancy.}
		\label{methods:fig_unitary_layer_7}
	\end {figure}

\begin{figure}[ht]
		\centering
		\includegraphics[width=0.95\textwidth]{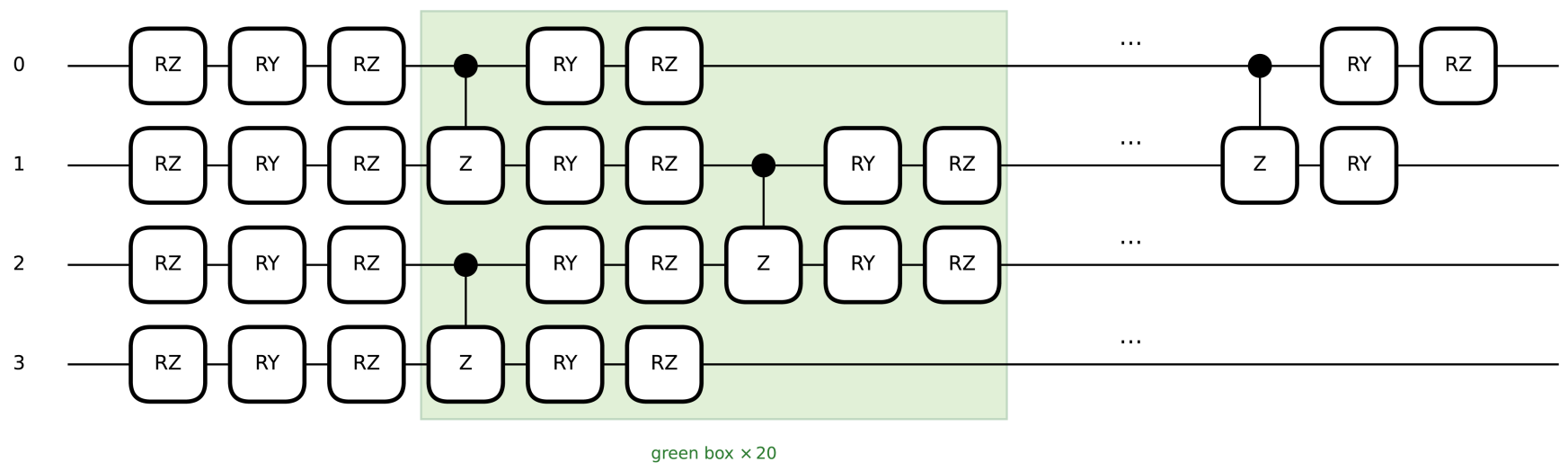}
		\caption{The structure of the $SU(16)$ circuit proposed by David Wierichs, et. al\cite{bib36}. The circuit has the lowest depth and fewest number of CNOT gates known currently.}
		\label{methods:SU16}
	\end {figure}

	Next, we construct the unitary part $U^{(i)}$ and $V^{\dagger(i)}$. They can be implemented by ordinary quantum circuits without any ancillary qubits and post-selection. The main challenge is to accommodate as much as trainable parameters while control the circuit depth as low as possible. To address the challenge, the quantum gates in the circuit should be parallel to the maximum extent, and the parameter redundancy should be carefully avoided. To maximize the quantum gates parallelism, we propose a quantum circuit structure that use 4-qubit general circuits ($SU(16)$ circuits) as components, inspired by the idea of modular designing of the classical integrated circuits. The circuit structures of $U^{(i)}$ and $V^{\dagger(i)}$ are illustrated in Fig. \ref{methods:fig_unitary_layer_7}. In a single $SU(16)$ circuit layer, the $SU(16)$ circuits are arranged to operate different qubits in parallel. Across the adjacent $SU(16)$ circuit layers, the $SU(16)$ circuits in the latter layer stride 2 qubits compared to the former layer to reduce the repeatedly operated qubits. The layer number $l_{SU}$ can be adjusted to control the circuit depth and the scale of parameter number. For the implementation of the $SU(16)$ circuits, we use the quantum circuit structure proposed by David Wierichs, Korbinian Kottmann and Nathan Killoran\cite{bib36}. Their circuit is able to perform arbitrary $SU(16)$ transformations on the quantum states of its input qubit with the lowest depth known currently. Meanwhile, their circuit only contains 61 CNOT gates, which is the theoretical minimum. This makes the circuit friendly for hardware implementation. The circuit structure is shown in Fig. \ref{methods:SU16}.

	The parameter redundancy is caused by the same kind duplicated quantum gates. For example, in Fig. \ref{methods:fig_unitary_layer_7}, if the $SU(16)$ circuits are applied on the qubits without the bridge layers between them, there will always be 2 qubits that are operated sequencially by the adjacent $SU(16)$ circuit, which will cause parameter redundancy: while the nominal parameter number is $((2^4)^2-1)\times2=510$, the actual effective parameter is $((2^4)^2-1)\times2-((2^2)^2-1)=495$. This can be checked strictly by Eq. \ref{methods:effective_parameter_original}. As a common result, sequentially applied general circuits with qubits operated multiple times always cause $(2^{n_{\text{shared\_qubit}}})^2-1$ parameter to be redundant. To avoid the parameter redundancy, we place bridge layers between the $SU(16)$ circuit layers to isolate the $SU(16)$ circuits. The bridge layer consists of a sequence of controlled parameterized $RX$, $RY$ and $RZ$ gates, with the controlling qubit and the target qubit of each rotation gate cross 6 qubits in the middle. This ensures that the controlled rotation gates will not be absorbed by the $SU(16)$ circuits while isolating the $SU(16)$ circuits effectively in minimum depth.

	To check the parameter redundancy is indeed removed, we regard a parametrized quantum circuit as a smooth map
	\begin{equation}
		U:\boldsymbol{\theta}\in\mathbb{R}^{P}\longmapsto U(\boldsymbol{\theta})\in SU(2^{n}),
	\end{equation}
	where $P$ is the nominal number of parameters and $n$ is the number of qubits. Let $\operatorname{vec}_{\mathbb{R}}(\cdot)$ denote the operation that stacks the real and imaginary parts of a matrix into a real column vector. The effective parameter number at a generic point $\boldsymbol{\theta}^{*}$ is defined as\cite{funcke2021expressivity}
	\begin{equation}
		P_{\mathrm{eff}}=\operatorname{rank}J(\boldsymbol{\theta}^{*}),\qquad J(\boldsymbol{\theta})=\frac{\partial\,\operatorname{vec}_{\mathbb{R}}\!\left(U(\boldsymbol{\theta})\right)}{\partial\boldsymbol{\theta}}\in\mathbb{R}^{\,2\cdot4^{n}\times P}.
		\label{methods:effective_parameter_original}
	\end{equation}
	Indeed, a parameter direction $\boldsymbol{c}\in\mathbb{R}^{P}$ is redundant if
	\begin{equation}
		\sum_{i=1}^{P}c_i\frac{\partial U}{\partial\theta_i}=0,
	\end{equation}
	which is equivalent to $J(\boldsymbol{\theta}^{*})\boldsymbol{c}=0$.

	\begin{proposition}
		\label{methods:prop:absence_parameter_redundancy}
		For all unitary-part circuits considered in this work with $6$ to $14$ qubits, the insertion of the bridge layers removes the continuous parameter redundancy of the overlapping $SU(16)$ circuit layers. In particular, at any generic parameter point,
		\begin{equation}
			P_{\mathrm{eff}}=\operatorname{rank}J(\boldsymbol{\theta}^{*})=P,
			\label{methods:eq:full_effective_parameter_number}
		\end{equation}
		where $P_{\mathrm{eff}}$ and $P$ represent the number of the effective and nominal parameters respectively.
	\end{proposition}

	\begin{proof}
		Directly evaluating Eq.~\eqref{methods:effective_parameter_original} requires materializing the full $2^{n}\times2^{n}$ unitary matrix. To avoid this cost, let $\{\lvert\psi_k\rangle\}_{k=1}^{K_{\mathrm{samp}}}$ be independently sampled random input states and define
		\begin{equation}
			F(\boldsymbol{\theta})=\operatorname{col}_{k=1}^{K_{\mathrm{samp}}}
			\begin{bmatrix}
				\operatorname{Re}\!\left(U(\boldsymbol{\theta})\lvert\psi_k\rangle\right)\\
				\operatorname{Im}\!\left(U(\boldsymbol{\theta})\lvert\psi_k\rangle\right)
			\end{bmatrix}
			\in\mathbb{R}^{\,2K_{\mathrm{samp}}\cdot2^{n}}.
			\label{methods:stacked_state_map}
		\end{equation}
		The corresponding sampled-state Jacobian is
		\begin{equation}
			J_{\psi}(\boldsymbol{\theta})=\frac{\partial F(\boldsymbol{\theta})}{\partial\boldsymbol{\theta}}\in\mathbb{R}^{\,2K_{\mathrm{samp}}\cdot2^{n}\times P}.
			\label{methods:effective_parameter_scalable}
		\end{equation}
		For any $\boldsymbol{c}\in\mathbb{R}^{P}$,
		\begin{equation}
			J(\boldsymbol{\theta}^{*})\boldsymbol{c}=0\quad\Longrightarrow\quad\left(\sum_{i=1}^{P}c_i\frac{\partial U}{\partial\theta_i}\right)\lvert\psi_k\rangle=0,\qquad k=1,\ldots,K_{\mathrm{samp}},
		\end{equation}
		and hence $J_{\psi}(\boldsymbol{\theta}^{*})\boldsymbol{c}=0$. Therefore,
		\begin{equation}
			\ker J(\boldsymbol{\theta}^{*})\subseteq\ker J_{\psi}(\boldsymbol{\theta}^{*}),
		\end{equation}
		which implies
		\begin{equation}
			\operatorname{rank}J_{\psi}(\boldsymbol{\theta}^{*})\leq\operatorname{rank}J(\boldsymbol{\theta}^{*})\leq P.
			\label{methods:eq:sampled_rank_inequality}
		\end{equation}
		Consequently, $\operatorname{rank}J_{\psi}(\boldsymbol{\theta}^{*})=P$ certifies that $\operatorname{rank}J(\boldsymbol{\theta}^{*})=P$ and that no nonzero redundant direction exists. As reported in Table \ref{methods:effective_parameter_check}, the sampled-state Jacobian attains rank $P$ for every tested unitary-part circuit containing the bridge layers, including the largest $14$-qubit unitary part. Equation~\eqref{methods:eq:sampled_rank_inequality} therefore yields $P_{\mathrm{eff}}=P$ for all these circuits, proving the proposition.
	\end{proof}

	In practice, $J_{\psi}(\boldsymbol{\theta}^{*})$ is computed by automatic differentiation at randomly selected parameter points. Its rank is determined through a singular value decomposition by counting the singular values above a prescribed numerical tolerance. In exact arithmetic, the right singular vectors associated with zero singular values span the redundant parameter directions. We choose $K_{\mathrm{samp}}$ such that
	\begin{equation}
		K_{\mathrm{samp}}\,2^{n+1}\geq P,
	\end{equation}
	so that the sampled-state Jacobian has sufficiently many rows to attain full column rank, and increase $K_{\mathrm{samp}}$ when necessary until the computed rank stabilizes. The maximum rank over several random parameter points is used to avoid accidentally evaluating the circuit at a nongeneric rank-deficient point. Since the circuit map is real analytic, its maximal Jacobian rank is attained almost everywhere in the parameter space.

	As a validation, the same procedure gives $\operatorname{rank}J=1$ for two sequential $RZ$ gates and $\operatorname{rank}J=495$ for two overlapping $SU(16)$ circuits, agreeing with the corresponding analytical values. Moreover, Eq.~\eqref{methods:effective_parameter_scalable} reduces the dimension of the differentiated output from $\mathcal{O}(4^{n})$ for the full unitary matrix to $\mathcal{O}(K_{\mathrm{samp}}2^{n})$ for the sampled states, making the effective-parameter check tractable for the $14$-qubit circuits used in this work.

	\begin{table}[htbp]
		\centering
		\begin{tabular}{lllll}
			\toprule
			$n$ & $SU(16)$ layer number & $K_{\mathrm{samp}}$ & $P_{\mathrm{nom}}$ & $P_{\mathrm{eff}}$ \\
			\midrule
			4 & 1 & \makecell{(checked directly\\by Eq. \ref{methods:effective_parameter_original})} & 255 & 255 \\
			8 & 2 & 16 & 773 & 773 \\
			8 & 5 & 32 & 2072 & 2072 \\
			8 & 9 & 32 & 3634 & 3634 \\
			8 & 13 & 64 & 5196 & 5196 \\
			10 & 2 & 8 & 1030 & 1030 \\
			10 & 5 & 16 & 2590 & 2590 \\
			10 & 9 & 16 & 4670 & 4670 \\
			10 & 13 & 16 & 6750 & 6750 \\
            12 & 2 & 4 & 1287 & 1287 \\
            12 & 5 & 8 & 3363 & 3363 \\
            12 & 9 & 16 & 5961 & 5961 \\
            12 & 13 & 16 & 8559 & 8559 \\
            14 & 2 & 4 & 1544 & 1544 \\
            14 & 5 & 4 & 3881 & 3881 \\
            14 & 9 & 4 & 6997 & 6997 \\
            14 & 13 & 8 & 10113 & 10113 \\
			\bottomrule
		\end{tabular}
		\caption{The results of the effective parameter number checks using Eq. \ref{methods:effective_parameter_scalable}. Here $n$, $K_{\mathrm{samp}}$, $P_{\mathrm{nom}}$, and $P_{\mathrm{eff}}$ denote the qubit number of the circuit, the number of randomly sampled input states, the nominal parameter number, and the lower bound of the effective parameter number, respectively. $K_{\mathrm{samp}}$ is chosen to ensure the condition $K_{\mathrm{samp}}2^{n+1}\geq P$. All the lower bounds of the effective parameter number reach the nominal parameter number, showing that the bridge layers are completely effective, and no parameter redundancy exists in the unitary part of $G_{Q_i}$, $G_{K_i}$, $G_{V_i}$, $1\leq i\leq 128$ and $G_{W_i}$, $1\leq i\leq 2$.}
		\label{methods:effective_parameter_check}
	\end{table}

	Next, we use the $G_{Q_i}, 1\leq i\leq 128$ to construct the gate $G_Q$ described in Eq. \ref{methods:attention-head}. To do this, we first need the controlled quantum gates $CG_{Q_i}$ and $\overline{C}G_{Q_i}$. we have:
	\begin{equation}
		\begin{matrix}
			CG_{Q_i} =
			\begin{bmatrix}
				I & \\
				& G_{Q_i}
			\end{bmatrix}, &
			\overline{C}G_{Q_i} =
			\begin{bmatrix}
				G_{Q_i} & \\
				& I
			\end{bmatrix}
		\end{matrix}
	\end{equation}
	where $CG_{Q_i}$ and $\overline{C}G_{Q_i}$ are the controlled form of $G_{Q_i}$, being active to the amplitude of $\ket{1}$ and $\ket{0}$ of the quantum state of the controlling qubit relatively. From this, we can derive the controlled-$G_{Q_i}$ gates with two control qubits. For example:
	\begin{equation}
		\begin{matrix}
			CCG_{Q_i} =
			\begin{bmatrix}
				I & & & \\
				& I & & \\
				& & I & \\
				& & & G_{Q_i}
			\end{bmatrix}, &
			C\overline{C}G_{Q_i} =
			\begin{bmatrix}
				I & & & \\
				& I & & \\
				& & G_{Q_i} & \\
				& & & I
			\end{bmatrix} \\
			\\
			\overline{C}CG_{Q_i} =
			\begin{bmatrix}
				I & & & \\
				& G_{Q_i} & & \\
				& & I & \\
				& & & I
			\end{bmatrix}, &
			\overline{C}\overline{C}G_{Q_i} =
			\begin{bmatrix}
				G_{Q_i} & & & \\
				& I & & \\
				& & I & \\
				& & & I
			\end{bmatrix}
		\end{matrix}
	\end{equation}
	These represent controlled-$G_{Q_i}$ gates controlled by two qubits, being active on the amplitude of $\ket{11}$, $\ket{10}$, $\ket{01}$ and $\ket{00}$ of the controlling qubits. Similarly, the following gate with 128 matrix blocks on the diagonal line
	\begin{equation}
		C_7^{(i)}G_{Q_i} = \begin{bmatrix}
			I \\
			& \dots \\
			& & I \\
			& & & G_{Q_i} \\
			& & & & I \\
			& & & & & \dots \\
			& & & & & & I
		\end{bmatrix}_{16384 \times 16384},
		\label{methods:C7GQ}
	\end{equation}
	where the matrix block $G_{Q_i}$ and the identity matrices are $128 \times 128$ in size, is a 7-bit controlled gate. There are $i-1$ identity matrices at the top left and $128-i$ identity matrices at the bottom right of $G_{Q_i}$. The derivation of the controlling logic of the 7 controlling qubits is easy: it forms exactly the binary form of $i-1$. For example, $C_7^{(1)}=\overline{C}\overline{C}\overline{C}\overline{C}\overline{C}\overline{C}\overline{C}$, $C_7^{(2)}=\overline{C}\overline{C}\overline{C}\overline{C}\overline{C}\overline{C}C$, and $C_7^{(128)}=CCCCCCC$. Once we derive all the proper 7-bit controlled circuits $C_7^{(1)}G_{Q_1}, C_7^{(2)}G_{Q_2}, \ldots, C_7^{(128)}G_{Q_{128}}$, We can use them to construct the circuit $G_Q$:
	\begin{equation}
		\begin{aligned}
			C_7^{(1)}G_{Q_1}C_7^{(2)}G_{Q_2}\ldots C_7^{(128)}G_{Q_{128}}&=
			\begin{bmatrix}
				G_{Q_1} \\ 
				& I \\ 
				& & \vdots \\ 
				& & & I \\ 
			\end{bmatrix}
			\begin{bmatrix}
				I \\
				& G_{Q_2} \\ 
				& & I \\ 
				& & & \vdots \\ 
				& & & & I \\ 
			\end{bmatrix}...
			\begin{bmatrix}
				I \\ 
				& \vdots \\ 
				& & I \\ 
				& & & G_{Q_{128}} \\ 
			\end{bmatrix} \\
			&=\begin{bmatrix}
				G_{Q_1} \\
				& G_{Q_2} \\
				& & \dots \\
				& & & G_{Q_{128}}
			\end{bmatrix}=G_Q.
		\end{aligned}
		\label{methods:ConstructGQ}
	\end{equation}

	Next, we consider the implementation of the quantum circuits $C_7^{(1)}G_{Q_1}, C_7^{(2)}G_{Q_2}, \ldots, C_7^{(128)}G_{Q_{128}}$. $G_{Q_1}$ to $G_{Q_{128}}$ is implemented by the SVD which decomposes the $G_{Q_i}$ into $G_{Q_i}=U^{(i)}D^{(i)}V^{(i)\dagger}$. To turn them into their controlled form, we turn the SVD components into their controlled form respectively:
	\begin{equation}
		\label{methods:C7Decom}
		C_7^{(i)}G_{Q_i}=C_7^{(i)}U^{(i)}C_7^{(i)}D^{(i)}C_7^{(i)}V^{(i)\dagger}.
	\end{equation}
	The implementation of $C_7^{(i)}U^{(i)}$ and $C_7^{(i)}V^{(i)\dagger}$ is straightforward, we simply need to add 7 controlling qubits to the quantum circuit shown in Fig. \ref{methods:fig_unitary_layer_7}. However, the implementation of $C_7^{(i)}D^{(i)}$ cannot be so straightforward due to the presence of measurement and post-selection operations. Let $c=i-1$ be the computational-basis index that activates $C_7^{(i)}$. Denoting the input state of the seven control qubits by $\ket{\phi}_{7}=\sum\limits_{j=0}^{127}\phi_j\ket{j}_7$, the ancillary qubit by $\ket{0}_{anc}$, and the target state by $\ket{\psi}_{tar}$, straightforwardly adding the $C_7^{(i)}$ control to the circuit in Fig. \ref{methods:fig_diag_circuit} gives
	\begin{equation}
		\begin{aligned}
			& \ket{\phi}_{7}\otimes\ket{0}_{anc}\otimes\ket{\psi}_{tar}\\ 
			& \rightarrow \phi_c \ket{c}_{7} (\ket{0}_{anc}( \cos(\theta) \ket{\psi})_{tar} + \ket{1}_{anc} (\sin(\theta) \ket{\psi})_{tar}) + \sum_{\substack{j=0\\j\neq c}}^{127} \phi_j \ket{j}_{7} \ket{0}_{anc} \ket{\psi}_{tar}\\
			& = \left(\phi_c \ket{c}_{7} (\cos \theta \ket{\psi})_{tar} + \sum_{\substack{j=0\\j\neq c}}^{127} \phi_j \ket{j}_{7} \ket{\psi}_{tar}\right) \ket{0}_{anc} + \left(\phi_c \ket{c}_{7} (\sin \theta \ket{\psi})_{tar}\right) \ket{1}_{anc},
		\end{aligned}
	\end{equation}
	Here $\ket{c}_{7}$ is the state of the seven control qubits that activates the controlled circuit. If measurement of the ancillary qubit gives 0, the control and target registers collapse onto $\phi_c \ket{c}_{7} (\cos \theta \ket{\psi})_{tar} + \sum_{j\neq c} \phi_j \ket{j}_{7} \ket{\psi}_{tar}$, which matches the expected operation of $C_7^{(i)}D^{(i)}$. If the result is 1, they collapse onto $\phi_c \ket{c}_{7} (\sin \theta \ket{\psi})_{tar}$, which does not preserve the inactive control branches. Therefore, $C_7^{(i)}D^{(i)}$ cannot be obtained by simply adding control qubits.

\begin{figure}
		\centering
		\includegraphics[width=0.6\linewidth]{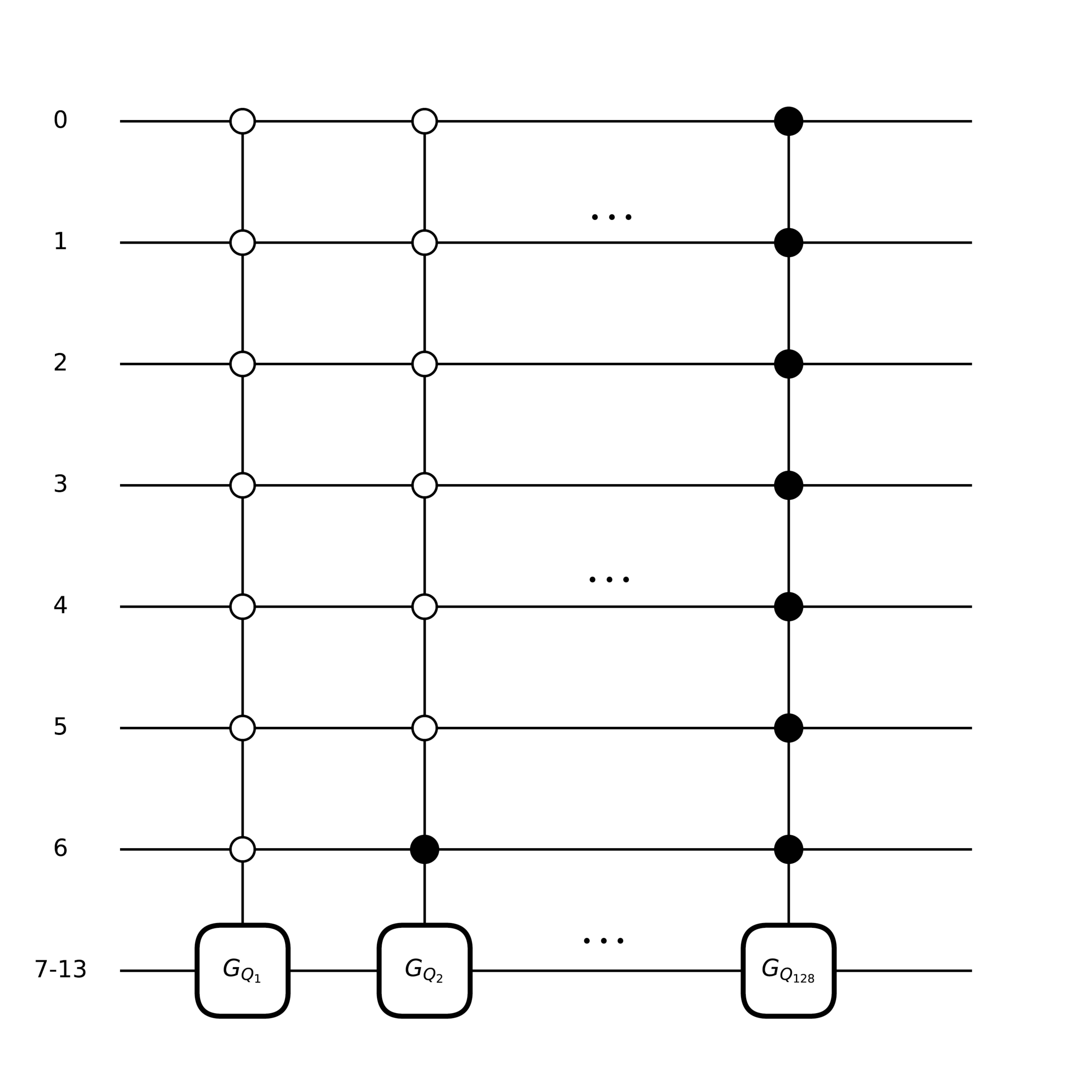}
		\caption{The quantum circuit diagram of the multi-head attention circuit $G_Q$. It contains 128 attention heads, realized by the 7-qubit sub-circuits $G_{Q_1}$ to $G_{Q_{128}}$. The controlled sub-circuit sequence implements the block-diagonal transformation of Eq. \ref{methods:attention-head}. Each of the sub-circuit is constructed by the SVD quantum circuit structure shown in Fig. \ref{methods:fig_attention_svd} so that the attention head matrices can be arbitrary, overcoming the unitary constraint of the quantum state evolution.}
		\label{methods:fig_attention_c7_sequence}
	\end{figure}

	Fortunately, we just need to make some small adjustment to implement the circuit $C_7^{(i)}D^{(i)}$ without changing the structure of the circuit $D^{(i)}$. The structure of the circuit is shown in Fig. \ref{methods:fig_controlled_non_unitary_diag}. The circuit performs the following operations:
	\begin{figure}[htbp]
		\centering
		\includegraphics[width=0.95\linewidth]{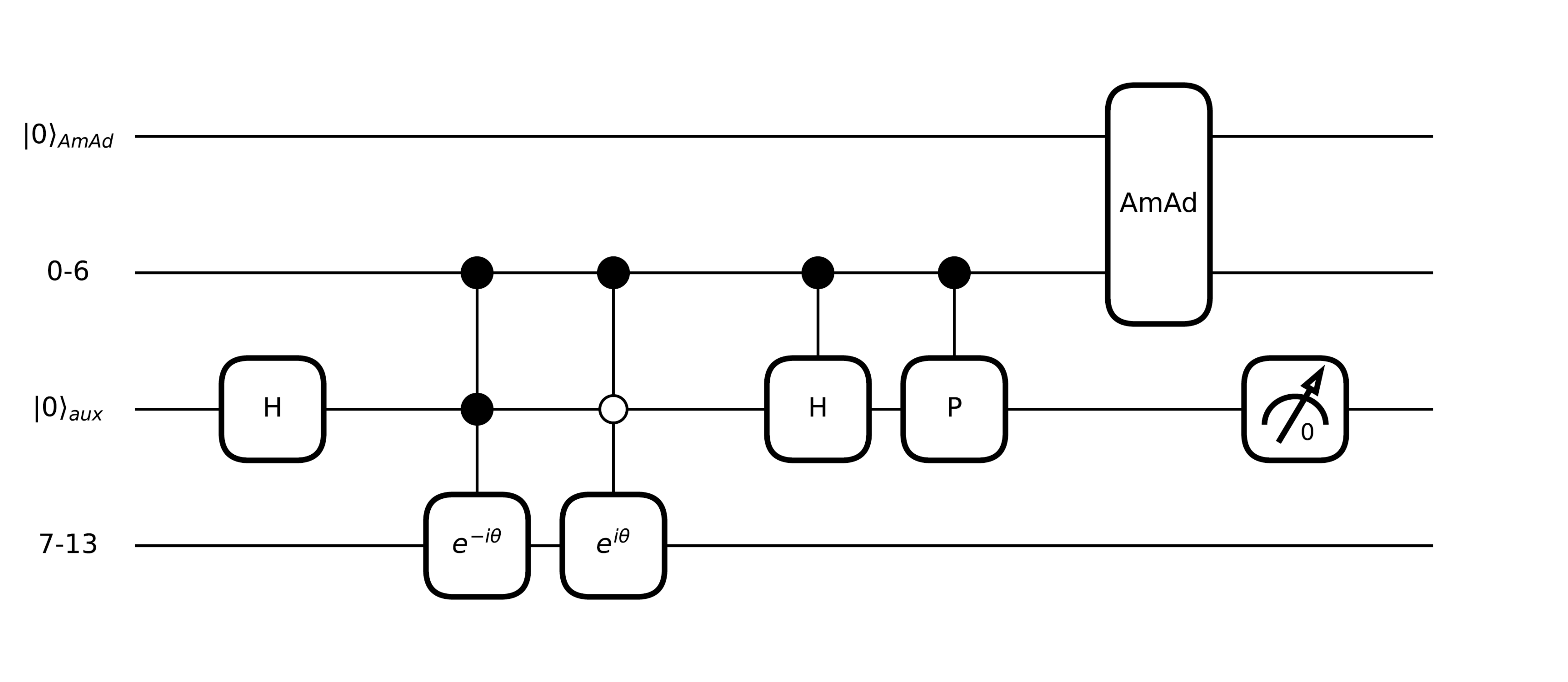}
		\caption{Quantum circuit implementing the controlled non-unitary diagonal operation $C_7^{(i)}D^{(i)}$. The seven control qubits select the diagonal transformation, while the ancillary qubits support the probabilistic diagonal operation and amplitude adjustment.}
		\label{methods:fig_controlled_non_unitary_diag}
	\end{figure}
	\begingroup
	\setlength{\jot}{1pt}
	\begin{equation}
		\begin{aligned}
			& \ket{\phi}_{7}\otimes\ket{0}_{anc}\otimes\ket{\psi}_{tar}\\
			\stackrel{H_{anc}}\rightarrow &\frac{1}{\sqrt{2}}\ket{\phi}_{7}\otimes(\ket{0}_{anc}+\ket{1}_{anc})\otimes\ket{\psi}_{tar}\\
			\stackrel{C_7^{(i)}Ce^{(-i\theta)}}\rightarrow &\frac{1}{\sqrt{2}}\left(\phi_c\ket{c}_{7}\otimes(\ket{0}_{anc}\otimes e^{-i\theta}\ket{\psi}_{tar}+\ket{1}_{anc}\otimes\ket{\psi}_{tar})\right.\\
			& \left.+\sum\limits_{\substack{j=0\\j\neq c}}^{127}\phi_j\ket{j}_7\otimes(\ket{0}_{anc}+\ket{1}_{anc})\otimes\ket{\psi}_{tar}\right)\\
			\stackrel{C_7^{(i)}\overline{C}e^{(i\theta)}}\rightarrow &\frac{1}{\sqrt{2}}\left(\phi_c\ket{c}_{7}\otimes(\ket{0}_{anc}\otimes e^{-i\theta}\ket{\psi}_{tar}+\ket{1}_{anc}\otimes e^{i\theta}\ket{\psi}_{tar})\right.\\
			& \left.+\sum\limits_{\substack{j=0\\j\neq c}}^{127}\phi_j\ket{j}_7\otimes(\ket{0}_{anc}+\ket{1}_{anc})\otimes \ket{\psi}_{tar}\right)\\
			\stackrel{C_7^{(i)}H_{anc}}\rightarrow &\frac{1}{2}\left(\phi_c\ket{c}_{7}\otimes(\ket{0}_{anc}+\ket{1}_{anc})\otimes e^{i\theta}\ket{\psi}_{tar}+(\ket{0}_{anc}-\ket{1}_{anc})\otimes e^{-i\theta}\ket{\psi}_{tar}\right.\\
			& \left.+\sqrt{2}\sum\limits_{\substack{j=0\\j\neq c}}^{127}\phi_j\ket{j}_7\otimes(\ket{0}_{anc}+\ket{1}_{anc})\otimes \ket{\psi}_{tar}\right)\\
			=&\phi_c\ket{c}_{7}\otimes\left( \ket{0}_{anc}\otimes \cos{\theta}\ket{\psi}_{tar}-\ket{1}_{anc}\otimes i\sin{\theta}\ket{\psi}_{tar}\right)\\
			&+\frac{1}{\sqrt{2}}\sum\limits_{\substack{j=0\\j\neq c}}^{127}\phi_j\ket{j}_7\otimes(\ket{0}_{anc}+\ket{1}_{anc})\otimes \ket{\psi}_{tar}\\
		\end{aligned}
	\end{equation}
	\endgroup
	Applying the ancillary phase gate and the amplitude-adjustment operation then gives
	\begin{equation}
		\begin{aligned}
			&\phi_c\ket{c}_{7}\otimes\left( \ket{0}_{anc}\otimes \cos{\theta}\ket{\psi}_{tar}-\ket{1}_{anc}\otimes i\sin{\theta}\ket{\psi}_{tar}\right)\\
			&\quad+\frac{1}{\sqrt{2}}\sum\limits_{\substack{j=0\\j\neq c}}^{127}\phi_j\ket{j}_7\otimes(\ket{0}_{anc}+\ket{1}_{anc})\otimes \ket{\psi}_{tar}\\
			\stackrel{C_7^{(i)}P_{anc}}\rightarrow &\phi_c\ket{c}_{7}\otimes\left( \ket{0}_{anc}\otimes \cos{\theta}\ket{\psi}_{tar}+\ket{1}_{anc}\otimes \sin{\theta}\ket{\psi}_{tar}\right)\\
			&+\frac{1}{\sqrt{2}}\sum\limits_{\substack{j=0\\j\neq c}}^{127}\phi_j\ket{j}_7\otimes(\ket{0}_{anc}+\ket{1}_{anc})\otimes \ket{\psi}_{tar}\\
			\stackrel{AmAd^{(i)}}\rightarrow &\frac{1}{\sqrt{2}}\phi_c\ket{c}_{7}\otimes\left( \ket{0}_{anc}\otimes \cos{\theta}\ket{\psi}_{tar}+\ket{1}_{anc}\otimes \sin{\theta}\ket{\psi}_{tar}\right)\\
			&+\frac{1}{\sqrt{2}}\sum\limits_{\substack{j=0\\j\neq c}}^{127}\phi_j\ket{j}_7\otimes(\ket{0}_{anc}+\ket{1}_{anc})\otimes \ket{\psi}_{tar}\\
			&\propto \left(\phi_c\ket{c}_7\otimes\cos{\theta}\ket{\psi}_{tar}+\sum\limits_{\substack{j=0\\j\neq c}}^{127}\phi_j\ket{j}_7\otimes\ket{\psi}_{tar}\right)\otimes\ket{0}_{anc}\\
			&+\left(\phi_c\ket{c}_7\otimes\sin{\theta}\ket{\psi}_{tar}+\sum\limits_{\substack{j=0\\j\neq c}}^{127}\phi_j\ket{j}_7\otimes\ket{\psi}_{tar}\right)\otimes\ket{1}_{anc}
		\end{aligned}
	\end{equation}
	where the amplitude adjustment gate $AmAd^{(i)}$ can be represented by Dirac notation as
	\begin{equation}
		AmAd^{(i)}=\left(\frac{1}{\sqrt{2}}\ket{c}_{7}\!\bra{c}_{7}+\sum\limits_{\substack{j=0\\j\neq c}}^{127}\ket{j}_7\!\bra{j}_{7}\right)\otimes I_{tar}.
	\end{equation}
	It is a non-unitary diagonal gate and can be implemented by the circuit shown in Fig. \ref{methods:fig_diag_circuit} with another ancillary qubit, denoted by $\ket{0}_{anc2}$ in Fig. \ref{methods:fig_controlled_non_unitary_diag}. The effect of the circuit $C_7^{(i)}D^{(i)}$ is exactly what we want. With the circuit $C_7^{(i)}U^{(i)}$ and $C_7^{(i)}V^{(i)\dagger}$, we construct the circuit $C_7^{(i)}G_{Q_i}$ by Eq. \ref{methods:C7Decom}, and construct the whole circuit $G_Q$ by Eq. \ref{methods:ConstructGQ}. The diagram of $G_Q$ is shown in Fig. \ref{methods:fig_attention_c7_sequence}. It consists 128 attention heads $G_{Q_1}$ to $G_{Q_{128}}$, each 7-qubit sub-circuit $G_{Q_i}$ is controlled by the rest 7 qubits under the controlling logic described above. $G_{K_i}$ and $G_{V_i}$ can be constructed in the exactly same way.

	\begin{figure}
		\centering
		\includegraphics[width=0.95\linewidth]{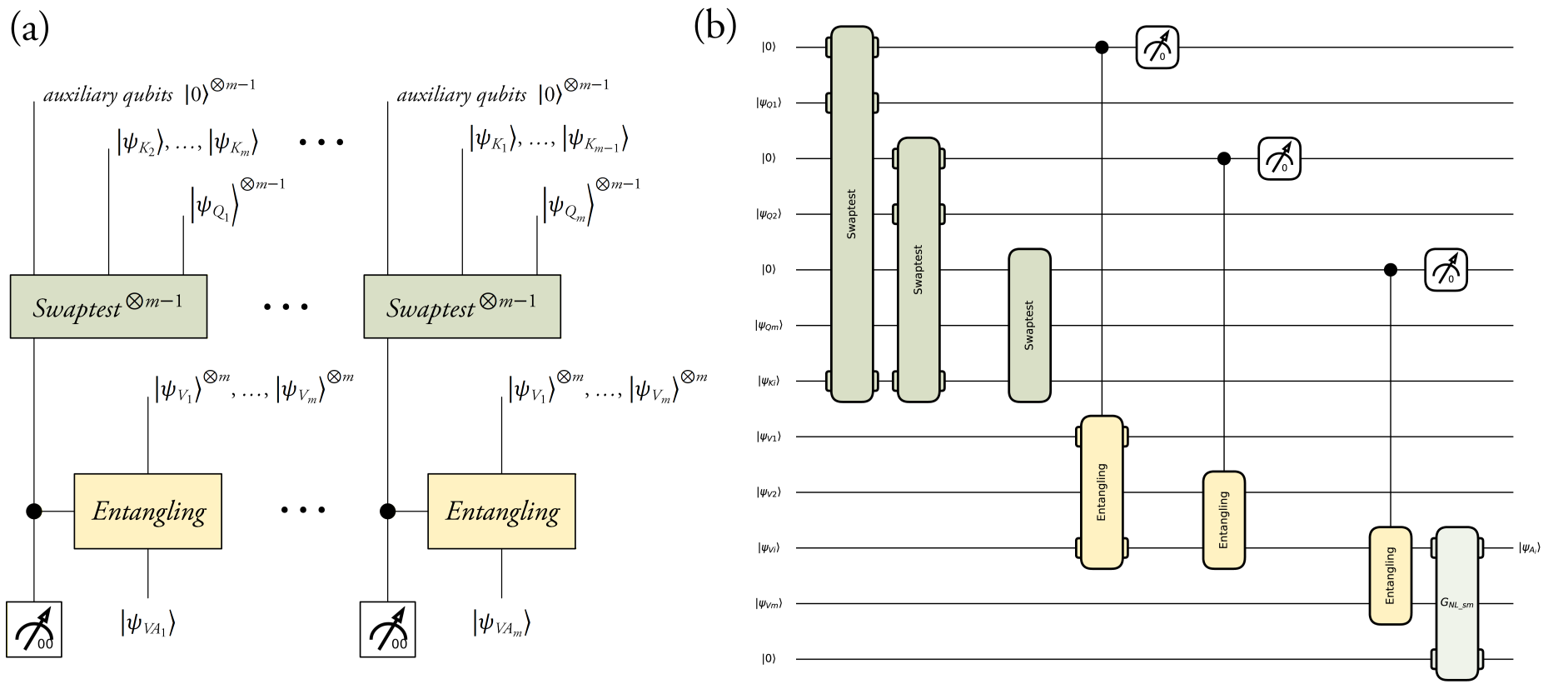}
		\caption{Detail of the quantum mixing circuits in the self-attention layer. They take the outputs of the attention-head circuits as input and pass them through a series of ordered SwapTest and entangling circuits. (a) General structure of the quantum mixing circuit. For token position $j$, the $m-1$ copies of the query state $\ket{\psi_{Q_j}}$ pass through SwapTest circuits with the key states $\ket{\psi_{K_\ell}}$, $\ell\neq j$. The fidelity between the query states and key states are encoded in the density operators of the auxiliary qubits. Copies of the value states $\ket{\psi_{V_\ell}}$, $1\leq\ell\leq m$, are then mixed by the entangling circuits under the control of these auxiliary qubits. Finally, the SwapTest auxiliary qubits are measured and reset. (b) Detailed arrangement of the SwapTest and entangling circuits in (a).}
		\label{methods:fig_attention_Mix}
	\end{figure}

	\subsubsection{Superposition of the attention states by quantum mixing circuits}
\begin{figure}
		\centering
		\includegraphics[width=0.7\linewidth]{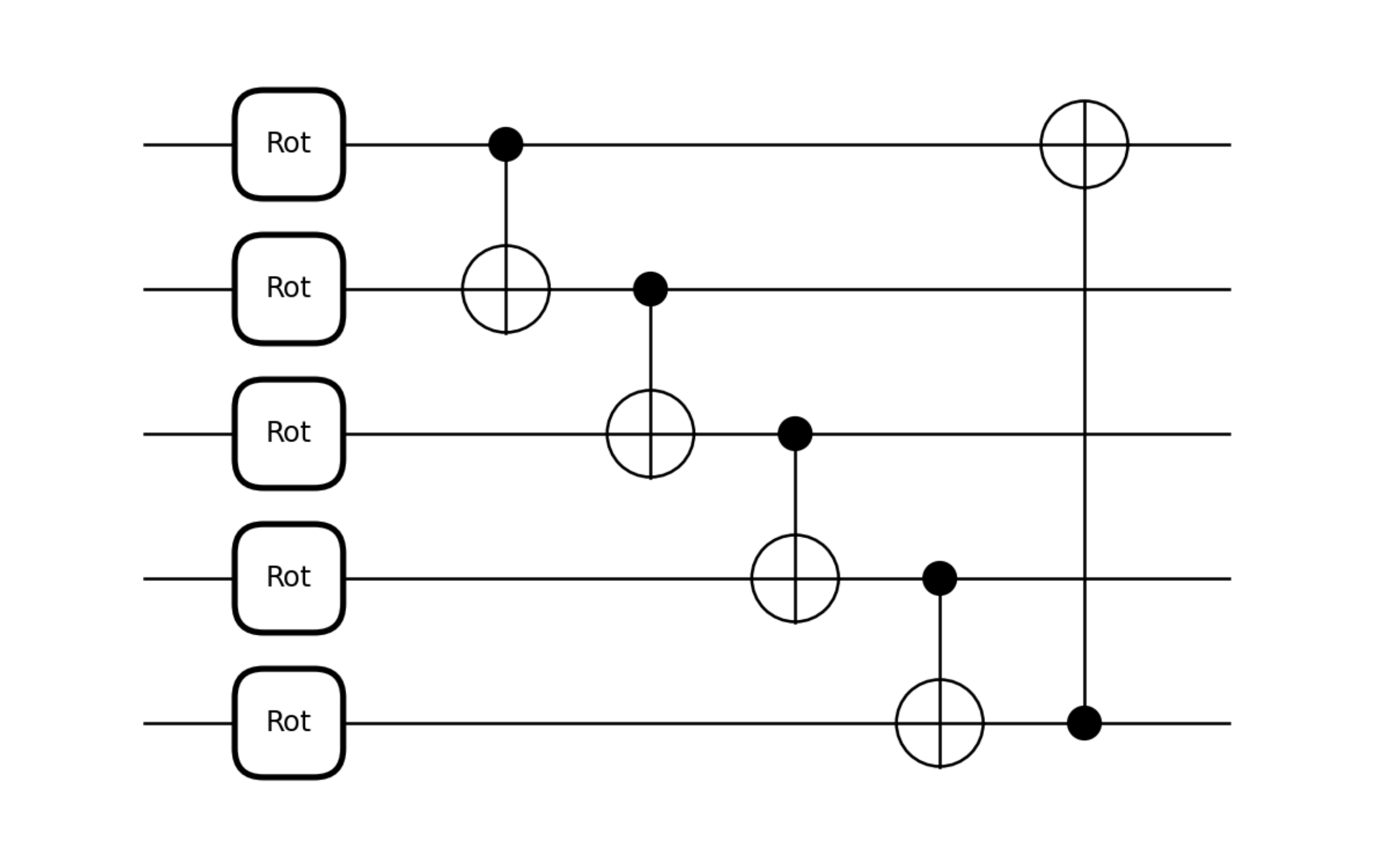}
		\caption{The quantum structure of the Strongly Entangling circuit from the paper \cite{bib11}. It consists single-bit general rotations gates on all the qubits at the beginning, followed by the CNOT gates perform on the adjacent qubits on sequence.}
		\label{methods:fig_Entangling}
		\includegraphics[width=0.5\linewidth]{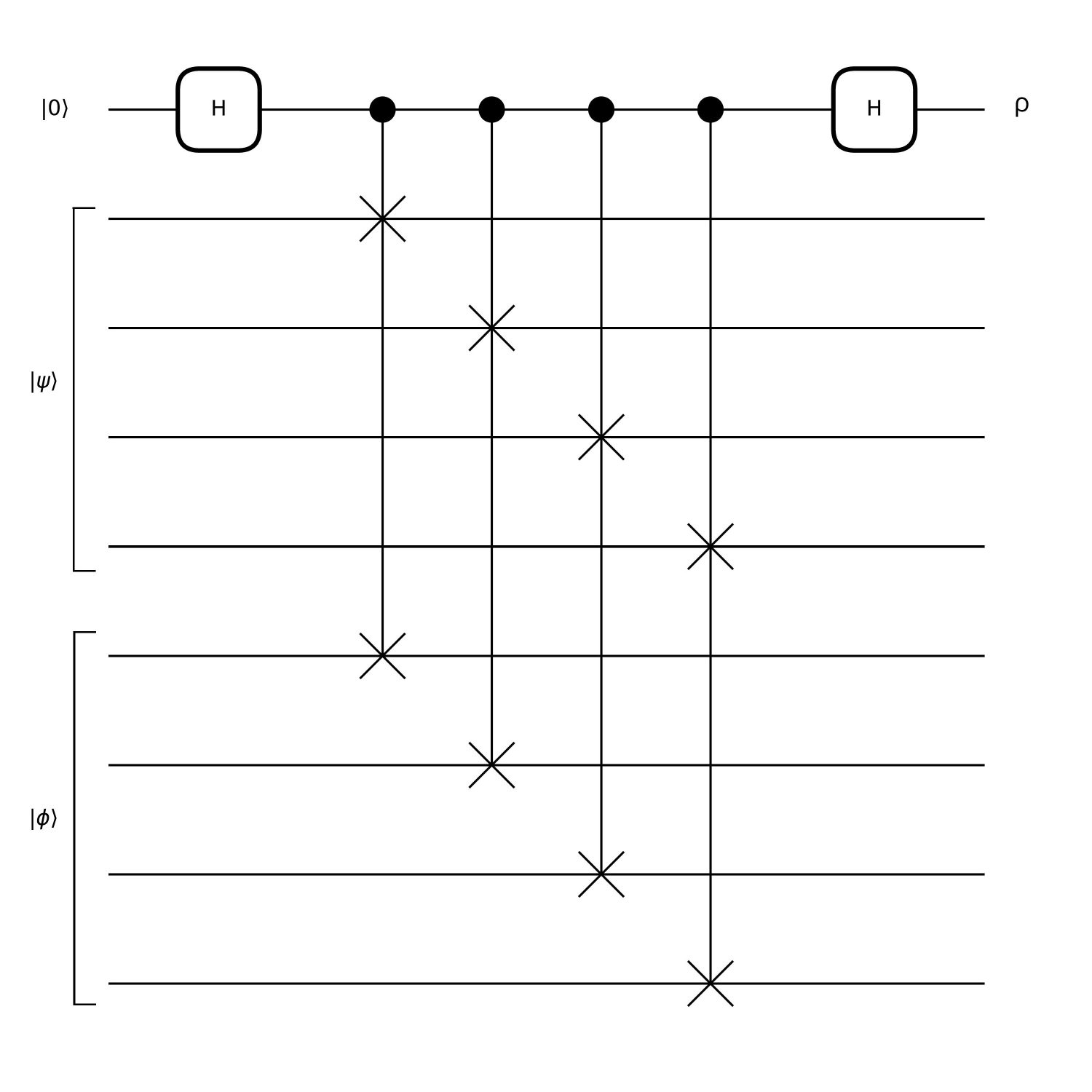}
		\caption{The quantum structure of the SwapTest circuit. Its input consists of an auxiliary qubit and two groups of qubits. After the controlled-Swap and Hadamard gates, the reduced density operator of the auxiliary qubit contains information about the overlap of the two input states. Measuring the auxiliary qubit gives the result 1 with probability $\frac{1}{2}\left(1-|\braket{\psi|\phi}|^2\right)$.}
		\label{methods:fig_SwapTest}
	\end{figure}

	After applying $G_Q$, $G_K$, and $G_V$, we obtain $m-1$ copies of each query state $\ket{\psi_{Q_j}}$ and key state $\ket{\psi_{K_j}}$, and $m$ copies of each value state $\ket{\psi_{V_j}}$, where $1\leq j\leq m$. The next step is to create a superposition of the value states using the query--key overlaps as weights. However, the mixing strategy used in classical models such as GPT-3 cannot be directly implemented in quantum computation, as Michał Oszmaniec et al.~\cite{bib9} showed that a circuit for acquiring a superposition of arbitrary unknown quantum states with arbitrary weights does not exist. To address this, we designed a quantum mixing method based on the SwapTest circuit\cite{bib12,bib13} and the Strongly Entangling circuit\cite{bib11}, as shown in Fig. \ref{methods:fig_attention_Mix}. For each token position $j$, every copy of $\ket{\psi_{Q_j}}$ passes through a SwapTest circuit with $\ket{\psi_{K_\ell}}$, where $1\leq\ell\leq m$ and $\ell\neq j$, together with an auxiliary qubit initialized to $\ket{0}$. The structure of the SwapTest circuit is shown in Fig.~\ref{methods:fig_SwapTest}. We denote the reduced density operator of the corresponding auxiliary qubit by $\rho_{ST_{j\ell}}$. The overlap between $\ket{\psi_{Q_j}}$ and $\ket{\psi_{K_\ell}}$ is encoded in $\rho_{ST_{j\ell}}$. The corresponding copies of the value states then pass through Strongly Entangling circuits controlled by these auxiliary qubits, thereby mixing the value states according to the query--key overlaps. After entanglement, the auxiliary qubits are measured, and we post-select the outcome $0$. Performing all SwapTests and entangling operations simultaneously requires $(m-1)^2$ auxiliary qubits. Since measured auxiliary qubits can be recycled, this cost can be reduced: if the process is divided into $g$ groups, where $1\leq g\leq(m-1)^2$, the number of auxiliary qubits is $\lceil(m-1)^2/g\rceil$. We denote the resulting mixed value states by $\ket{\psi_{VA_1}},\ket{\psi_{VA_2}},\ldots,\ket{\psi_{VA_m}}$.

	\subsubsection{Approximated nonlinear transformation by pre-trained quantum circuits}
	Finally, we perform the softmax transformation on the mixed value states $\ket{\psi_{VA_1}}, \ket{\psi_{VA_2}}, \ldots, \ket{\psi_{VA_m}}$. The softmax function is defined as
	\begin{equation}
		\mathrm{Softmax}(\vec{x}) = \frac{\left(e^{\beta x_1}, e^{\beta x_2}, \ldots, e^{\beta x_{2^n}}\right)}{\sum\limits_{k=1}^{2^n} e^{\beta x_k}},
	\end{equation}
	where $\beta$ is a fixed parameter that cannot be trained. Due to its nonlinear nature, implementing the softmax function directly on quantum states is challenging. To address this, we design a quantum circuit capable of approximating the softmax function using one auxiliary qubit. We first perform a unitary layer shown in Fig. ~\ref{methods:fig_unitary_layer_14} on the data qubits, then a bridge layer on all the qubits, including the auxiliary qubit. Next, we apply another unitary layer of Fig. ~\ref{methods:fig_unitary_layer_14} to the auxiliary qubit and the first $(n-1)$ data qubits, again modifying the last multi-controlled X gate to act on all qubits. Finally, we measure the auxiliary qubit and post-select the outcome $0$. The outputs of the softmax circuit, which also serve as the final outputs of the self-attention layer, are denoted as $\ket{\psi_{A_1}}, \ket{\psi_{A_2}}, \ldots, \ket{\psi_{A_m}}$. After the execution of the softmax circuit, the auxiliary qubit can be recycled for further use. 

\begin{figure}
		\centering
		\includegraphics[width=0.6\linewidth]{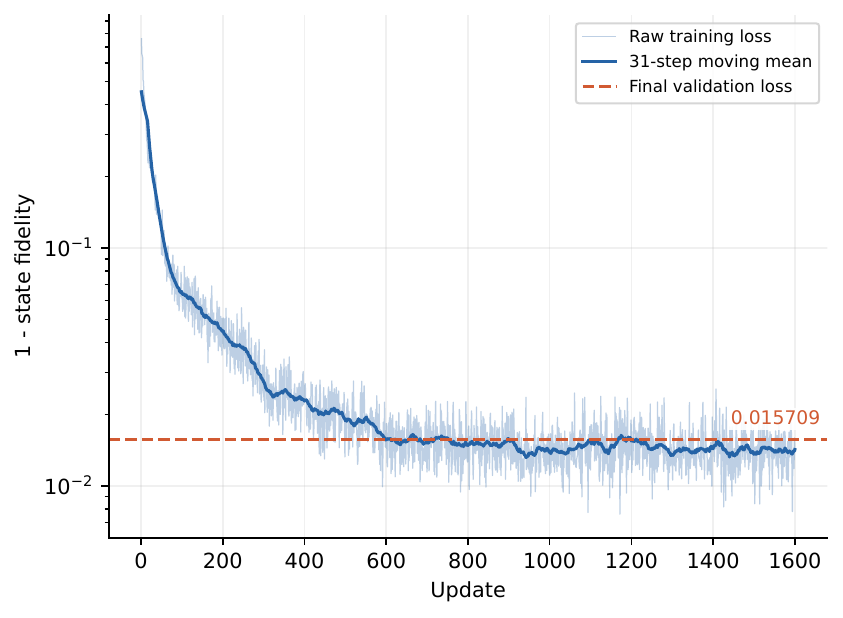}
		\caption{The training of a 4-qubit Softmax Approximation Circuit. Parameter-shift is used to acquire the parameter gradients, and Adam optimizer is used to update the circuit parameters. After 1600 parameter updatings, the fidelity of the circuit can reach 0.984, effectively approximates the Softmax function.}
		\label{methods:fig_train_softmax}
	\end{figure}

	The softmax circuit must be trained independently before being incorporated into the full quantum model. To generate the training data, we create $1600 \times 16 = 25600$ pairs of initial and target states. The initial states $\vec{s}_{\mathrm{ini}}^{(i)}$, $1\leq i\leq25600$, are randomly generated normalized $2^n$-dimensional real vectors, while the target states $\vec{s}_{\mathrm{tar}}^{(i)}$ are obtained by applying the Softmax function to the corresponding initial states. For training, we initialize all quantum-gate parameters to small random values. We prepare $n$ qubits in $\ket{s_{\mathrm{ini}}^{(i)}}$ and pass them through the softmax circuit. Another set of $n$ qubits is prepared in the corresponding target state $\ket{s_{\mathrm{tar}}^{(i)}}$, and a SwapTest circuit measures the fidelity between the target and output states. The infidelity is minimized using the Adam optimizer \cite{bib14} with learning rate 0.01 and parameters $\beta_1=0.9$ and $\beta_2=0.99$. The parameter-shift rule \cite{bib15,bib16} is used to compute the gradient. The circuit is trained for 1600 iterations using batches of 16 state pairs. The training results for a 4-qubit (plus 1 ancillary qubit) Softmax Approximation Circuit is shown in Fig \ref{methods:fig_train_softmax}. The final fidelity of the circuit after 1600 parameter updatings can reach 0.984, which confirm the feasibility of approximating the Softmax transformation on quantum states with this circuit structure.

	\subsection{The Feed-forward Layer Using Non-unitary Parameterized Quantum Circuits}

	In the classical Transformer architecture, the feed-forward network (FFN) module is positioned near the output of the decoder-only Transformer units. It is a fully connected neural network with a single hidden layer. The number of neurons in both the input and output layers matches the model dimension $d=12288$, while the hidden layer contains four times as many neurons, namely $4d=49152$. The FFN output is a vector with the same dimension as its input. Structurally, the feed-forward network consists of a two-layer perceptron, and its function can be expressed as
	\begin{equation}
		FFN(x)=\phi(xW_1+b_1)W_2+b_2,
	\end{equation}
	where $W_1$, $W_2$ are trainable weight matrices, and $b_1$, $b_2$ are trainable bias parameters. The $\phi(x)$ is the activation function of the neuron, which is unually the Sigmoid funciton $\phi(x)=\frac{1}{1+e^{-x}}$.

	Running and training such a neural network is computationally expensive. Each neuron in the hidden layer is fully connected to all neurons in both the input and output layers. As a result, there are $2\times4d\times d$ connections, leading to $O(d^2)$ multiplication and addition operations. Meanwhile, the number of element-wise operations for the $6d$ neurons is $O(d)$. Overall, the time complexity of the network is $O(d^2)$.

	Fortunately, quantum computing can significantly speed up this process. This is because the primary effect of the neural network consists of linear transformations together with the activation function. These linear transformations can be performed by quantum circuits. Since only $n=\lceil\log_2d\rceil$ qubits are required to represent the $d$-dimensional input and output vectors, an exponential speed-up can be achieved by implementing a quantum circuit whose effect mirrors that of the classical neural network. However, constructing the nonlinear Sigmoid transformation is challenging because quantum gates are linear.

\begin{figure}[ht]
		\centering
		\includegraphics[width=0.7\textwidth]{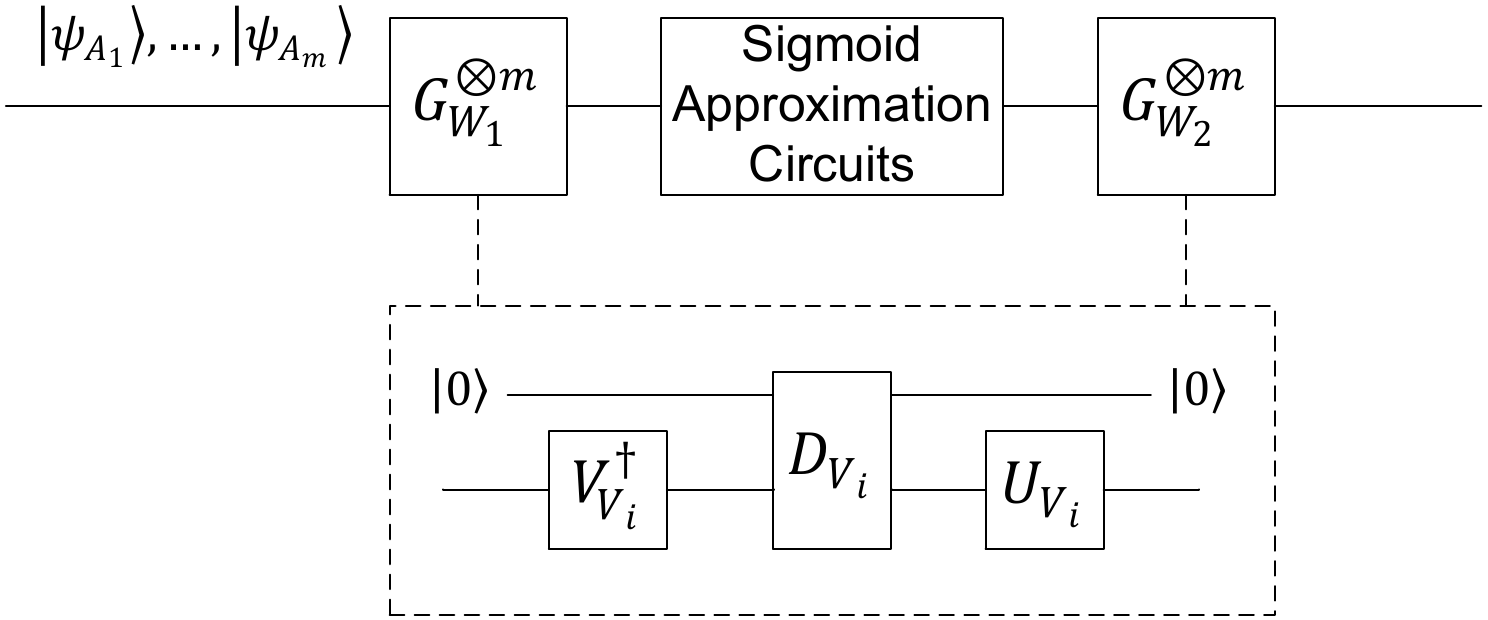}
		\caption{The diagram of the structure of the FFN layer. The output qubits $\ket{\psi_{A_1}}, \ldots, \ket{\psi_{A_m}}$ from the self-attention layer go into a $G_{W_1}$ - Sigmoid - $G_{W_2}$ - Sigmoid circuit to form the output of the FFN layer. The $G_{W_1}$ and $G_{W_2}$ are the quantum circuits that perform the equivalent transformation of a single classical neural network layer on the quantum states of the qubits. They are constructed by the SVD decomposition, illustrated in the lower diagram. The Sigmoid circuits are pre-trained quantum circuits that approximate the Sigmoid function transformation.}
		\label{methods:fig_FFN}
	\end {figure}

\begin{figure}[ht]
		\centering
		\includegraphics[width=1.0\textwidth]{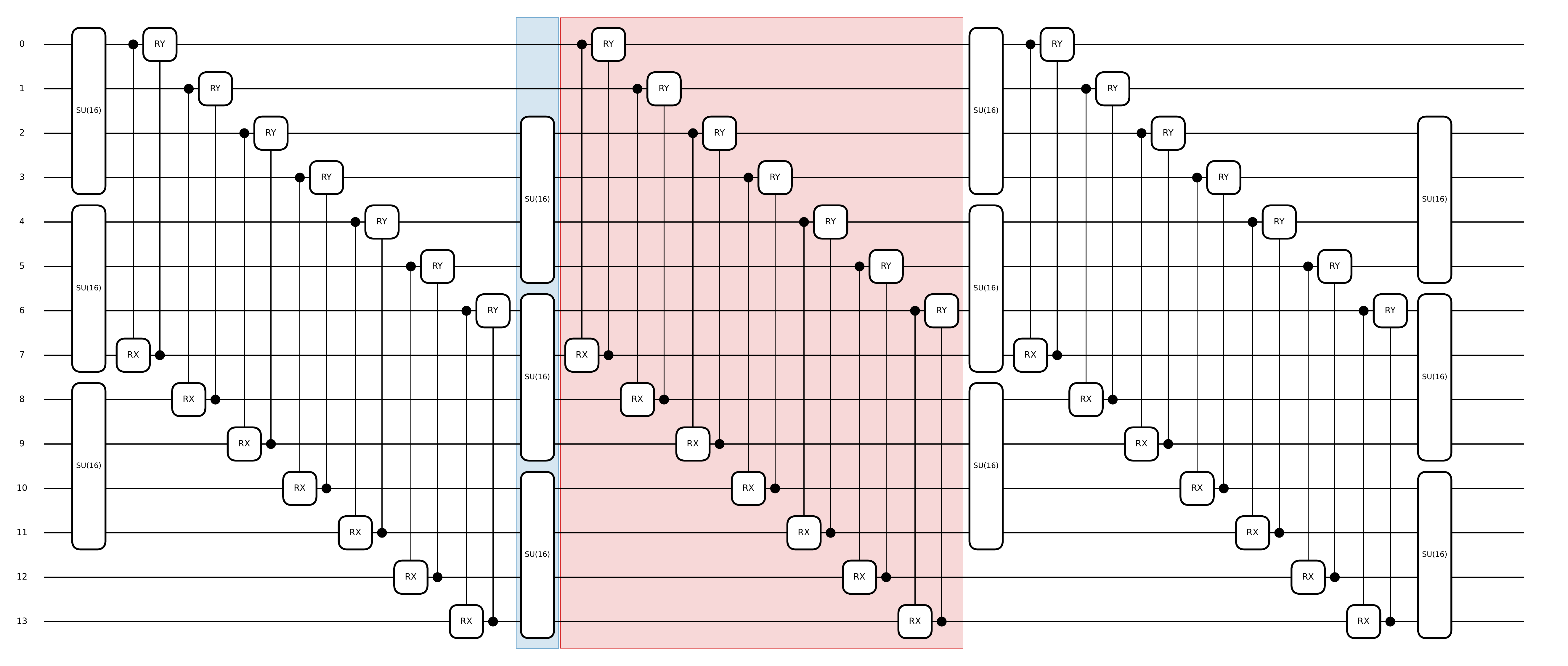}
		\caption{The quantum circuit diagram of the unitary part in the circuit $G_{W_i}$, $1\leq i\leq 2$. The circuit follows the pattern of $SU(16)$ circuit layers and bridge layers, marked by the blue box and the red box on the diagram respectively. The $SU(16)$ circuits in a single layer are arranged to operate in parallel to accommodate more trainable parameters in lower circuit depth, and bridge layers locate at the middle of the $SU(16)$ circuit layers to isolate the $SU(16)$ circuits, thus avoiding any parameter redundancy.}
		\label{methods:fig_unitary_layer_14}
	\end {figure}

	Here we present the equivalent quantum circuit of the classical FFN network of GPT-3. The gerenal structure of the circuit is shown in Fig. \ref{methods:fig_FFN}. The aim is to construct two quantum circuits $G_{W_1}$ and $G_{W_2}$, corresponding to the matrices $W_1$ and $W_2$, that operate the states of the qubits as the way that the classical FFN network operates the state vectors restored in classical bits. To take fully use of the potential of quantum circuits, we use the quantum circuit structure based on the singular value decomposition, which is also used in the quantum self-attention heads, to implement the quantum circuit, that is
	\begin{equation}
		G_{W_i}=U_{W_i}D_{W_i}V_{W_i}^\dagger, 1\leq i\leq 2.
	\end{equation}
	The structure of the non-unitary diagonal circuit $D_{W_i}$ is shown in Fig \ref{methods:fig_diag_circuit}, and the structure of the unitary parts $U_{W_i}$ and $V_{W_i}$ are shown in Fig. \ref{methods:fig_unitary_layer_14}.

\begin{figure}
		\centering
		\includegraphics[width=0.6\linewidth]{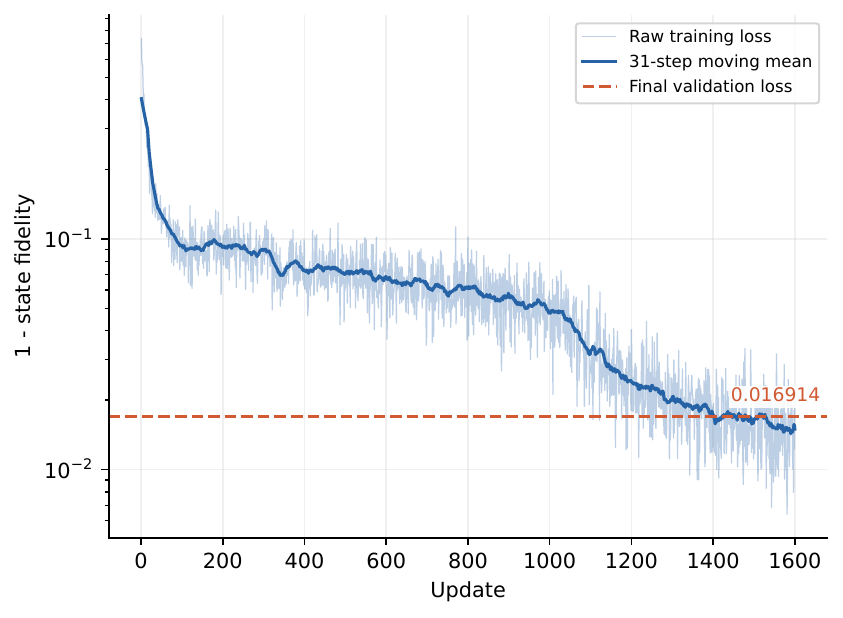}
		\caption{The training of a 4-qubit Sigmoid Approximation Circuit. Parameter-shift is used to acquire the parameter gradients, and Adam optimizer is used to update the circuit parameters. After 1600 parameter updatings, the fidelity of the circuit can reach 0.983, effectively approximates the Sigmoid function.}
		\label{methods:fig_train_sigmoid}
	\end{figure}

	Despite for the matrix transformations, the Sigmoid function also need to be performed on the quantum states. GPT-3 uses the following Sigmoid function as the nonlinear activation function of its neurons: 
	\begin{equation}
		\sigma(x)=\frac{1}{1+e^{-x}}.
	\end{equation}
	As the nonlinear nature of the Sigmoid function, the operation cannot be done directly on the quantum states of the qubits without using auxiliary qubits. To implement the Sigmoid operation on the qubits, we use the same structure of the circuits that perform Softmax function mentioned in Chapter \ref{methods:section_self_attention}, and train the Sigmoid circuits independently before putting them into the whole model. To generate the training data, we also use the same method of the training data generation of the Softmax circuit, and simply replace the target states with the results of the Sigmoid function of the corresponding initial states. The reward function, the optimizer and its parameters, the number of iterations and the batch size all remain the same with the training of the Softmax circuit. The training result of a 4-qubit Sigmoid Approximation Circuit is shown in Fig. \ref{methods:fig_train_sigmoid}. After 1600 parameter updatings, the fidelity of the circuit can reach 0.983, approximates the Sigmoid function effectively.

	\subsection{Training Algorithm based on Multi-agent Reinforcement Learning}

	The training dataset of the GPT-3 model is a combination of data from a modified version of the Common Crawl dataset, some texts from other datasets and some webpage texts\cite{bib8}. A logistic regression classifier based on HashingTF and the standard tokenizer of the PySpark library is used to filter the Common Crawl dataset to re-sample it so that the high quality documents will be prioritized. Before training the GPT-3 model, Text classifier is trained first. For the dataset of the training process of the classifier, the positive examples contains texts from WebText, Wikipedia and the web books by the authors of \cite{bib8}, and the negative examples are from the unfiltered Common Crawl dataset. After the classifier is trained, it is used to filter the Common Crawl dataset. It generates a score of a text in the dataset by the classifier, denoted as $document\_score$, and then uses the following inequality to decide whether to keep or discard the text in the dataset:
	\begin{equation}
		\text{np.random.pareto}(9)>1-\text{document\_score}
	\end{equation}
	The text is kept when the inequality is True. The texts in the dataset are applied by the classifier and checked by the inequality one by one.

	The training data, including the filtered Common Crawl dataset, also go through a fuzzy deduplication to remove the texts that are highly similar to the other texts in the dataset. The fuzzy deduplication is done by a 10-hash MinHashLSH implementation by the PySpark library. The deduplication is also applied on the filtered Common Crawl dataset and WebText to remove the similar texts between the two datasets. About $10\%$ of the texts in the datasets are removed after the deduplications, as mentioned in \cite{bib8}. 

	Due to the auto-regressive characteristic of GPT-3, the cross-entropy loss is chosen as the cost function. Let $V$ denote the vocabulary size. For token position $t$ in a sequence, the cross-entropy loss is
	\begin{equation}
		L_t=-\sum\limits_{i=1}^{V}y_{ti}\log_2\hat{y}_{ti},
	\end{equation}
	where $\boldsymbol{y}_t=(y_{t1},y_{t2},\ldots,y_{tV})$ is the correct one-hot distribution and $\hat{\boldsymbol{y}}_t=(\hat{y}_{t1},\hat{y}_{t2},\ldots,\hat{y}_{tV})$ is the probability distribution predicted by the model. The total loss of a sequence is
	\begin{equation}
		L=\frac{1}{m}\sum\limits_{t=1}^mL_t
	\end{equation}
	where $m$ is the length of the sequence. The training process of the GPT-3 model uses an Adam optimizer to minimize the cost function $L$, and the parameters of the optimizer are $\beta_1=0.9, \beta_2=0.95$ and $\epsilon=10^{-8}$. To make the cost function decreases more smoothly, the learning rate of the optimizer has a Linear warm-up and a consine decay . For the first $3.75\times10^8$ tokens, a linear LR warm-up is used by the authors to make the learning rate grows up gradually, and a cosine decay is also used to generally decreases the learning rate to make the cost function converge to a smaller value. The learning rate matches the following equation:
	\begin{equation}
		\eta_t=
		\begin{cases}
			\frac{0.9}{3.75\times10^8}\eta_0t & t\leq3.75\times10^8 \\
			\frac{1}{2}\left(1+0.8\cos{\frac{t\pi}{2.6\times10^{11}}}\right)\eta_0 & t>3.75\times10^8
		\end{cases}
		\label{methods:equ_learning_rate}
	\end{equation}
	where $\eta_0$ is a constant, and $t$ is the number of the tokens that have been trained. After $2.6\times10^{11}$ tokens are used for training, the learning rate decreases to $0.1\eta_0$. The batch size gradually grows up during the training process, from $3.2\times10^4$ at the beginning to $1.2\times10^{10}$ in the end. 

	For each training loop, the tokens for training are randomly picked up from the datasets without replacement. The length of the sequence is fixed by $n_{ctx}=2048$ tokens, and the sequences that are shorter than 2048 tokens are combined with other short sequences during the training. It is no need to mask the different sub-sequences in a combined sequence, as mentioned in \cite{bib8}, only a special token need to be defined to represent the end of a sequence to delimit the different sub-sequences. The authors claimed that these operations can improve the computing efficiency during the training process, because there is no need to design any specialized training method for the sequences in different length.

	We have defined the circuit structures of our quantum large model in the previous chapters. The parameters in these circuits must also be trained. In this work, we maximize the fidelity between the target state and the output state of the model. For the $k$-th token position in the sequence $Y$, the corresponding fidelity is
	\begin{equation}
		F_Y^{(k)}=|\langle y_k\vert\hat{y}_k\rangle|^2=\operatorname{Tr}(\rho_k\hat{\rho}_k),
	\end{equation}
	where $\ket{y_k}$ is the amplitude encoding of the correct one-hot target distribution and $\ket{\hat{y}_k}$ is the output state of the quantum model. The density operators $\rho_k=\ket{y_k}\!\bra{y_k}$ and $\hat{\rho}_k=\ket{\hat{y}_k}\!\bra{\hat{y}_k}$ correspond to the target and predicted states, respectively. Here, $d=2^n$ denotes the Hilbert-space dimension of the $n$ qubits and therefore the model dimension. Fidelity can be expressed as a matrix trace and evaluated as the expectation value of the Hermitian operator $\rho_k$, making it a natural objective in quantum machine learning. Maximizing fidelity plays a role analogous to minimizing cross-entropy, with both objectives favoring the correct output state. The total fidelity for a sequence is
	\begin{equation}
		F_Y = \frac{1}{m} \sum\limits_{k=1}^{m} F_Y^{(k)}.
	\end{equation}

	The training objective is to maximize the reward of the entire sequence. However, the well-known backpropagation algorithm, which is widely used in classical machine learning, cannot be directly implemented on quantum hardware in a reasonable time and qubit resource consuming. When computing the gradients of the parameters in a given layer of a neural network, backpropagation requires access to the outputs of that layer, which correspond to intermediate states of the network. In quantum settings, however, the intermediate states of parameterized quantum circuits cannot be accessed without destroying the quantum state and interrupting the execution of the circuit. According to the measurement postulate of quantum mechanics, any measurement collapses the quantum state. Although repeatedly interrupt the circuit and reconstruct the intermediate states by large number of measurements is theoretically feasible under the assumption of being able to access numerous copies of the quantum circuit and not caring about the time consuming \cite{bib39}, it is not applicable for large quantum circuits like quantum large models. Therefore, gradients of parameters in quantum circuits must be estimated without relying on intermediate states.

	\subsubsection{Reinforcement Learning Training mechanism without gradient back-propagation}
	\label{methods:MARL-QCT1}
	To optimize the previously defined circuit without accessing its intermediate quantum states, we formulate the parameter adjustment as a Reinforcement Learning problem \cite{bib37}. SPSA\cite{spall1992spsa} tunes the parameters to a moderate fidelity before Reinforcement Learning takes part. At decision step $t$, $\boldsymbol{\theta}_t=(\theta_{t,1},\ldots,\theta_{t,N_\theta})^{\mathsf T}\in[-\pi,\pi)^{N_\theta}$ denotes the current values of all $N_\theta$ trainable circuit parameters, and a sequence $Y_t$ is sampled from the dataset. We write the previously defined sequence fidelity evaluated under $\boldsymbol{\theta}_t$ as $F_{Y_t}(\boldsymbol{\theta}_t)$. The state of the Reinforcement Learning problem is defined as
	\begin{equation}
        \begin{split}
            S_t &=\left[\sin\boldsymbol{\theta}_t,\cos\boldsymbol{\theta}_t,\boldsymbol{d}^{\mathrm{fresh}}_t,\boldsymbol{d}^{\mathrm{mom}}_t,\boldsymbol{z}_t\right],
        \end{split}
		\label{methods:single_agent_RL_state}
	\end{equation}
	where the sine and cosine functions encode the periodic parameters without introducing an artificial discontinuity between $-\pi$ and $\pi$, $\boldsymbol{d}^{\mathrm{fresh}}_t = \mathrm{sgn}(\widehat{\boldsymbol{g}}_t)$ is the gradient direction estimated by modified simultaneous perturbation, $\widehat{\boldsymbol{g}}_t$ is the modified simultaneous perturbation gradient estimation and can be computed by
    \begin{equation}
        \widehat{\boldsymbol{g}}_t=\frac{F_{Y_t}\!\left(\boldsymbol{\theta}_t+2\rho_t\boldsymbol{\Delta}_t\right)-F_{Y_t}\!\left(\boldsymbol{\theta}_t\right)}{2\rho_t}\boldsymbol{\Delta}_t.
    \end{equation}
    Here, $\boldsymbol{\Delta}_t$ is the random perturbation vector, and $\rho_t$ is the perturbation size. The modification of simultaneous perturbation is to let $F_{Y_t}\!\left(\boldsymbol{\theta}_t\right)$ be reused later to save circuit callings. $\boldsymbol{d}^{\mathrm{mom}}_t=\operatorname{clip}\!\left(\frac{\widehat{\boldsymbol{m}}_t}{\sqrt{\widehat{\boldsymbol{v}}_t}+\varepsilon_m},-1,1\right)$ is the gradient direction smoothed by Adam optimizer from $\boldsymbol{d}^{\mathrm{fresh}}_t$, where 
    \begin{equation}
        \widehat{\boldsymbol{m}}_t=\frac{\beta_1\boldsymbol{m}_{t-1}+(1-\beta_1)\widehat{\boldsymbol{g}}_t}{1-\beta_1^{t+1}}, \qquad \widehat{\boldsymbol{v}}_t=\frac{\beta_2\boldsymbol{v}_{t-1}+(1-\beta_2)\left(\widehat{\boldsymbol{g}}_t\odot\widehat{\boldsymbol{g}}_t\right)}{1-\beta_2^{t+1}}.
	\end{equation}
    $\boldsymbol{z}_t$ is a context vector containing the current fidelity $F_{Y_t}$, the normalized update scale $\alpha_t>0$, the normalized perturbation magnitude $\rho_t$, the previous action $a_{t-1}$ and reward $r_{t-1}$ explained later. The action set is
	\begin{equation}
		\mathcal{A}_t=\left\{(\boldsymbol{\kappa},s)| \boldsymbol{\kappa}\in\{\boldsymbol{0},\boldsymbol{d}^{\mathrm{fresh}}_t,\boldsymbol{d}^{\mathrm{mom}}_t\}, s\in\{0.25,0.5,1,2\}\right\}.
		\label{methods:RL_action_set}
	\end{equation}
	For an action $a_t=(\boldsymbol{\kappa}(a_t),s(a_t))\in\mathcal{A}_t$, $\boldsymbol{\kappa}({a_t})$ denotes the selected direction and $s({a_t})$ denotes its step multiplier. The candidate parameters are therefore
	\begin{equation}
		\boldsymbol{\theta}'_t(a_t)=\operatorname{wrap}_{[-\pi,\pi)}\!\left(
			\boldsymbol{\theta}_t+\alpha_t s(a_t)\boldsymbol{\kappa}(a_t)\right),
		\label{methods:single_agent_RL_action}
	\end{equation}
	where $\operatorname{wrap}_{[-\pi,\pi)}$ maps every updated angle into $[-\pi,\pi)$. We judge whether $a_t$ is a beneficial action by comparing the fidelity before and after the action: if $F_{Y_t}(\boldsymbol{\theta}'_t(a_t))>F_{Y_t}(\boldsymbol{\theta}_t)$, we accept $a_t$ and the new parameters; otherwise, we reject $a_t$ and keep the original parameters. Explicitly, the parameters at time $t+1$ are
	\begin{equation}
		\boldsymbol{\theta}_{t+1}=
		\begin{cases}
			\boldsymbol{\theta}'_t(a_t), &  F_{Y_t}(\boldsymbol{\theta}'_t(a_t))>F_{Y_t}(\boldsymbol{\theta}_t)\\
			\boldsymbol{\theta}_t, & \text{otherwise}
		\end{cases},
		\label{methods:RL_guarded_transition}
	\end{equation}
	Here $F_{Y_t}\!\left(\boldsymbol{\theta}_t\right)$ reuses the former measured value during modified simultaneous perturbation. It rejects harmful actions and prevents the parameter trajectory from moving away from a better observed state. The single step reward is
	\begin{equation}
		r(a_t)=
		\begin{cases}
			\tanh\!\left[c_R\dfrac{F_{Y_t}(\boldsymbol{\theta}'_t(a_t))-F_{Y_t}(\boldsymbol{\theta}_t)}{\alpha_t}\right], & a_t\neq \boldsymbol{0}\\[6pt]
			-c_{\mathrm{idle}}, & a_t=\boldsymbol{0},\ F_{Y_t}(\boldsymbol{\theta}_t)<F_{\mathrm{tar}}\\
			0, & a_t=\boldsymbol{0},\ F_{Y_t}(\boldsymbol{\theta}_t)\geq F_{\mathrm{tar}}
		\end{cases},
		\label{methods:RL_reward}
	\end{equation}
	where $c_R>0$ rescales the fidelity change, $c_{\mathrm{idle}}>0$ penalizes premature inactivity, and $F_{\mathrm{tar}}$ is the desired stopping fidelity. For a finite decision horizon $T$, the learning objective is maximizing
	\begin{equation}
		J(\pi)=
		\mathbb{E}_{\pi}\!\left[
		\sum_{t=0}^{+\infty}\gamma^tr(a_t)\right],
		\label{methods:RL_objective}
	\end{equation}
	where $\gamma$ is the discount factor. A dueling DQN\cite{wang2016dueling} is used to find the best parameter tuning policy. The state transitions $(S_t,a_t,r_t(a_t), S_{t+1})$ are stored in replay memory $\mathscr{R}$. For a random mini-batch $\mathcal{M}\subset\mathscr{R}$, the DQN is trained with
	\begin{equation}
		\mathcal{L}=
		\frac{1}{|\mathcal{M}|}
		\sum_{(S_t,a_t,r,S_{t+1})\in\mathcal{M}}
		\ell_{\mathrm{H}}\!\left(
		Q(S_t,a_t;\pi),(r+\gamma Q(S_{t+1},a_{t+1};\pi)
		\right),
		\label{methods:contextual_DQN_loss}
	\end{equation}
	where $Q(S_t,a_t;\pi)$ represents the estimation of the future cumulative reward 
    \begin{equation}
        Q(S_t,a_t;\pi)=\mathbb{E}_\pi[\sum\limits_{i=0}^{+\infty}\gamma^ir(a_{t+i})\mid S_t,a_t],
    \end{equation}
    and $\ell_{\mathrm{H}}$ is the smooth $L_1$ loss. Minimizing $\mathcal{L}$ trains the DQN to rank the keep, fresh-direction, and momentum-direction actions according to their measured fidelity improvements.

	\subsubsection{Multi-agent expansion for large scale circuits}
	\label{methods:MARL-QCT2}
	To scale this formulation to circuits with large parameter sets, we partition the parameter indices into $M$ disjoint blocks $\mathcal{B}_1,\ldots,\mathcal{B}_M$ and assign one independent dueling DQN to each block. Agent $i$ owns block $\mathcal{B}_i$, and the active index $i_t\in\{1,\ldots,M\}$ rotates through the blocks. The simultaneous perturbation gradient estimation is restricted to the parameters in the active block $\boldsymbol{\theta}_{t,\mathcal{B}_{i_t}}$. The state for agent $i_t$ is
	\begin{equation}
		S_t^{(i_t)}=\left[
		\sin\boldsymbol{\theta}_{t,\mathcal{B}_{i_t}},
		\cos\boldsymbol{\theta}_{t,\mathcal{B}_{i_t}},
		\boldsymbol{d}^{\mathrm{fresh}}_{t,\mathcal{B}_{i_t}},
		\boldsymbol{d}^{\mathrm{mom}}_{t,\mathcal{B}_{i_t}},
		\sin\boldsymbol{\theta}_t,
		\cos\boldsymbol{\theta}_t,
		\boldsymbol{z}^{(i_t)}_t
		\right],
		\label{methods:contextual_DQN_state}
	\end{equation}
	where $\boldsymbol{z}^{(i_t)}_t$ not only contains the information mentioned in the previous chapter but also the normalized active block identity. Each agent uses the same action set in Eq.~\eqref{methods:RL_action_set}. The candidate parameters generated by an action is
	\begin{equation}
		\begin{split}
			\boldsymbol{\theta}'_{t,\mathcal{B}_{i_t}}(a_{t,\mathcal{B}_{i_t}})
			&=\operatorname{wrap}_{[-\pi,\pi)}\!\left(
				\boldsymbol{\theta}_{t,\mathcal{B}_{i_t}}+
				\alpha_t s(a_{t,\mathcal{B}_{i_t}})\boldsymbol{\kappa}(a_{t,\mathcal{B}_{i_t}})\right),\\
			\boldsymbol{\theta}'_{t,\overline{\mathcal{B}}_{i_t}}(a_{t,\mathcal{B}_{i_t}})
			&=\boldsymbol{\theta}_{t,\overline{\mathcal{B}}_{i_t}},
		\end{split}
		\label{methods:block_DQN_action}
	\end{equation}
	where $\overline{\mathcal{B}}_{i_t}=\{1,\ldots,N_\theta\}\setminus\mathcal{B}_{i_t}$ is the complement of the active block, and all parameters outside the active block remain unchanged. The new parameters are determined by Eq.~\eqref{methods:RL_guarded_transition}. Dueling DQN agent $i$ stores its own replay memory $\mathscr{R}_i$, and applies Eq.~\eqref{methods:contextual_DQN_loss} with $Q$, $\boldsymbol{\phi}$, $\mathscr{R}$, and $\mathcal{M}$ replaced by $Q_i$, $\boldsymbol{\phi}_i$, $\mathscr{R}_i$, and $\mathcal{M}_i$, respectively, thereby defining its loss $\mathcal{L}_i$. The $M$ block policies collectively implement the overall policy $\pi$ in Eq.~\eqref{methods:RL_objective}. The agents act sequentially on the common circuit, so an accepted update by one agent becomes part of the global parameter context observed by subsequent agents. Counterfactual evaluations train the active DQN to compare its block actions. The complete process forms the Multi-Agent Reinforcement Learning for Quantum Circuit Training (MARL-QCT) algorithm, and is summarized in Algorithm~\ref{methods:algorithm-RLPQC}.

	\begin{algorithm}[htbp]
		\caption{Multi-Agent Reinforcement Learning for Quantum Circuits Training}
		\begin{algorithmic}
			\Require {Initial parameters $\boldsymbol{\theta}_{\mathrm{ini}}$, training dataset $\mathscr{Y}$, block number $M$, decision horizon $T$, replay mini-batch size $B$, counterfactual period $K_{\mathrm{cf}}$, and initial scales $\epsilon_0$, $\alpha_0$}
			\Ensure {Final circuit parameters $\boldsymbol{\theta}_{f}$}
			\State $\boldsymbol{\theta}_0\gets\operatorname{SPSA}(\boldsymbol{\theta}_{\mathrm{ini}},\mathscr{Y})$, $\epsilon\gets\epsilon_0$
			\State Partition $\{1,\ldots,N_\theta\}$ into $\mathcal{B}_1,\ldots,\mathcal{B}_M$
			\For{$i=1$ to $M$}
			\State Initialize $Q_i(\cdot,\cdot;\boldsymbol{\phi}_i)$ and replay memory $\mathscr{R}_i$
			\EndFor
			\For{$t=0$ to $T-1$}
			\State $i_t\gets 1+(t\bmod M)$
			\State Sample $Y_t$ from $\mathscr{Y}$
			\State Update $\boldsymbol{d}^{\mathrm{fresh}}_{t,\mathcal{B}_{i_t}}$ and $\boldsymbol{d}^{\mathrm{mom}}_{t,\mathcal{B}_{i_t}}$ by modified local simultaneous perturbation
			\State Construct $S_t^{(i_t)}$ using Eq.~\eqref{methods:contextual_DQN_state}
			\State Select $a_{t,\mathcal{B}_{i_t}}\in\mathcal{A}_t$ with the $\epsilon$-greedy policy of $Q_{i_t}$
			\State $F_{Y_t}(\boldsymbol{\theta}_t)\gets F_{Y_t}(\boldsymbol{\theta}_t)$
			\State Construct $\boldsymbol{\theta}'_t(a_{t,\mathcal{B}_{i_t}})$ using Eq.~\eqref{methods:block_DQN_action}
			\State $F_t^{a_{t,\mathcal{B}_{i_t}}}\gets F_{Y_t}(\boldsymbol{\theta}'_t(a_{t,\mathcal{B}_{i_t}}))$
			\State Compute $r_t(a_{t,\mathcal{B}_{i_t}})$ using Eq.~\eqref{methods:RL_reward}
			\State Store $(S_t^{(i_t)},a_{t,\mathcal{B}_{i_t}},r_t(a_{t,\mathcal{B}_{i_t}}))$ in $\mathscr{R}_{i_t}$
			\If{$a_{t,\mathcal{B}_{i_t}}=\boldsymbol{0}$}
			\State $\boldsymbol{\theta}_{t+1}\gets\boldsymbol{\theta}_t$
			\ElsIf{$F_t^{a_{t,\mathcal{B}_{i_t}}}>F_{Y_t}(\boldsymbol{\theta}_t)$}
			\State $\boldsymbol{\theta}_{t+1}\gets\boldsymbol{\theta}'_t(a_{t,\mathcal{B}_{i_t}})$
			\Else
			\State $\boldsymbol{\theta}_{t+1}\gets\boldsymbol{\theta}_t$
			\EndIf
			\If{$t\bmod K_{\mathrm{cf}}=0$}
			\State Select $\mathcal{C}_t\subset\mathcal{A}_t\setminus\{a_{t,\mathcal{B}_{i_t}},\boldsymbol{0}\}$
			\ForAll{$\widetilde{a}\in\mathcal{C}_t$}
			\State Construct $\boldsymbol{\theta}'_t(\widetilde{a})$
			\State $F_t(\widetilde{a})\gets F_{Y_t}(\boldsymbol{\theta}'_t(\widetilde{a}))$
			\State Compute $r_t(\widetilde{a})$ and store $(S_t^{(i_t)},\widetilde{a},r_t(\widetilde{a}))$ in $\mathscr{R}_{i_t}$
			\EndFor
			\EndIf
			\If{$|\mathscr{R}_{i_t}|\geq B$}
			\State Sample $\mathcal{M}_{i_t}\subset\mathscr{R}_{i_t}$ with $|\mathcal{M}_{i_t}|=B$
			\State Compute $\mathcal{L}_{i_t}$ using Eq.~\eqref{methods:contextual_DQN_loss}
			\State Update $\boldsymbol{\phi}_{i_t}$ using $\nabla_{\boldsymbol{\phi}_{i_t}}\mathcal{L}_{i_t}$
			\EndIf
			\State Update $\epsilon$ and $\alpha_{t+1}$
			\EndFor
			\State $\boldsymbol{\theta}_{f}\gets\boldsymbol{\theta}_T$
		\end{algorithmic}
		\label{methods:algorithm-RLPQC}
	\end{algorithm}
	\FloatBarrier

	\subsection{Analysis of the Time Complexity}
    \label{methods:time_comp_analy}

	We examined the time complexity of our quantum large model and compare it with that of the classical GPT model. The time complexity of our model is subject to the quantum circuit depth. When computing the circuit depth, we first decompose our circuit into the universal quantum gate set, including the single qubit gates like rotation gates, phase-shift gates and Hadamard gates, and the CNOT gate. We also regard the depth of the controlled rotation gates as 1 because it is explicitly marked as supported by the quantum chips of Tianyan platform (see Chapter \ref{methods:Tianyan_exps}). This ensures that the time complexity analysis approximate to the real condition of executing our model on physical quantum hardwares. Table \ref{methods:tab_time_complexity} shows the quantum circuit depth and the corresponding time complexity magnitude of the 3 layers of our quantum large model, compared with that of the classical Transformer.
	\begin{table}[htbp]
		\centering
		\begin{tabular}{lllll}
			\toprule
			Layer & Circuit depth & Time complexity & \makecell{Time complexity of \\ classical Transformer}\\
			\midrule
			Encoding & \makecell{Subject to the \\ quantum hardware} & $O(m + (\log_2{d})^2)$ in worst case & $O(md)$ \\
			Self-Attention & Eq. \ref{methods:depth_of_self_attention} & $O((m+p)\log_2d-p\log_2p)$ & $O(m^2 d)$ \\
			FFN & Eq. \ref{methods:depth_of_FFN} & $O(\log_2d)$ & $O(md^2)$ \\
			\bottomrule
		\end{tabular}
		\caption{The time complexity of the encoding layer, the self-attention layer and the FFN layer of our quantum large model, compared with the classical Transformer. Our quantum large model achieves an exponential acceleration in the self-attention layer and the FFN layer while keeping the relatively same time consuming magnitude with the classical ones.}
		\label{methods:tab_time_complexity}
	\end{table}

	First is the encoding layer. It consists of three components: querying the lookup table, quantum state preparation, and positional encoding. For a sequence of length $m$ and $n=\log_2d$ qubits per token, the querying process involves $m$ queries to classical storage devices, such as memory or hard drives, which results in a time complexity of $O(m)$. Here, $d$ is the dimension of the quantum model, equivalent to the dimension of the state vector of the qubits. Quantum state preparation varies depending on hardware support and the use of dynamic quantum circuits. If the quantum computer supports the operation of the state to be prepared, the process requires only $O(1)$ time. Otherwise, the state preparation must be decomposed into a quantum circuit with only the supported quantum gates, which typically has a time complexity ranging from $O(n)$ to $O(n^2)$, depending on the quantum gate capabilities. Here, $n = \log_2{d}$ represents the number of qubits for each token. In general, the worst-case time complexity for preparing a single state is about $O(n^2)$. As quantum states for all tokens in a sequence can be prepared simultaneously, the time complexity for preparing the entire sequence is also $O(n^2)$ in the worst case. Positional encoding is executed by a quantum circuit containing simultaneous quantum rotation gates, yielding a time complexity of $O(1)$ for the entire sequence. Due to the Quantum No-Cloning Theorem, the quantum state preparation and positional encoding must be executed $3m-2$ times. However, since these processes can be performed in parallel, they do not affect the overall time complexity. Therefore, the time complexity of the encoding layer is $O(m + n^2) = O(m + (\log_2{d})^2)$. In comparison, the encoding layer of the classical GPT model has a time complexity of $O(md)$, with the positional encoding process being the bottleneck.

	Next is the self-attention layer. It consists of three parts: the attention head circuits $G_Q$, $G_K$, $G_V$, the quantum mixing circuits, and the Softmax circuit. The attention head circuits $G_Q$, $G_K$, and $G_V$ share the same structure and can be executed simultaneously, as the encoding layer outputs $3m-2$ copies of the encoded qubits. Thus, the total complexity is equivalent to the complexity of $G_Q$. For a $G_Q$ with $p$ attention heads, the circuit contains $p$ controlled circuits $C_7^{(i)}G_{Q_i}$, $1\leq i\leq p$, which can only be executed sequentially. Each $C_7^{(i)}G_{Q_i}$ consists of the three subcircuits $C_7^{(i)}U_{Q_i}$, $C_7^{(i)}D_{Q_i}$ and $C_7^{(i)}V_{Q_i}$. $U_{Q_i}$ and $V_{Q_i}$ are two $\lceil n - \log_2{p} \rceil$-qubit circuits. By choosing $p$ as a power of 2, we eliminate the ceiling function, optimizing the efficiency of $G_{Q_i}$. As shown in Fig. ~\ref{methods:fig_unitary_layer_7} and Chapter \ref{methods:section_self_attention}, $U_{Q_i}$ and $V_{Q_i}$ consist $n_{SUa}$ $SU(16)$ gate layers and $n_{SUa}-1$ bridge layers. The depth of each $SU(16)$ gate is 126, and the depth of a single bridge layer is $2\lceil\frac{n - \log_2{p}}{2}\rceil=n - \log_2{p}+\frac{1-(-1)^{n - \log_2{p}}}{2}$ when $n - \log_2{p}$ is larger than (not including) 4 and smaller than (and not including) 8, so $U_{Q_i}$ and $V_{Q_i}$ have a depth of $126n_{SUa}+\left(n - \log_2{p}+\frac{1-(-1)^{n - \log_2{p}}}{2}\right)(n_{SUa}-1)$. $D_{Q_i}$ consists of 3 quantum gates and a measurement operation on the auxiliary qubit, and 2 controlled diagonal unitary gates. The controlled diagonal unitary gates have the depth of $n+1$\cite{bib17,bib18,bib19}, so the depth of $D_{Q_i}$ is $2n+6$. The controlled diagonal circuit $C_7^{(i)}D_{Q_i}$ shown in Fig. \ref{methods:fig_controlled_non_unitary_diag} has twice the depth of $D_{Q_i}$. Meanwhile, it takes two runs in expectation to make the circuit succeeded under the RUS protocol, so the depth of $C_7^{(i)}D_{Q_i}$ is $8n+24$ in expectation. Therefore, the total depth of the circuit $G_Q$ is $\left(252n_{SUa}+\left(2n - 2\log_2{p}+1-(-1)^{n - \log_2{p}}\right)(n_{SUa}-1)+8n+24\right)p$, that is $\bigl(2(n_{SUa}+3)n+2(1-n_{SUa})\log_2p+\left(253-(-1)^{(n-\log_2p)}\right)n_{SUa}+\allowbreak(-1)^{(n-\log_2p)}+23\bigr)p$, as well as $G_K$ and $G_V$. The quantum mixing circuits consist of SwapTest circuits \cite{bib12,bib13} and Strong Entangling circuits \cite{bib11}. The depth of each SwapTest circuit is $n + 2$, while the depth of each Strong Entangling circuit is $n + 3$. All SwapTest circuits can be executed simultaneously, as the attention head circuits output $m - 1$ copies of the key and query states, and $m$ copies of the value states. Thus, the SwapTest circuits contribute a depth of $n + 2$ to the self-attention layer. However, only Strong Entangling circuits that prepare different $\ket{\psi_{VA_j}}$ states can be executed simultaneously, meaning that $m - 1$ Strong Entangling circuits must be executed sequentially, adding a depth of $(m - 1)(n + 3)$. Finally, the Softmax circuit consists of two sequentially executed unitary layer with $n_{SUs}$ $SU(16)$ layers and $n_{SUs}-1$ bridge layers respectively, each with the structure shown in Fig.~\ref{methods:fig_unitary_layer_14}, apart by a bridge layer across all the data qubits and the ancillary qubit, followed by a measurement operation with post-selection. The depth of the bridge layer is $2\lfloor \frac{n}{2}\rfloor=n-\frac{1-(-1)^n}{2}$ when $n$ is larger than (and including) 8, so the depth of the Softmax circuit is $\left(252n_{SUs}+\left(2n-1+(-1)^n\right)\right)(n_{SUs}-1)+n+1-\frac{1-(-1)^{(n+1)}}{2}+1$, alternatively $(2n_{SUs}-1)n+(251+(-1)^n)n_{SUs}+\frac{5-3(-1)^n}{2}$. Therefore, the total depth of the self-attention layer is the sum of the $G_Q$, SwapTest circuits, Strong Entangling circuits and the Softmax circuit:
	\begin{equation}
		\label{methods:depth_of_self_attention}
		\begin{aligned}
            &\left[m+2(n_{SUa}+3)p+2n_{SUs}-1\right]n+3m+2(1-n_{SUa})p\log_2 p\\
            &\quad+\left[\left(253-(-1)^{n-\log_2 p}\right)n_{SUa}
            +(-1)^{n-\log_2 p}+23\right]p\\
            &\quad+\left(251+(-1)^n\right)n_{SUs}
            +\frac{3}{2}\left(1-(-1)^n\right), n>4,4<n-\log_2p<8
        \end{aligned}
	\end{equation}
	which results in a time complexity of approximately $O(mn+pn-p\log_2p) = O((m+p)\log_2d-p\log_2p)$ since $n_{SUa}$ and $n_{SUs}$ are constants larger than 1. In comparison, the self-attention layer of the classical GPT-3 model has a time complexity of $O(m^2 d)$, driven by the process of multiplying the attention head matrices $W_Q$, $W_K$, $W_V$ and mixing the value vectors according to the inner products of the query and key states.

	Finally is the feed-forward layer. It has 2 SVD decomposition structure, each containing 2 unitary layers with the depth of $126n_{SUf}+\left(n-\frac{1-(-1)^n}{2}\right)(n_{SUf}-1)$ and a non-unitary diagonal layer with the depth of $2(2n+6)$ by expectation. Beside that, it has 2 Sigmoid circuits with the depth of $(2n_{SUs}-1)n+(251+(-1)^n)n_{SUs}+\frac{5-3(-1)^n}{2}$. As a result, the total depth of the feed-forward layer is:
	\begin{equation}
		\label{methods:depth_of_FFN}
		\begin{aligned}
            (4n_{SUf}+4n_{SUs}+2)n+2\left(251+(-1)^n\right)(n_{SUf}+n_{SUs})+31-5(-1)^n,n>4.
        \end{aligned}
	\end{equation}
	results in a time complexity of approximately $O(n)=O(\log_2d)$. In comparison, the time complexity of the feed-forward layer of GPT-3 model (and other models that base on the Transformer architecture) has a time complexity of $O(md^2)$.

	\subsection{Experiments}

	We did 3 parts of experiments to verify the effectiveness and advantage of our quantum large model. First, we successfully implemented a small scale model on a quantum computer of the Tianyan Quantum Computing Platform, and compared its time consuming with the classical Transformer in similar scale. Second, we used our MARL-QCT algorithm to train some component circuits in our quantum large model, and compared the training performance between SOTA quantum circuit training algorithm. Finally, we trained a small scale quantum model to learn 2 tasks that are widely used to verify the expressivity and generalization ability of generative large model architectures under small scale. 

	\subsubsection{Inference Speed Test on Quantum Hardware}
    \label{methods:Tianyan_exps}

    We tested the inference time consuming on the quantum computer Tianyan-287 of Tianyan Quantum Computing Platform~\cite{bib50}. It is the most advanced quantum chip in China, containing the largest amount of data qubit number (105 qubits) among the quantum chips that being able to execute universal quantum operations. We implemented a quantum large model with the largest scale capable with Tianyan-287: 4 qubits per token with sequence length $m=3$, consuming 92 qubits. To scale down the quantum large model into the capability of Tianyan-287, we set 4 attention heads in the self-attention layer. In the attention-head circuits $G_{Q_i}$, $G_{K_i}$ and $G_{V_i}$, we use the controlled $SU(4)$ circuits as the unitary parts in the SVD circuit structure shown in Fig. \ref{methods:fig_attention_svd}. In the FFN layer, $SU(16)$ circuits act as the unitary parts of the SVD circuit structure in $G_{W_i}$. We tested the time consuming of the self-attention layer and the FFN layer of our model on Tianyan-287.  We collected 2 types of data: the usage time of the quantum circuit tasks on Tianyan-287, and the quantum circuit execution time. Table \ref{methods:tab_tianyan_time} shows the chip occupation time and the circuit execution time of the self-attention layer and FFN layer. We also timed the self-attention layer and the FFN layer of a classical Transformer with model dimension $2^4=16$ and sequence length $m=3$, corresponding to the scale of the quantum large model implemented on Tianyan-287, on a classical server equipped with 2 Intel Xeon Gold 5218 CPU, 768GB RAM and an NVIDIA H200 NVL GPU.
	\begin{table}[htbp]
		\centering
		\begin{tabular}{llllll}
			\toprule
			Layer & \multicolumn{2}{c}{Theoretical calculation number} & \multicolumn{2}{c}{Measured execution time} & \makecell{Chip usage\\time (s)}\\ 
             & classical & quantum & classical & quantum &\\
            \midrule
			Self-Attention & 206208 & 315 & 0.321 & 0.452 & 202.706 \\
			FFN & 385536 & 1044 & 0.183 & 0.303 & 83.482 \\
			\bottomrule
		\end{tabular}
		\caption{The time complexity and time consuming of the implemented quantum large model on the Tianyan-287 quantum chip, with the scale of $d=16$, $m=3$, compared with the classical Transformer with the same scale. For the chip usage time, Each task contains 10 shots of circuit executions, and we repeat all the experiments 3 times. The table shows the average.}
		\label{methods:tab_tianyan_time}
	\end{table}

    The quantum circuit depth of the self-attention layer and the FFN layer of our quantum large model after scale reduction is 315 and 1044 respectively. In comparison, in the classical Transformer with the same scale, the self-attention layer and the FFN layer contains 206208 and 385536 times of calculation respectively, excluding simultaneous calculations. The number of calculations of our quantum large model is thus 300 to 1000 times fewer than the classical under in the same scale. In Tab. \ref{methods:tab_tianyan_time}, the chip usage time is directly acquired from the Tianyan Platform. It takes all the time consuming related to the tasks into account, including the platform's internal computation of qubit topology, the resource scheduling, the initialization of the qubits, the encoding of the parameterized quantum operations, the execution of the quantum circuits, etc.. The actual execution time of the quantum circuits is derived from the aggregation of the duration of the longest quantum gate of each layer. The duration of the quantum gates are updated in real time by Tianyan Platform. The durations of a single qubit gate varies from 28ns to 56ns, subject to the parameter value of the gates. The duration of 2 qubit gates are longer in average, CNOT gate varies from 25ns to 38ns with the average of 30.81ns. Since a single logical gate in classical modern integrated circuits takes only about 1-20ps, which is about $10^3$ times faster, the quantum circuit execution time and the classical Transformer execution time are similar.

	\subsubsection{Training Quantum Circuits with MARL-QCT}

    Tianyan-287 is an NISQ device that cannot handle the task of training of deep quantum circuits by MARL-QCT due to its noise. To evaluate the proposed training approach and its scalability, we simulated applying the MARL-QCT algorithm to some important component circuits in the self-attention layer and the FFN layer on a classical server, and compared the performance with 2 SOTA algorithms, SPSA~\cite{spall1992spsa} and parameter-shift~\cite{bib16}. Specifically, we tuned the parameters of a general four-qubit circuit ($\mathrm{SU}(16)$ circuit) and an attention projection circuit $G_Q$ with the scale of qubit number $n=6$ and attention head number $p=4$, respectively, using our MARL-QCT algorithm to let the circuits approximate a given $SU(16)$ transformation and a 64-dimensional transformations with 4 diagonal blocks. For each experiment, The datasets consist of batches of input quantum states generated randomly, paired with the corresponding target states transformed from the initial states by the transformation. These experiments verified the scaling ability and the distinct advantage of MARL-QCT algorithm we proposed, compared with the SOTA algorithms.

	In these experiments, the three compared algorithms (MARL-QCT, SPSA and parameter-shift) start tuning the parameters from a rough tuned parameter with the fidelity of about 0.8. We compared 2 matrices: the final fidelity the algorithms reach under a fixed tuning iteration number, and the iteration numbers the algorithms need to reach a given target fidelity. When tuning the $SU(16)$ circuit, 17 agents control the interleaved blocks of the 255 parameters of the $\mathrm{SU}(16)$ circuit in cyclic order. In the $G_Q$ task, 51 agents tune the parameters in similar way.

    After training, MARL-QCT managed to reach the fidelity of 0.95 in 24000 iterations starting from rough tuned parameters in the $SU(4)$ task. In comparison, MARL-QCT made 143.99\% and 57.87\% more fidelity improvement compared with SPSA and parameter-shift in the same iteration number of 24000 beginning from the rough tuned parameter. Meanwhile, to reach the same fidelity 0.95, MARL-QCT only takes 33.71\% circuit callings of parameter-shift, while SPSA never reached this fidelity througout the experiments. These results confirmed the advantage of MARL-QCT compared to SPSA and parameter-shift. Parameter-shift evaluates the gradients of the parameters one-by-one. The sequential gradient evaluation makes it extremely slow when tuning large scale of parameters. In contrast, SPSA is able to estimate the gradients of the parameters in parallel and is fast in low fidelity conditions, but it has difficulty to convergent to a high fidelity because of the noise in its gradient estimations which grows quickly with the scale of the parameter number. In the $G_Q$ task, MARL-QCT reached the fidelity of 0.956 in 36000 iterations. In comparison, MARL-QCT made 42.41\% and 34.72\% more fidelity improvement compared with SPSA and parameter-shift in the same iteration of 36000, and reached the fidelity of 0.95 by only 22.78\% iterations of the parameter-shift, while SPSA still never reached this fidelity througout the experiments. These results verify the stability of the speed advantage of MARL-QCT to the parameter-shift: although the growth of the parameter number slows down all the 3 algorithms, making the advantage of MARL-QCT under fixed iterations less apparent, the advantage of the circuit calling time of MARL-QCT compared to the parameter-shift becomes stronger. As the comparison experiment results of the 2 tasks show, MARL-QCT managed to achieve a high fidelity without the large time consuming of the parameter-shift, addressing the exclusiveness of high fidelity and low time consuming in SOTA algorithms, and strengthen its speed advantage in reaching a fidelity target with the growth of the parameter number.

	\begin{figure}[htbp]
		\centering
		\includegraphics[width=\linewidth]{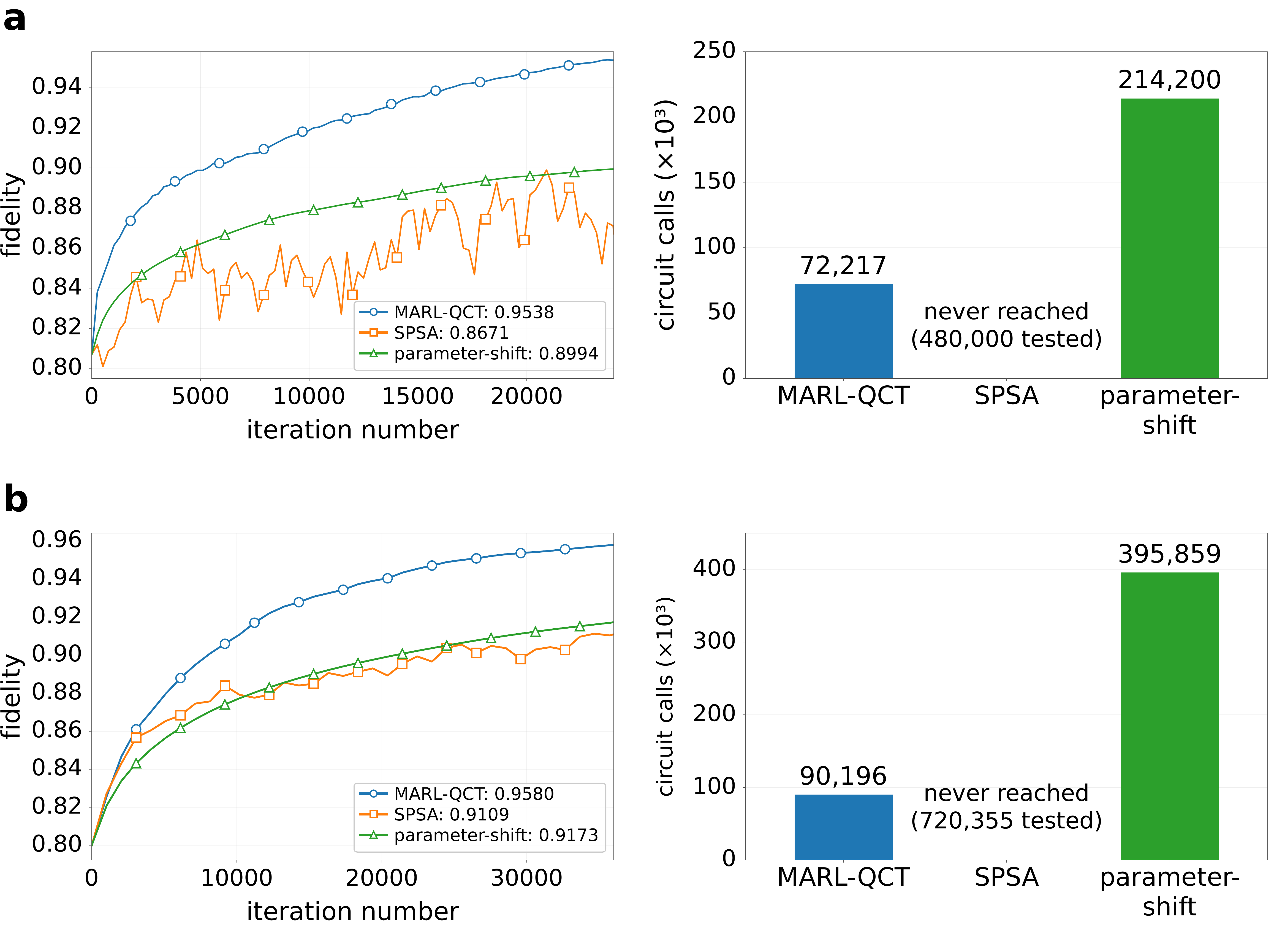}
		\caption{\textbf{Comparisons of the performance of MARL-QCT, SPSA and parameter-shift on quantum circuit parameter tuning}. The experiment results show that MARL-QCT is able to reach a higher fidelity with sufficiently lower time consuming, establishing a distinct advantage compared to the SOTA algorithm. (a) Results for the $\mathrm{SU}(16)$ circuit task, tuning 255 parameters. Starting from the rough tuned parameters with fidelity 0.806949, MARL-QCT reaches the fidelity 0.953760 in 24000 iterations, while SPSA and parameter-shift reach 0.867119 and 0.899941. MARL-QCT thus gives a 143.99\% and 57.87\% more fidelity improvement. Meanwhile, MARL-QCT takes only 33.71\% circuit calls of the parameter-shift to reach the target fidelity 0.95, while SPSA cannot reach the target fidelity because of the gradient estimation noise. (b) Results for the $G_Q$ circuit task, tuning 1020 parameters. Starting from the rough tuned parameters with fidelity 0.800053, MARL-QCT can reach the fidelity 0.957973 in 36000 iterations, while SPSA and parameter-shift reach 0.910945 and 0.917277. MARL-QCT thus gives a 42.41\% and 34.72\% more fidelity improvement. Meanwhile, MARL-QCT takes only 22.78\% circuit callings of the parameter-shift to reach the target fidelity 0.95, while SPSA still cannot reach the target fidelity.}
		\label{methods:fig_training_comparison}
	\end{figure}
	\FloatBarrier

	\subsubsection{Sequence-to-Sequence Task Tests}

\begin{figure}[htbp]
		\centering
		\includegraphics[width=\linewidth]{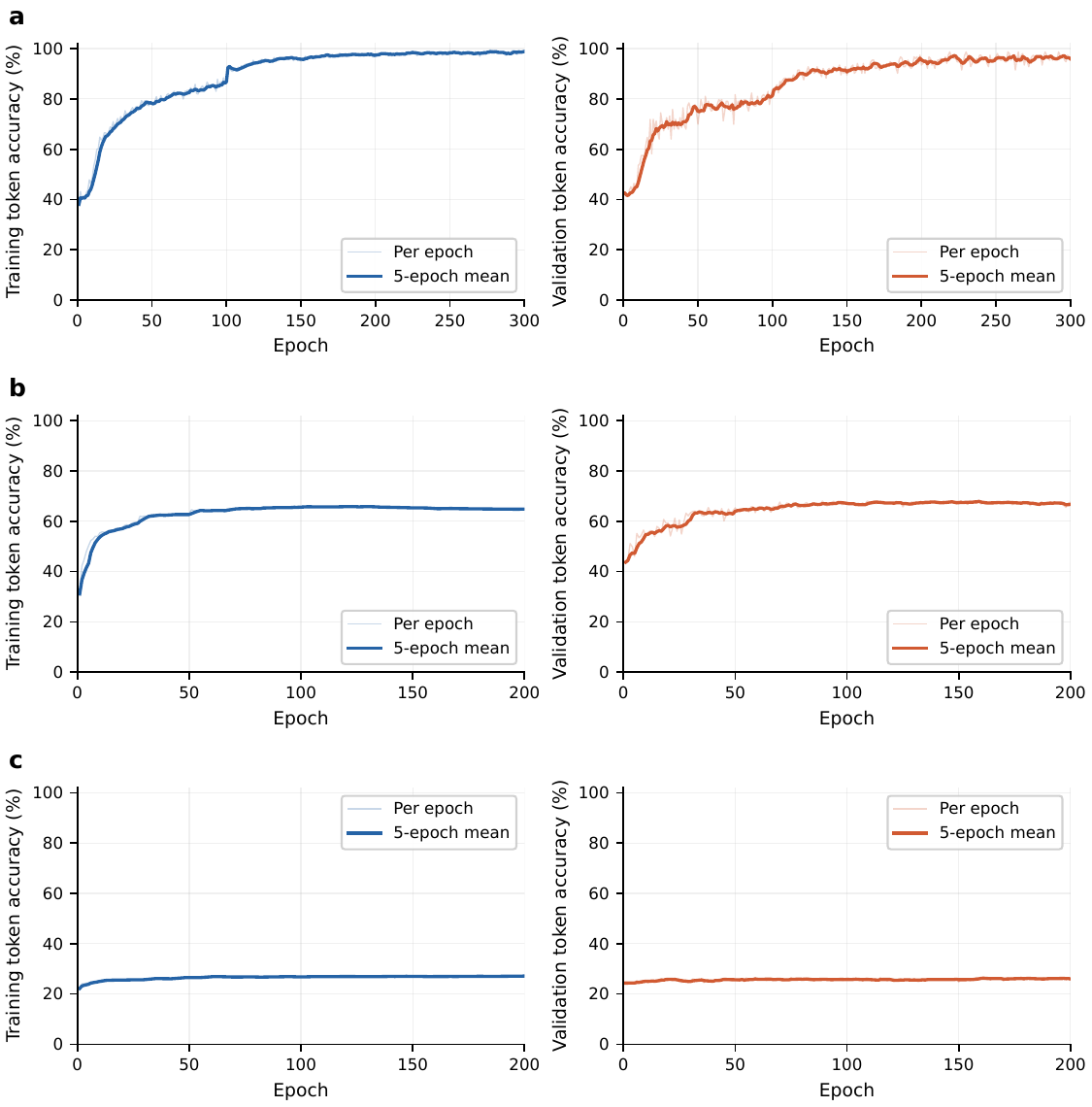}
		\caption{Teacher-forced token accuracy for Dyck, SCAN-length and PCFG task. Figure a, b and c show Dyck-4, SCAN-length and PCFG, respectively, with training on the left and validation on the right. The validation token accuracy for the 3 tasks reaches 96.69\%, 67.56\% and 25.76\% respectively. The results shows that our quantum large model under the tested scale can learn the Dyck task effectively, sometimes produce error when encountering complex input sequence analysis in SCAN-length task, while not able to learn the PCFG task due to a small model scale.}
		\label{methods:fig_sequence_task_results}
	\end{figure}

	To verify the expressivity and generalization ability, we test the proposed quantum large model with 3 qubits per token and sequence length $m=6$ on the Dyck-4 Balanced Brackets task, the SCAN-length task and the Probabilistic Context-Free Grammar task. The model uses one self-attention layer with 2 heads, one FFN layer and three entangling layers, with 877 trainable parameters. The Balanced Brackets task tests the model's ability to track nested dependencies and generate structurally valid completions. Successful completion requires remembering which brackets remain open and closing them in a reversed order, as in a stack. We test our model on short sequences within fixed length and bounds of the nesting depth. Specifically, we use a Dyck-4 dataset with four bracket types: parentheses, square brackets, braces and angle brackets. Each input is a valid but incomplete bracket sequence, and the target is the sequence of brackets needed to close it, followed by the end-of-sequence token. The dataset contains 210 training examples, 49 validation examples and 49 test examples, with no identical input shared between them. Input sequences contain at most four bracket tokens, targets contain at most three, and the maximum nesting depth is three. As shown in Fig.~\ref{methods:fig_sequence_task_results}(a), with training on the left and validation on the right, it obtains 96.69\% teacher-forced token accuracy, which shows that the model can learn the bounded bracket-completion mapping and generate the corresponding structured outputs autoregressively.

	The SCAN-length task\cite{lake2018scan} tests the model's ability to generalize learned rules of command sequences to action sequences, even longer than those observed during training. The model must reuse the meanings of learned actions, modifiers and conjunctions to produce the required action order and repetitions at longer output length. The task maps commands similar to natural language to action sequences, like `walk', `jump and walk' and `turn left twice after jump' map to `I\_WALK', `I\_JUMP I\_WALK' and `I\_JUMP I\_TURN\_LEFT I\_TURN\_LEFT', respectively. We use the SCAN-length split dataset with 2,000 training examples, 200 validation examples and 500 test examples. Inputs contain at most six tokens. Training and validation targets contain at most five actions, whereas every test target contains exactly six actions, excluding the end-of-sequence token. The validation set therefore remains the maximal sequence length of the training set, while the test set is longer. As shown in Fig.~\ref{methods:fig_sequence_task_results}(b), the model can reach 67.56\% teacher-forced token accuracy, which is lower than that of the Dyck task since the task requires analysis to the input sequences that are more complex than those in the Dyck task, like the quantity analysis to the words like "twice", “three times", etc., and re-ordering the actions when the words like "after" appear.

	The Probabilistic Context-Free Grammar (PCFG) task tests the model's ability to learn symbolic sequence transformations and execute nested operations while retaining symbol identities and producing the required output order. For a nested expression, the model must evaluate the innermost function and use its result as the input to the enclosing function. The task maps expressions containing function tokens and atomic symbols to the symbol sequences obtained by executing their operations. We use short expressions from the train, development and test sets of standard PCFG SET\cite{hupkes2020compositionality}, using 2,000 training examples, 120 validation examples and 150 test examples. Input sequences contain three to five tokens, and target sequences contain two to five symbols, excluding the end-of-sequence token. The subset contains six unary functions, \texttt{copy}, \texttt{echo}, \texttt{repeat}, \texttt{reverse}, \texttt{shift} and \texttt{swap\_first\_last}, with up to three nested functions per expression. For example, \texttt{reverse S099 S080 S001} maps to \texttt{S001 S080 S099}. The original atomic vocabulary is mapped to 116 symbols while avoiding collisions between distinct atoms within an entry. Binary functions are absent under these length constraints. All sets use the same length limits. As shown in Fig.~\ref{methods:fig_sequence_task_results}(c), the model can reach 25.76\% teacher-forced token accuracy, which does not match the requirements to solve this task, probably resulted by the difficulty of the task and the small scale of the parameter number.
	\FloatBarrier
	
\section{Data availability}
The data supporting the findings of this study, including the chip usage data of Tianyan-287, the data of the quantum circuit tuning experiments by MARL-QCT (including the comparisons to the 2 SOTA algorithms), and the data of the training simulations of the 3 sequence-to-sequence tasks, have been deposited in an unpublished Zenodo record. Editors and reviewers can access and download these files during peer review through the following private link: \href{https://zenodo.org/records/23720084?preview=1&token=eyJhbGciOiJIUzUxMiJ9.eyJpZCI6IjE5NjY2ZWJlLTRjY2EtNGQ2Ni1iN2FlLTk0OGYxYjYxYWE0OSIsImRhdGEiOnt9LCJyYW5kb20iOiIzYzQ3NDAyMjNlZmU5MDNmYWQxOTVhNDBmMWE5YzZhOCJ9.6Dh6XkDRySTdHAl4_rtmPb94_vEyz87cY3SvV3JfaBOxgo2Y5jLI1egs__O5AicgP7Mr5OPe9BMOZATY3ugWAQ}{here}. The dataset will be made publicly available on Zenodo no later than publication of the article, and the persistent DOI will be provided in the final manuscript.

\section{Code availability}
The codes for the experiments of MARL-QCT, the Tianyan-287 inference executions, and the sequence-to-sequence tasks are also available in the unpublished Zenodo record: \href{https://zenodo.org/records/23720084?preview=1&token=eyJhbGciOiJIUzUxMiJ9.eyJpZCI6IjE5NjY2ZWJlLTRjY2EtNGQ2Ni1iN2FlLTk0OGYxYjYxYWE0OSIsImRhdGEiOnt9LCJyYW5kb20iOiIzYzQ3NDAyMjNlZmU5MDNmYWQxOTVhNDBmMWE5YzZhOCJ9.6Dh6XkDRySTdHAl4_rtmPb94_vEyz87cY3SvV3JfaBOxgo2Y5jLI1egs__O5AicgP7Mr5OPe9BMOZATY3ugWAQ}{here}.

\section{Funding}
This work was supported in part by the National Natural Science Foundation of China (Grant Nos. 62173315, 62373342, 62101525, 62341113, 62201543, and U23A20275), the Fundamental Research Funds for the Central Universities (Grant No. WK2102026004), and the China Environment for Network Innovations (CENI; Grant No. 2016-000052-73-01-000515).

\section{Acknowledgments}
The authors would like to thank ‌CHINA TELECOM QUANTUM TECHNOLOGY CO.,LTD.‌ for providing access and technical support for the quantum chips of Tianyan Platform.

\section{Author contributions}
Tianze Zhu designed the whole architecture of the quantum large model and the experiments. Xiaofeng Jiang suggested this research, as well as the idea of the nonlinear function approximation circuits and combining quantum circuit training with multi-agent Reinforcement Learning. He also revised the article. Min Liao wrote the codes for the quantum circuit structure of the self-attention layer, and implemented the sequence-to-sequence task experiments. Luying Zhang wrote the codes for the quantum circuit structure of the feed-forward layer, and part of the codes for the Tianyan inference executions. Ruixi Liu collected and processed the datasets for the sequence-to-sequence task experiments. Sen Kuang gave some constructive advices on the designing of the model structure. Other authors gave ideas or supports on some points of the model designing and experiments respectively.

\section{Competing interests}
The authors declare no competing interests.


\clearpage
\renewcommand{\refname}{Methods references}


\end{document}